\documentclass[twocolumn,showpacs,preprintnumbers,amsmath,amssymb,pra]{revtex4}
\usepackage{graphicx}
\usepackage{dcolumn}
\usepackage[tight]{subfigure}
\usepackage{amsmath}
\usepackage{verbatim}
\usepackage[usenames,dvipsnames]{color}
\usepackage{bm} 
\usepackage{bbm}
\newcommand{\assign}{:=}

\newcommand{\Figure}[2]{
  \begin{figure}[ht]
    \includegraphics[width=0.9\linewidth]{#1}
    \caption{#2}
  \end{figure}
}
\expandafter\newcommand\csname Figure2\endcsname[3]{
\begin{figure}[ht]
  \includegraphics[width=0.9\linewidth]{#1}
  \\
  \includegraphics[width=0.9\linewidth]{#2}
  \caption{#3}
\end{figure}
}

\newcommand{\wideeq}[1]{\begin{widetext}
    #1
  \end{widetext}}

\newcommand{\Fig}[1]{Fig.~\ref{#1}}

\newcommand{\Eq}[1]{Eq.~(\ref{#1})}
\newcommand{\eq}[1]{(\ref{#1})}
\newcommand{\Sec}[1]{Sec.~\ref{#1}}
\renewcommand{\sec}[1]{\ref{#1}}
\newcommand{\App}[1]{Appendix~\ref{#1}}

\newcommand{\Cite}[1]{Ref.~\onlinecite{#1}}

\newcommand{\new}[1]{{\textcolor{blue}{#1}}}

\newcommand{\newer}[1]{{\textcolor{blue}{#1}}}
\newcommand{\braggio}[1]{{\textcolor{blue}{#1}}}
\newcommand{\todo}[1]{ \textbf{\textcolor{Bittersweet}{TODO: #1}} }
\newcommand{\cut}[1]{ \textbf{\textcolor{Bittersweet}{CUT: #1}} }
\newcommand{\hide}[1]{ \textbf{\textcolor{Gray}{#1}} }
\newcommand{\tmtextit}[1]{ \textit{#1}}
\newcommand{\nocomma}{}
\newcommand{\tmem}[1]{{\em #1\/}}
\newcommand{\tmop}[1]{\ensuremath{\operatorname{#1}}}
\newcommand{\nobracket}{}

\renewcommand{\new}[1]{{#1}}
\renewcommand{\braggio}[1]{{#1}}
\renewcommand{\newer}[1]{{#1}}
\renewcommand{\cut}[1]{}
\renewcommand{\todo}[1]{}
\renewcommand{\hide}[1]{}
\newcommand{\op}[1]{\hat{#1}}
\renewcommand{\vec}[1]{\mathbf{#1}}
\newcommand{\vecg}[1]{{\bm #1}}

\begin{document}
\title{
  Qubit quantum-dot sensors: Noise cancellation by coherent backaction,\\
  initial slips, and elliptical precession
}

\author{M. Hell$^{(1,2)}$}
\author{M. R. Wegewijs$^{(1,2,3)}$}
\author{D. P. DiVincenzo$^{(1,2,4)}$}
\affiliation{
  (1) Peter Gr{\"u}nberg Institut,
      Forschungszentrum J{\"u}lich, 52425 J{\"u}lich,  Germany
  \\
  (2) JARA- Fundamentals of Future Information Technology
  \\
  (3) Institute for Theory of Statistical Physics,
      RWTH Aachen, 52056 Aachen,  Germany
  \\
  (4) Institute for Quantum Information,
      RWTH Aachen, 52056 Aachen,  Germany
}
\date{\today}

\begin{abstract}
  We theoretically investigate the backaction of a sensor quantum dot with strong local Coulomb repulsion on the
  transient dynamics of a qubit that is probed capacitively. We show that
  the measurement backaction induced by the noise of electron cotunneling
  through the sensor is surprisingly mitigated by the recently identified
  \tmtextit{coherent} backaction [M. Hell, M. R.
Wegewijs, and D. P. DiVincenzo, Phys. Rev. B {\textbf{89}}, 195405 (2014)] arising from
  quantum fluctuations. This indicates that a sensor with quantized states may
  be switched off better than naively expected. This renormalization
  effect is missing in semiclassical stochastic fluctuator models and
  typically also in Born-Markov approaches, which try to avoid the calculation
  of the nonstationary, nonequilibrium state of the qubit  \tmtextit{plus}
  sensor. Technically, we integrate out the current-carrying electrodes to obtain
  kinetic equations for the joint, nonequilibrium detector-qubit dynamics. We
  show that the sensor current response, level renormalization, cotunneling
  broadening, and leading non-Markovian corrections always appear together and
  cannot be turned off individually in an experiment or ignored theoretically.
  We analyze the backaction on the reduced qubit state -- capturing the full
  non-Markovian effects imposed by the sensor \newer{quantum dot} on the qubit -- by applying a Liouville-space
  decomposition into quasistationary and rapidly decaying modes. Importantly, the
  sensor cannot be eliminated completely even in the simplest
  high-temperature, weak-measurement limit since the qubit state experiences
  an initial slip depending on the initial \newer{preparation} of
 qubit\tmtextit{plus} sensor quantum dot.
 The slip persists over many qubit cycles, i.e., also on the
  time scale of the qubit decoherence induced by the backaction.
  \cut{Although a \emph{quantum}-dot sensor has the advantage of suppressing backaction,}
  \newer{A \emph{quantum}-dot sensor can thus} not be modeled as usual as a ``black box'' without accounting
 for its dynamical variables;
  it is part of the quantum circuit.
  We furthermore find that the
  Bloch vector relaxes (rate $1 / T_1$) along an axis that is \emph{not}
  orthogonal to the plane in which the Bloch vector dephases (rate $1 / T_2$),
  blurring the notions of relaxation and dephasing times. Moreover, the
  precessional motion of the Bloch vector is distorted into an {\tmem{ellipse}} in
  the tilted dephasing plane.
\end{abstract}
\pacs{73.63.Kv, 73.63.-b, 03.65.Yz} \maketitle




\section{Introduction}\label{sec:intro}

The ongoing effort to mitigate the qubit decoherence due to environmental noise
detrimental to quantum computing has made substantial progress by identifying
well-isolated two-level systems {\cite{Cirac95,Tyryshkin03,Petta05,Koch07}}
and developing efficient decoupling techniques
{\cite{Bluhm10,Bylander11,Rong11}}. However, active readout elements must be
integrated into any quantum computer and ``noise'' from such sensors may soon
become a relevant source of errors. Therefore, the
unavoidable disturbance of the qubit evolution during a readout process gains
importance, both for single-shot qubit readout {\cite{Hanson05,Barthel10}} as
well as continuous qubit monitoring {\cite{Ilichev03}}. \newer{Another
question of central
importance for qubit manipulations is} which parameters should be varied to
switch off a sensor most effectively when no measurement is intended to be made.

In a single-shot measurement, the goal is to achieve a strong, projective
measurement that dephases the qubit state as quickly as possible. However, any
measurement still takes a finite time and relaxation processes
{\cite{Barthel10}}, excitation processes {\cite{Onac06}}, and incoherent
detector dynamics limit the detector efficiency
{\cite{Korotkov01,Gurvitz05,Clerk04,Korotkov08}}. For continuously monitoring
the qubit evolution, by contrast, one has to realize a {\tmem{weak}}
measurement. The aim is here to disturb the qubit evolution as weakly as
possible to retain the (partial) purity of the quantum state {\cite{Oxtoby06}},
while avoiding quantum jumps and the quantum Zeno effect {\cite{Goan01}}.
Understanding the backaction of a weak measurement is therefore of great
interest.

In this paper, we focus on the noninvasive, weak-measurement backaction
exerted on a qubit by a capacitively coupled sensor quantum dot (SQD)
{\cite{Schoelkopf98,Barthel10}}. SQDs are attractive qubit detectors due to
their strong tunability and higher signal-to-noise ratio as compared to
quantum point contacts {\cite{Elzerman04,Reilly07}} and dispersive readout
schemes {\cite{Colless13}}.
This derives from the fact that an SQD is an {\tmem{interacting}} quantum
system. \newer{For smaller quantum dots (QDs) than typically used for readout, the
electrons may even occupy {\tmem{discrete}} energy
levels.}
The required readout current easily leads to a strong nonequilibrium and nonstationary sensor state.
This altogether makes the description of the backaction arising from an SQD on a qubit challenging.
Thus, typical weak-coupling
approaches to decoherence assuming an environment in equilibrium with a
continuous spectrum
{\cite{Paladino02,Grishin05,Bergli09,Shnirman05,Ithier05,Chirolli08,Galperin04}}
cannot be applied here and naive extensions are prone to errors as we will illustrate. Understanding
the intertwined evolution of SQD and qubit is of key importance to understand
the measurement backaction
{\cite{Shnirman98,Makhlin01a,Gurvitz05,Gurvitz08,Hell14a}}.

Out of these challenges arises the question which physical effects have to be
included for a consistent description of the measurement backaction of a
sensor QD on a qubit. For the qubit, we consider the simplest model that
involves capacitive readout, a charge qubit, realized as a double quantum dot.
An important aspect lies in the type of setup considered, namely that of
indirect detection: one measures the conductance of a sensor QD in the attached
electrodes, while only the SQD capacitively interacts with the qubit, see \Fig{fig:model}. To maximize
the sensitivity, the SQD is operated at the threshold to the Coulomb blockade
regime. Here, the conductance of the SQD shows the strongest response to small
qubit-induced level shifts. Previous studies addressing the backaction of an
SQD on a qubit in this regime focused mainly on the lowest-order approximation
in the tunnel coupling $\Gamma$ of the SQD to the attached electrodes
{\cite{Shnirman98,Makhlin00,Makhlin01a,Gurvitz05,Gurvitz08,Oxtoby06}},
strictly valid \newer{when operated} in the single-electron tunneling (SET) regime. There
are two main reasons for going beyond this approximation. 

The first reason
relates to the backaction on the qubit and its dependence on the level
position, experimentally controlled by gate voltages. This is important
since the level position is one of the key experimental control parameters by which one can try to
switch off the sensor backaction. In this approximation, the leading-order rates (SET $\propto
\Gamma$) become exponentially small when the level position ($\varepsilon$) of
the SQD is tuned away from the electrochemical potentials of the electrodes $(\mu_r)$.
Thus, when one is interested in the backaction at the onset of Coulomb
blockade, next-to-leading order $\propto \Gamma^2$ cotunneling
processes should also be accounted for because they are only algebraically
suppressed, scaling as $1 / (\varepsilon - \mu_r)$. One would thus naively
expect that the backaction is suppressed only inversely proportional to the
detuning from resonance. Yet, level-renormalization effects should also be
considered; they lead to level shifts that depend logarithmically on the level
position $\varepsilon$ and therefore the \tmtextit{response} of the
\tmtextit{level renormalization} to the measurement perturbation also scales
as $1 / (\varepsilon - \mu_r)$. A central finding of our study is that
level-renormalization effects in fact \tmtextit{mitigate} the naively expected cotunneling
decoherence.

A second reason \newer{for going beyond the lowest-order approximation} comes in view when one takes into account the experimentally used sensor
signal: if
one accounts for the first nonvanishing contributions $\propto \lambda \Gamma
/ T$ that produce a nonzero sensor signal, one has to include also
renormalization effects since they appear in the same order. Basically, if one
has time to measure, one has time to fluctuate as well. As already emphasized
in {\color{black} Ref. {\cite{Hell14a}}}, incorporating terms $\propto \lambda
\Gamma / T$ is another reason that forces us to keep also cotunneling
processes $\propto \Gamma^2 / T$ since in the weak-measurement limit $\lambda
\ll \Gamma$ the latter are larger. Only when one is \emph{not} interested in the sensor
current, one can consistently neglect cotunneling and renormalization effects by taking the high-temperature limit:
as we show, they must either be kept or neglected together.

The above-mentioned processes combine in a nontrivial way to give three types
of backaction on the qubit {\cite{Hell14a}}. First, both SET and cotunneling
processes contribute to a stochastic switching of the SQD charge state. This
switching generates a randomly fluctuating effective magnetic field acting on
the qubit Bloch or isospin vector {\cite{Ithier05,Chirolli08}}. This ``noise''
-- called here the {\tmem{stochastic backaction}} -- leads to a shrinking of
the Bloch vector, i.e., to decoherence. In addition, there is a
{\tmem{dissipative backaction}}, which is the flip-side of the measurement
action: it arises whenever one accounts for a nonzero response of the sensor SET tunnel rates to
the qubit state and therefore a nonzero sensor signal. Finally, there
is a {\tmem{coherent backaction}}, the most striking finding of {\color{black}
Ref. {\cite{Hell14a}}}. It arises from the above-mentioned
level-renormalization response and translates into torque terms involving the
qubit Bloch vector. These torques and related precession effects are similar
to those emerging in various other QD transport setups: It is well-known that
tunneling processes can produce exchange fields leading to an (iso)spin
precession in the context of spintronics {\cite{Koenig03,Braun04set}}, double
dots {\cite{Wunsch05}}, molecular quantum dots {\cite{Donarini06,Schultz10}},
and superconducting devices {\cite{Governale08}}. All these
level-renormalization effects arise from quantum fluctuations of electrons by
tunneling into the attached electrodes. In this respect, a qubit coupled to a sensor QD is not different.

An interesting question is how these different types of backaction
relate to the information gained during the measurement process.
Clearly, the quasistationary time-dependent current through the sensor contains
information about the qubit state and at the same time causes
decoherence of the qubit. However, besides this fundamentally unavoidable
backaction, the decoherence induced by the detector can be
stronger. 
This can be formulated in terms of general inequalities relating the noise of the
measured operator (here the position of the qubit electron) to the noise of the measurement
signal (here the current) and additional noise cross terms
{\cite{Clerk10}}.
The situation considered in this paper is far away from the quantum limit (meaning
the above-mentioned inequality is far from being satisfied with
equality).\\ 
A part of our work actually
focuses on a simple limit when only the stochastic backaction is
accounted for but dissipative and coherent backaction are neglected.
The SQD then acts rather as an ``ordinary''
environment and no information is obtained during the operation -- a
situation relevant when the detector is supposed to be switched
off. Yet, even in this simple limit, there are effects beyond the scope
of the picture \new{developed} in {\Cite{Clerk10}}: the fast relaxation due to switching on the sensor may
affect the qubit state also in a way that depends on the initial \emph{dynamical variables}
of the sensor. This is not captured at all by the noise inequalities mentioned
above. The explicit time evolution of the sensor, which is not considered in {\Cite{Clerk10}}, is thus also crucial to understand
the backaction on the qubit. An interesting question arising from
this insight is how this
backaction effect has to be assessed in view of the information gain. Our
work could thus spur new
activity on the topic of information gain during the measurement.\\
\newer{The impact of the three backaction effects on the full \tmem{transient
dynamics} of the qubit so far remained an outstanding question that we address in
this article (the analysis in {\color{black} Ref. {\cite{Hell14a}}} was
restricted to the stationary state). Our analysis is divided into two parts.}\\
\tmtextit{Part 1}. The main point of this paper is to eliminate the electrodes'
degrees of freedom {\cite{Makhlin00}} and  to analyze the transient dynamics within the resulting physical
picture \new{of} the coupled SQD-qubit dynamics. In this way, we can deal
with the nontrivial interplay of the SQD-qubit coherence, strong local Coulomb
interaction in the SQD, nonequilibrium conditions imposed by the attached
electrodes, as well as both leading (SET) and next-to-leading order effects in the
tunneling (cotunneling). The necessary inclusion of the latter furthermore
forces us to extend {\color{black} Ref. {\cite{Hell14a}}} by including also
the leading memory effects on the sensor-qubit system due to the tunneling to the electrodes, which is necessary for the
study of the transient qubit dynamics. (For the stationary state, which we
studied in {\color{black} Ref. {\cite{Hell14a}}}, they can be be ignored
without making additional approximations.)
\braggio{The importance of memory effects for the dynamics when going beyond
weak coupling is known especially since \Cite{Braggio06}, see also Refs.
\cite{Splettstoesser06,Flindt08PRL,Flindt08PRB,Braggio09,Splettstoesser10,Goorden04}
and progress for strong backaction
\cite{Hartmann07,Kennes13}. Non-Markovian corrections have also been
studied in related contexts, such as the
backaction of a quantum point contact on a double dot \cite{Hartmann07,Braggio09}
or quantum-feedback control based on quantum measurements
\cite{Rebentrost09,Floether12}. The various effects of
non-Markovian processes remain, however, an uncharted territory \cite{Glaser15}.
\\
The central result in our case are the kinetic equations
(\ref{eq:kineq}) for the system of qubit plus quantum-dot sensor. The equations} reveal the above three-fold nature of the backaction of the sensor
QD on the qubit; importantly, the relevant energy scale for this backaction is
{\tmem{not}} simply the internal capacitive interaction $\lambda$ (SET-induced
stochastic backaction) but additionally involves the energy scale $\Gamma \lambda / T \nocomma$
(dissipative and coherent backaction involving transport processes).

Our kinetic equation furthermore allows us to identify slowly evolving
quasistationary modes -- containing the qubit evolution -- and faster evolving
decay modes reflecting the dissipative SQD dynamics due to its coupling to the electrodes.
The coupling between
these modes generates the total backaction on the qubit and is mediated by
all three types of backaction. To account for all these backaction effects, it
is indispensable to keep the capacitive interaction ($\lambda \neq
0$) when integrating out the electrodes. In this aspect, our work differs from the
otherwise closely related approach of {\color{black} Ref. {\cite{Emary08}}},
which starts out from the assumption that the electrodes affect exclusively the
SQD. There, all backaction effects derive only from the internal interaction,
i.e., the stochastic backaction.

A surprising finding of our analysis is that the total backaction exhibits a
strong reduction when tuning the SQD \new{towards} the Coulomb blockade regime: we find
that the coherent backaction actually cancels the cotunneling (``broadening'')
corrections in the coupling of the quasistationary to the decay modes. This
eliminates the naively expected leading power-law dependence $\propto 1 /
(\varepsilon - \mu_r)$ of the backaction, \newer{affecting also the}
decoherence time scales.
This indicates that a sensor with quantized orbital states can be switched off more efficiently
by controlling its gate voltage than naively expected,
the first important experimental implication of this article. \new{This
requires, however, to prepare
the sensor state in a controlled way to avoid a slip of the qubit state
(see part 2 below).}

It is important to emphasize already here that the coherent backaction,
which is responsible for this mitigation, is not an independent
mechanism that can be ``added'' to counteract cotunneling noise. Instead, it
arises together with cotunneling as an integral part of quantum fluctuation
effects of the qubit-sensor system when consistently describing all types of
backaction. Notably, we show that this mitigation is not captured by widely-used classical
stochastic fluctuator models and can also be easily overlooked in
Born-Markov approaches \newer{that integrate out the entire environment
of the qubit (i.e., including the SQD)}. The effect of the exchange of electrons between the
SQD and the electrodes can thus not be fully captured by classical switching of the SQD
charge state. \newer{We also review and compare in detail our results with earlier
works and pinpoint a number of limitations of standard approaches.}

The prominent role of renormalization effects underlying the coherent backaction
distinguishes a QD sensor with few, discrete energy levels from what can be
expected for a sensor with a continuous energy spectrum. The recent study
{\cite{Lindebaum11}} showed that \newer{similar} torque terms
appearing in a spintronic context
are much suppressed in single-electron transistors (continuous spectrum) as opposed to QDs
(discrete spectrum). In the former case, renormalization effects tend to
nullify when averaging over their continuous energy spectrum.
This motivates the extensive analysis of the
detection of a qubit state by a sensor QD undertaken in this paper. Our
work raises the interesting question \new{to which extent backaction
effects due to renormalization effects
are suppressed} in a single-electron transistor.

\tmtextit{Part 2.} One might think that following the above description one
can in a second step eliminate the sensor QD from the description to
obtain an effective theory for the qubit only. However, already on general
grounds, this is questionable: specific to our indirect detection problem is
that the environment of the qubit is \tmtextit{not stationary}. Moreover, initial
correlations between SQD and qubit -- both microscopic systems -- might exist. When integrating out the
environment, the factorizability and the stationarity of the environment are, however,
often invoked to eliminate the so-called slip of the initial condition
for the subsystem {\cite{Gaspard99}} coming from the initial state of the
environment \newer{and its short-time transient evolution}. To illustrate this point, we analyze in more detail the simpler
high-temperature limit where the complications due to cotunneling and coherent backaction can be
consistently ignored. Even in this high-temperature, weak-measurement
limit the qubit develops a slip of order $\lambda / \Gamma$ on a time scale $t
< 1 / \Gamma$, which is beyond the control over the qubit system
alone. The slip
depends explicitly on the initial \tmtextit{qubit-sensor state}.
The second experimental implication of our work is that the dynamical state of the sensor \emph{and} its correlations with the qubit cannot be ignored and must be brought under experimental control.
This slip effect is cumulative, e.g., it results in phase shifts that still affect the
qubit on much longer time scales $t \sim 1 / \lambda \gg 1 /
\Gamma$ relevant to the readout. By contrast, the relevant time scales
$T_1$ (relaxation) and $T_2$ (dephasing) of the transient qubit dynamics for times $t \gg 1 / \Gamma$
do {\tmem{not}} depend on the initial sensor QD state.\\
The precession axis of the qubit Bloch vector also turns out to be independent of the initial state.
In the simple high-temperature limit, we furthermore identify an additional effect
of the (purely stochastic) backaction, which is to induce a tilt of the Bloch
vector precession axis. We find that the circular isospin precession becomes
slightly elliptical in the presence of the detector, adding as a fingerprint
oscillations to the exponential decay of the qubit-state purity.
\newer{This mixes the notions of relaxation and dephasing as we will
see.} \new{It is an interesting question how these effects behave at low
temperature and strong qubit-sensor coupling where they have
received little attention so far.}

\emph{Outline.} After this topical outline, we now present the organization of  the sections of the paper and the key equations.
In {\Sec{sec:model}}, we briefly review the generic indirect
readout model of \ {\color{black} Ref. {\cite{Hell14a}}} and discuss the
dynamical variables that are needed to describe the mixed quantum state of the
joint qubit-sensor system. This requires \tmtextit{two} isospins, which capture both the reduced qubit state as well as the correlations with the sensor QD.\\
After this, we outline the key technical challenges of our approach in {\color{black} Sec.
\ref{sec:kineq}}, deferring details to the App. \ref{sec:impnm}, and we present the
time-local kinetic equation (\ref{eq:kineq}) for the coupled sensor QD
plus qubit system. \newer{In App. \ref{sec:impnm}, we further \new{discuss} the
importance of including non-Markovian corrections to retain the
positivity of the reduced density operator.}
Without further approximations, we identify the relevant
unperturbed modes ($\lambda = 0$) \newer{with the electrodes integrated
out.}
From the
representation of the kinetic equation (\ref{eq:kineq2}) in these modes we
prove the exact cancellation between the coherent backaction and the
cotunneling (``broadening'') noise [\Eq{eq:r}]. We furthermore study the
implications for the dependence of the total backaction on various
experimentally relevant parameters (tunneling-rate asymmetries, bias, and gate
voltages).
From the formal solution of the effective quasistationary mode evolution
[Eqs. \eq{eq:xqt} and \eq{eq:xqefft}], with details given in Appendix \ref{app:liouville},
we infer that the qubit evolution is non-Markovian and exhibits a slip
of the initial condition \newer{that we characterize in
Appendix \ref{app:slip}}. \new{Initial slips generally go hand in hand with non-Markovian dynamics as Refs.
\cite{Haake83,Geigenmueller83,Haake85,Gaspard99,Flindt08PRL,Wilhelm08} and
the references therein point out. Initial entanglement between a qubit
and its environment, one cause of initial slips, can drastically affect the qubit coherence \cite{Wilhelm08}.}
\\
In {\Sec{sec:results}}, we attempt to integrate
out the sensor QD to derive an effective Liouvillian [\Eq{eq:leff}] that
effectively incorporates its fast switching dynamics. We then
focus on the analytically tractable case of high temperature. \new{This suffices to
illustrate the general importance of initial slips 
in the context of detector backaction} [\Eq{eq:taueff0}],
the breakdown of orthogonality of relaxation and dephasing qubit modes,
and the exponentially damped oscillatory but \emph{elliptical} precession of
the qubit Bloch vector. In this high-temperature limit, we obtain tangible expressions for the qubit
relaxation and dephasing rates, expanded to leading order in $\lambda /
\Gamma$. In {\Sec{sec:comparison}}, we compare our results
with semiclassical stochastic \new{fluctuator models} as well as Born-Markov and exact
quantum approaches \new{to provide} further insight into the origin of the coherent
backaction. In the accompanying Appendix {\color{black}
\ref{app:renormalization}} we show how the coherent backaction affects the
qubit phase evolution in a way that is not accounted for by semiclassical stochastic
fluctuator approaches. We summarize our findings in {\color{black} Sec.
\ref{sec:summary}}.




\begin{center}
  {\Figure{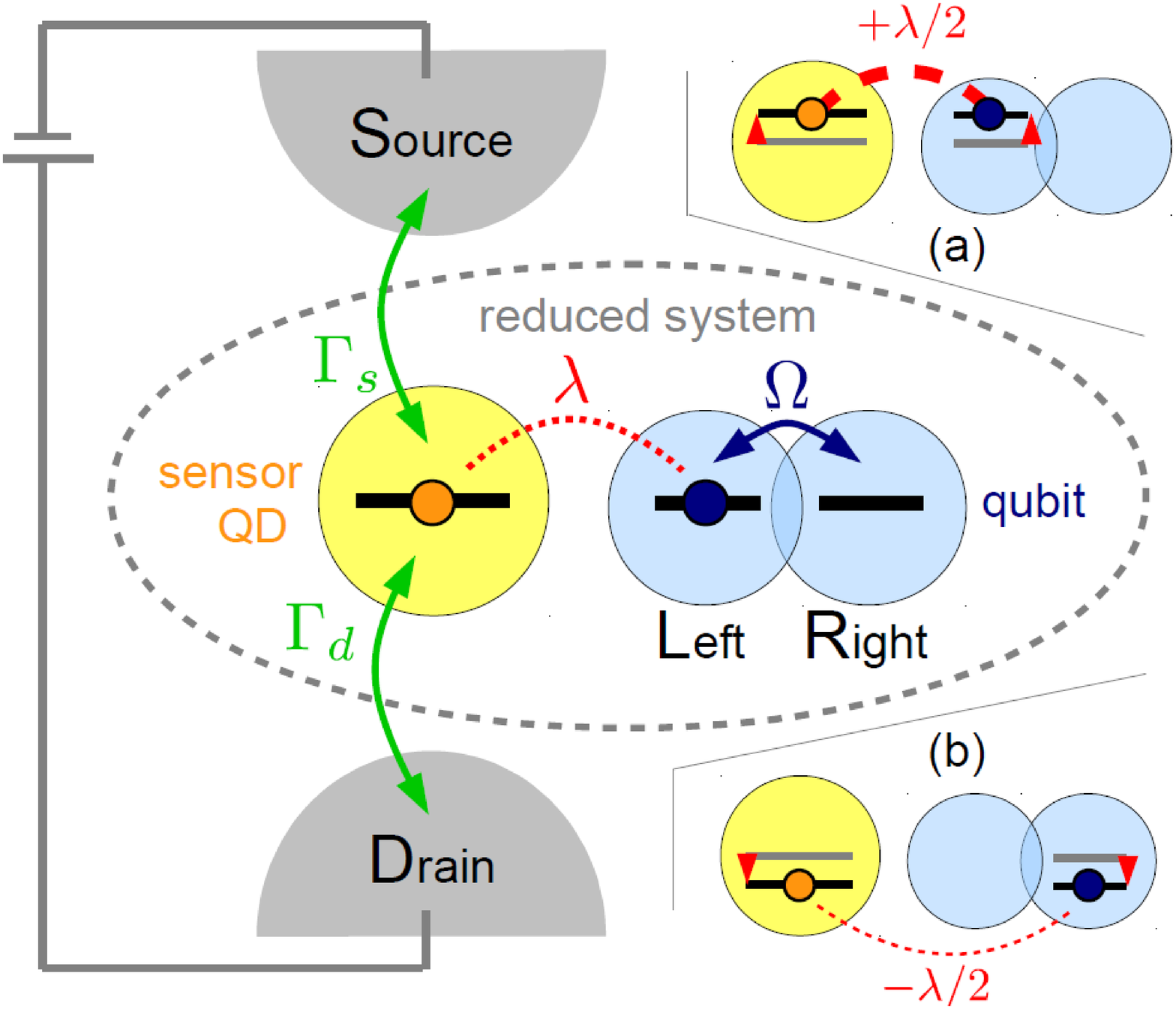}{Sensor
  quantum dot (SQD) tunnel-coupled to source and drain electrodes and
  capacitively coupled to a qubit, whose different logical states involve two
  possible positions, left and right, in a double quantum dot. If the qubit
  electron is left (a) or right (b), the Coulomb repulsion to the SQD electron
  is larger or smaller, respectively, compared to the full delocalization of the
  qubit electron.\label{fig:model} \ }}
\end{center}

\section{Indirect detection}\label{sec:model}

\subsection{Model}
We analyze the indirect detection setup sketched in {\color{black} Fig.
\ref{fig:model}}: the readout ($H_I$) of a double-quantum dot charge qubit
($H_Q$) by a proximal sensor quantum dot ($H_S$), which in turn is read out
($H_T$) by the conductance of the transport current in one of the attached
electrode reservoirs ($H_R$). The model we use, discussed in detail in
{\color{black} Ref. {\cite{Hell14a}}}, thus consists of three ``layers'' with
their respective interactions:
\begin{eqnarray}
  H & = & H_Q + (H_I + H_S) + (H_T + H_R) .  \label{eq:H}
\end{eqnarray}
This models the essential physics found in many experiments on QD qubits and
can be extended to superconducting qubits {\cite{Devoret05}} as well as to
spin qubits if measured by spin-to-charge conversion
{\cite{Elzerman04,Barthel10}}.

\emph{Qubit.}
The qubit is realized as a charge qubit, a single electron occupying a double
quantum dot. This electron can reside either on the left dot, denoted by the
state ${\color{black} |} L {\color{black} \rangle}$, or on the right dot,
denoted by the state ${\color{black} |} R {\color{black} \rangle}$. The qubit
state is represented by the ensemble average ${\color{black} \vecg{\tau}} =
{\color{black} \langle} \op{{\color{black} \vecg{\tau}}} {\color{black}
\rangle}$ of an operator $\op{{\color{black} \vecg{\tau}}}$ (corresponding
to an isospin $\op{{\color{black} \vecg{\tau}}} / 2$) with components
\begin{eqnarray}
  \hat{\tau}_i & = & \sum_{l, l' = L, R} (\sigma_i)_{l l'}  {\color{black} |}
  l {\color{black} \rangle} {\color{black} \langle} l' {\color{black} |} . 
  \label{eq:tau}
\end{eqnarray}
{\color{red} }Here, $\sigma_i$ denotes the Pauli matrix for $i = x, y, z$. The
average $z$ component $\tau_z$ quantifies the imbalance between the
probabilities for finding the qubit electron in the left orbital rather
than in the
right orbital, while $\tau_x$ and $\tau_y$ quantify coherences between the
left and right occupation. The general form of the Hamiltonian of the isolated
qubit is
\begin{equation}
  \begin{array}{lll}
    H_Q & = & \vecg{\Omega} \cdot \op{\vecg{\tau}} \text{/} 2, \label{eq:hqubit}
  \end{array}
\end{equation}
in which the qubit field $\vecg{\Omega}$ is used in applications to control the
qubit evolution: it induces coherent tunneling of the qubit electron between
the two dots ($\Omega_x$, $\Omega_y$) combined with a detuning ($\Omega_z$).
In our analysis, $\vecg{\Omega}$ is constant in time and later on, when we
discuss tangible results, we will chose $\vecg{\Omega} = \Omega \vec{e}_x$.
However, unless stated otherwise, we first keep $\vecg{\Omega}$ general.

\emph{Sensor quantum dot.} The sensor quantum dot (SQD or sensor QD) is modeled as a single, interacting,
spin-degenerate orbital level with Hamiltonian $H_S = \varepsilon \hat{n} + U
\hat{n}_{\uparrow} \hat{n}_{\downarrow}$. Here, $\hat{n}_{\sigma} =
d^{\dag}_{\sigma} d_{\sigma}$ is the number operator for electrons with spin
$\sigma = \uparrow, \downarrow$, where $d_{\sigma}$ denotes the corresponding
field operator, and $\hat{n} = \hat{n}_{\uparrow} + \hat{n}_{\downarrow}$ is
the total electron number operator. We take the Coulomb repulsion energy $U$ here to be
the largest energy scale (except for the bandwidth $2 W$ of the electrodes), in
accordance with typical experimental situations. We therefore exclude the
double occupation of the sensor QD orbital in the following. This \ allows us
to reduce the model to
\begin{equation}
  \begin{array}{lllllll}
    H_S & = & \varepsilon \hat{P}^1, &  & \hat{P}^1 & = & \sum_{\sigma} 
    {\color{black} |} \sigma {\color{black} \rangle} {\color{black} \langle}
    \sigma {\color{black} |},
  \end{array}
\end{equation}
if we accordingly adjust the high-energy cutoffs in the electrodes [discussed below
\Eq{eq:phipsi}]. In the considered subspace, we can replace $\hat{n} =
\hat{P}^1$. For the readout one tunes the level position $\varepsilon = - V_g$
by a gate voltage $V_g$ close to the electrochemical potentials of one of the
electrodes. While the spin of the qubit electron is irrelevant (it was not
written above since the readout couples to the charge, see below), it is
important to include the spin degree of freedom of the SQD because the
spin degeneracy enters into the tunneling rates.

\emph{Electrodes.} The final stage of the readout involves the electrodes, treated as
noninteracting reservoirs of electrons with spin:
\begin{equation}
  \begin{array}{lll}
    H_R & = & \sum_{r, k, \sigma} \omega_{r k \sigma} c^{\dag}_{r k \sigma}
    c_{r k \sigma},
  \end{array}
\end{equation}
with the field operators $c_{r k \sigma}$ ($c^{\dag}_{r k \sigma}$) acting on the
electrons in orbital $k$ with spin $\sigma$ in the source ($r = s$) and the drain ($r
= d$), respectively. These are each held in equilibrium with a common
temperature $T$, but at different electrochemical potentials $\mu_s = V_b
\text{/} 2$ and $\mu_d = - V_b / 2$ by a applying a bias voltage $V_b$.

\emph{Readout.} The indirect readout of the qubit state using the sensor QD involves two
couplings: the first one is the capacitive interaction of the SQD electron
charge $\hat{n}$ with the charge polarization $\hat{\tau}_z$ of the qubit:
\begin{equation}
  \begin{array}{lll}
    H_I & = & \hat{n}  \vecg{\lambda} \cdot \op{\vecg{\tau}} \text{/} 2.
  \end{array}
\end{equation}
The measurement vector $\vecg{\lambda}$ specifies both the basis in which one
measures and the measurement strength $\lambda$:
\begin{equation}
  \begin{array}{lll}
    \vecg{\lambda} & = & \lambda \vec{e}_z \label{eq:lambdadef}.
  \end{array}
\end{equation}
Thus, depending on the qubit state, the sensor QD level experiences an energy/gate-voltage offset of at most $\pm \lambda / 2$ [see {\color{black} Figs.
\ref{fig:model}} (a) and \ref{fig:model}(b)]. This in turn affects the conductance measured
in one of the electrodes due to tunneling to and from the SQD:
\begin{equation}
  \begin{array}{lll}
    H_T & = & \sum_{r, k, \sigma} t_r d^{\dag}_{\sigma} c_{r k \sigma} +
    \text{H.c.}
  \end{array}
\end{equation}
The strength of this second coupling involved in the readout is quantified by
the tunnel rates $\Gamma_r = 2 \pi | t_r |^2 \nu_r$, where we take both the
tunneling amplitude $t_r$ and the density of states $\nu_r$ to be spin
$(\sigma)$- and energy $(k)$-independent within the electrode bandwidth $2 W$.

The threefold layered structure of this indirect detection (negligible direct
coupling of the electrodes to the qubit) is reflected in our theoretical analysis
 of the measurement backaction. In \Sec{sec:projection}, we first
eliminate the ``outer'' detection layer -- the electrodes -- in favor of
effective equations describing the joint SQD-qubit dynamics {\footnote{The
first step of our approach is in principle generally possible at least for
weak tunnel coupling $\Gamma$, i.e., not limited to the weak-measurement
situation we study in this paper.}}. In \Sec{sec:results}, we then attempt to integrate out
the ``inner'' detection layer -- the SQD -- to find the effective qubit evolution.

\subsection{Weak-measurement and weak-tunneling limit}\label{sec:weakcoupling}

We consider a sensor with a fast response, i.e., the internal dynamics of
qubit plus sensor QD is slow as compared to the electron tunneling dynamics
induced by the attached electrodes:
\begin{eqnarray}
  \Delta & \sim & \lambda, \Omega \ \  \ll  \ \ \Gamma . 
  \label{eq:weakmeasure}
\end{eqnarray}
This condition means physically that many electrons pass through the SQD
during its interaction time with the qubit ($\lambda \ll \Gamma$) --- a
{\tmem{weak measurement}} is performed. Moreover, if $\vecg{\Omega}$ lies in the $x$-$y$ plane, the internal qubit evolution describes a coherent tunneling
of the qubit electron with dwell times of electrons in the SQD that are much
smaller than the period of a qubit cycle $(\Omega \ll \Gamma)$. Each electron
sees a ``snapshot'' of the SQD-plus-qubit state. If we assumed $\Gamma
\lesssim \Omega$ instead (but still weak measurement $\lambda \ll \Gamma$),
the readout would be too slow to resolve any qubit evolution.

As we see below in {\Sec{sec:kineq}}, the leading-order
response of the tunnel rates to a measurement-induced gate voltage offset
$\sim \lambda$ is given by $\Gamma \lambda / T$. This, in return, induces a
dissipative backaction affecting the {\tmem{polarization}} of the qubit that
also scales as $\Gamma \lambda / T$. Moreover, when condition
(\ref{eq:weakmeasure}) holds, the tunneling also affects the isospin
{\tmem{coherences}} of the qubit-SQD state. This is well-known from the
analysis of two-level systems coupled to a reservoir: Here, the
density-operator coherences in the energy basis matter when the levels (here
split by $\Delta$) are degenerate on the scale of the coupling (here $\Gamma$)
to the environment. In our case, the two levels correspond to the sensor QD
and the qubit, each being a two-state system. One has to carefully identify
which coherences are relevant, which is done below in {\color{black} Sec.
\ref{sec:charge-specific-isospin}}. These coherences are affected by tunneling
processes: The simple physical intuition behind this is that if an electron on
the sensor has time to interact with the qubit and change the current, it
certainly has time to fluctuate into the electrodes. This leads to a response
of the level renormalization and results in a coherent backaction which scales
as $\Gamma \lambda / T$, i.e., in the same way as the response of the sensor
tunneling rates resulting in the dissipative backaction. This is a central result of
{\color{black} Ref. {\cite{Hell14a}}} and here we explore its effect on transient dynamics.

It should thus be noted that the energy scale for backaction on the qubit is not
simply $\lambda$ (from the internal interaction $H_I$) but also $\lambda
\Gamma / T$, the scale of effective dissipative \tmtextit{and} coherent
coupling between sensor QD and qubit, which are induced by tunneling
processes. In a way, these couplings account for an indirect interaction of
the qubit with the electrodes extending the approach of {\color{black} Ref.
{\cite{Emary08}}}. This is thus another relevant perturbative scale for a
weak-measurement expansion, besides the scale $\lambda$ itself, see
{\color{black} Ref. {\cite{Hell14a}}} for a detailed exposition.

Another crucial point for this work is that if terms of order $\Gamma \lambda
/ T$ are taken into account and $\lambda \ll \Gamma$, then \emph{at least} cotunneling terms
scaling as $\Gamma^2 / T$ must also be accounted for \newer{(if not even
higher-order tunneling terms).} We
can neglect higher-order tunneling processes beyond cotunneling if we restrict
the temperatures such that
\begin{eqnarray}
  \Gamma^2 / T & \ll & \lambda .  \label{eq:hight}
\end{eqnarray}
This condition means that the cotunneling-induced noise imposes only a weak
perturbation of the qubit. Taken together, we employ here a weak-coupling
limit in {\tmem{two}} ways, namely that of \tmtextit{weak measurement} and
{\tmem{weak tunneling}}:
\begin{equation}
  \begin{array}{lllll}
    \Gamma / T & \ll & \Delta / \Gamma & \ll & 1
  \end{array} .
\end{equation}
Note that by \Eq{eq:hight} this imposes a stronger condition than the
usual weak-tunneling assumption $\Gamma / T \ll 1$ alone. We next discuss the
dynamical variables needed to describe the measurement backaction.

\subsection{Charge-specific isospins and qubit
decoherence}\label{sec:charge-specific-isospin}

To describe both the backaction of the sensor on the qubit as well as to compute
the signal current through the sensor QD, one needs at least the reduced
density operator $\rho (t)$ of the combined qubit \tmtextit{plus} SQD system
obtained by tracing over the electrodes. Even though we do not analyze the
sensor signal here, the signal is of course of high interest to gain insight
into, e.g., the efficiency of the measurement
{\cite{Korotkov01,Gurvitz05,Clerk04,Korotkov08}}. Studying the
backaction in a situation where the sensor signal current is not negligible,
is an experimentally highly relevant situation, which we pursue in this
paper.\\ 
The relevant part of the SQD-qubit density operator $\rho (t)$ can be
expanded as follows {\cite{Hell14a}}:
\begin{eqnarray}
  \rho (t) & = & \frac{1}{2}  \sum_n \op{P}^n \otimes \left[ p^n (t)
  \op{\mathbbm{1}} + \sum_i \tau^n_i (t)  \op{\tau}_i \right] . 
  \label{eq:reddens}
\end{eqnarray}
Here, $\op{P}^n$ denotes the projector onto the charge states $n = 0, 1$ of
the sensor QD. The numbers $p^n (t) = \tmop{tr} ( \hat{P}^n \rho (t))$ give
the probability for the sensor QD to be in the respective charge state $n = 0,
1$, which for any time $t$ sum up to one due to the probability conservation: $p^0
+ p^1 = 1$. The only irrelevant coherences (off-diagonal \ matrix elements in
the energy basis) of $\rho$ are those involving different charge states on the
sensor. These can be shown to decouple from the relevant part due to the charge
conservation by the tunneling. However, all remaining qubit-SQD density matrix
elements including their coherences must be kept in (\ref{eq:reddens}).
These are the six numbers $\tau^n_i (t) = \tmop{tr} ( \op{P}^n \op{\tau}_i \rho
(t))$, which are the averages of the isospin components $i = x$, $y$, and $z$
for the two sensor QD charge states $n = 0$ or $1$, respectively. \newer{To
describe the correlated SQD-qubit system, one thus needs two {\tmem{charge-specific}} isospins ${\color{black} \vecg{\tau}}^0$ and ${\color{black} \vecg{\tau}}^1$.}
Based on \Eq{eq:reddens}, it is convenient to introduce the following
column representation of the density operator:
\begin{eqnarray}
  \vecg{\rho} & = & \left(\begin{array}{c}
    p^0\\
    p^1\\
    {\color{black} \vecg{\tau}}^0\\
    {\color{black} \vecg{\tau}}^1
  \end{array}\right).  \label{eq:rhorep}
\end{eqnarray}
In this form, one distinguishes the charge and isospin part, respectively, but
the isospin part is still kept basis independent. In other words, we may
represent ${\color{black} \vecg{\tau}}^0$ and ${\color{black}
\vecg{\tau}}^1$ in a different orbital basis than the one used in definition
(\ref{eq:tau}) if we transform the directions of $\vecg{\lambda}$ and
$\vecg{\Omega}$ accordingly.

\newer{We now further explore the physical meaning and importance of the two charge-specific isospins.}
By construction, they sum up to the total
isospin,
\begin{eqnarray}
  {\color{black} \vecg{\tau}} & = & {\color{black} \vecg{\tau}}^0 +
  {\color{black} \vecg{\tau}}^1,  \label{eq:tautot}
\end{eqnarray}
which is often of main interest. This is the usual Bloch vector that describes
the state of the qubit, i.e., its reduced density operator. One can easily show that a single Bloch vector can describe the joint SQD-qubit state only if it is factorizable.
In other words, if
\begin{eqnarray}
  \rho & = & \rho_S \otimes \rho_Q = \left( \sum_n p^n \op{P}^n \right)
  \otimes \frac{1}{2} \left( \op{\mathbbm{1}} + \sum_i \tau_i (t)  \op{\tau}_i
  \right), \nonumber\\
  &  &  \label{eq:rhosep}
\end{eqnarray}
then the charge-specific isospins are given by $\vecg{\tau}^0 = p^0 \vecg{\tau}$
and $\vecg{\tau}^1 = p^1 \vecg{\tau}$, respectively, by comparing Eq.
(\ref{eq:rhosep}) with \Eq{eq:reddens}.
The other combination of the two isospins,
\begin{eqnarray}
  \vecg{\delta} := p^0  \vecg{\tau}^1 - p^1 \vecg{\tau}^0,
\end{eqnarray}
quantifies the nonfactorizability of the qubit-sensor density operator:
\begin{eqnarray}
  \rho = \rho_S \otimes \rho_Q & \Leftrightarrow & \vecg{\delta} =  0.  \label{eq:faccond}
\end{eqnarray}
We emphasize that the nonfactorizability is
crucial to describe the readout and its backaction on the qubit state
${\color{black} \vecg{\tau}}$. We will see in \Sec{sec:zerocoupling} below
\Eq{eq:vda} that the deviation
$\vecg{\delta}$ \newer{has no impact on} the qubit evolution \emph{only} if the qubit and sensor are strictly
decoupled ($\lambda = 0$). For the coupled case ($\lambda \neq 0$),
which is of interest, we have to
keep the individual dynamics of $\vecg{\tau}^0$ and $\vecg{\tau}^1$.\\ 
The
relevance of the two isospins for the decoherence can be seen explicitly from the equation of
motion for $| {\color{black} \vecg{\tau}} (t) |^2$, which characterizes the
purity of the isospin state:
\begin{eqnarray}
  & \begin{array}{lllll}
    \frac{d}{d t} [| {\color{black} \vecg{\tau}} |^2] & = & 2
    \dot{\vecg{\tau}} \cdot \vecg{\tau} & = & - 2 {\color{black}
    \vecg{\lambda}} \cdot ({\color{black} \vecg{\tau}}^0 \times
    {\color{black} \vecg{\tau}}^1),
  \end{array} &  \label{eq:puritydecay}
\end{eqnarray}
where we inserted after the first equality
\begin{eqnarray}
  & \begin{array}{lll}
    \dot{\vecg{\tau}} & = & \vecg{\Omega} \times \vecg{\tau} + \vecg{\lambda}
    \times \vecg{\tau}^1 .
  \end{array} &  \label{eq:taudotsumrule}
\end{eqnarray}
Equation (\ref{eq:puritydecay}) is an exact result which can be obtained from
the Heisenberg equation of motion for $\vecg{\tau}$ with respect to the full
Hamiltonian (\ref{eq:H}). The only essential assumption on which Eq.
(\ref{eq:puritydecay}) relies is the indirect readout structure of our setup,
i.e., $[H_T, H_Q] = 0$. It thus holds generally for any $\vecg{\Omega}$ and \emph{any}
tunneling $\Gamma$. Preservation of this exact equation
imposes an exact isospin sum rule [discussed below \Eq{eq:kineq}], which
any kinetic equation for $\rho$ should satisfy: This was only realized recently, see
Refs. {\cite{Salmilehto12}} and {\cite{Hell14a}}, in particular, see Appendices D and E.

Equation (\ref{eq:puritydecay}) shows that the reduction of the purity of the
qubit state appears only due to noncollinearities of ${\color{black}
\vecg{\tau}}^0$ and ${\color{black} \vecg{\tau}}^1$. These
noncollinearities develop because of the readout: the isospins ${\color{black}
\vecg{\tau}}^0$ and ${\color{black} \vecg{\tau}}^1$ are subject to
different effective ``magnetic'' fields depending on the charge state $n$ of
the sensor QD. To see this, we rewrite $H_Q + H_I = \op{\vec{B}}_{\tmop{eff}}
(\hat{n}) \cdot \op{\vecg{\tau}} / 2$ with an effective field acting on the
isospin $\op{{\color{black} \vecg{\tau}}}$,
\begin{eqnarray}
  \op{\vec{B}}_{\tmop{eff}} & = & \tilde{\vecg{\Omega}} + \vecg{\lambda}
  \delta \hat{n} ,  \label{eq:beff}
\end{eqnarray}
where \ $\delta \hat{n} = \hat{n} - \langle \hat{n} \rangle$ and
${\color{black} \langle} \hat{n} {\color{black} \rangle} = \tmop{tr}
(\hat{P}^1 \rho) = p^1$. Here, the first part is the {\tmem{mean field}},
\begin{eqnarray}
  \tilde{\vecg{\Omega}} & = & \vecg{\Omega} + \langle \hat{n} \rangle
  \vecg{\lambda},  \label{eq:field}
\end{eqnarray}
which the isospin experiences due to the internal isospin field $\vecg{\Omega}$ and
the average field caused by the mean charge ${\color{black} \langle} \hat{n}
{\color{black} \rangle} = p^1$ on the sensor QD with respect to the exact
total density operator $\rho$. The mean-field contribution
$\tilde{\vecg{\Omega}}$ to $\op{\vec{B}}_{\tmop{eff}}$ is the same for both
charge-specific isospins, $\vecg{\tau}^0$ and $\vecg{\tau}^1$, and therefore not
responsible for the qubit-state decay. The mean SQD occupation $\langle
\hat{n} \rangle = p^1$ merely tilts the qubit precession axis and changes its
frequency contributing to the detuning of the qubit. Note that the average is here an ensemble
average but not a time average since $p^1 = p^1 (t)$ can change in time with
the state $\rho (t)$. The qubit decay is induced by the second, fluctuating contribution
$\vecg{\lambda} \delta \hat{n}$ to \Eq{eq:beff}, where $\delta \hat{n}
= \hat{n} - \langle \hat{n} \rangle$ is the charge-state dependent deviation
from the mean field. This generates a noncollinearity of ${\color{black}
\vecg{\tau}}^0$ and ${\color{black} \vecg{\tau}}^1$, which reduces the
purity of the qubit state by \Eq{eq:puritydecay}.

Even though our approach does not make use of the decomposition into a
mean-field and a fluctuating part, we can identify both effects in our results in {\Sec{sec:effliouville}}. We will first identify $p^1 = p^1_{\tmop{st}}$ with the state
of the SQD in the stationary limit, in which the ensemble and the time average are
equal.  We
further connect more precisely the decoherence rates to the components of the
fluctuating part $\vecg{\lambda} \delta \hat{n}$ along and perpendicular to the
mean field $\tilde{\vecg{\Omega}}$ in {\color{black} Sec.
\ref{sec:leffexpand}} (in accordance with the literature
{\cite{Ithier05,Chirolli08}}). This accounts for what we call
\tmtextit{stochastic backaction} on the qubit by the sensor QD. This effect is
also present for single-electron transistor sensors with a continuum of
electronic levels, but (classically) quantized charge states.

However, there are also a {\tmem{dissipative}} and a \tmtextit{coherent
backaction} effect {\cite{Hell14a}} (see {\color{black} Sec.
\ref{sec:weakcoupling}}). As we discuss below, they modify the relative
orientations of ${\color{black} \vecg{\tau}}^0$ and ${\color{black}
\vecg{\tau}}^1$ and therefore affect the qubit decay as well. This mechanism ---
first noted in {\color{black} Ref. {\cite{Hell14a}}} --- derives from a
renormalization effect induced by the interplay of the readout interactions
($\lambda$) and the tunneling on and off the sensor QD ($\Gamma$) as discussed
in {\Sec{sec:kineq}}. It results in isospin torques
similar to those encountered in spintronic QD setups.
As mentioned in the Introduction (Sec. {\sec{sec:intro}}), the prominent role of renormalization effects distinguishes a QD sensor with few, discrete energy levels from sensors with a continuous energy spectrum.

Finally, we note from \Eq{eq:puritydecay} that the Bloch vector may
not just shrink exponentially. We will find that ${\color{black} \vecg{\tau}}^0$ and
${\color{black} \vecg{\tau}}^1$ perform different precessional motions (due to both
$\op{\vec{B}}_{\tmop{eff}}$ \tmtextit{and} the coherent backaction
\newer{being dependent on the charge state of the sensor}),
which implies that the component of ${\color{black} \vecg{\tau}}^0 \times {\color{black}
\vecg{\tau}}^1$ along the measurement vector ${\color{black}
\vecg{\lambda}}$ also oscillates in time. Thus, the rate of decay of the
purity is not purely exponential but additionally oscillates in time as
explicitly confirmed by our analysis in {\color{black} Sec.
\ref{sec:elliptical}}. This illustrates that the motion of the charge-specific
isospins is closely related to the qubit decay. Accounting for the interplay
of their dynamics turns out to be the key to set up a correct description of
the transient qubit dynamics that includes \newer{all the different types of backaction}.




\section{Qubit-sensor quantum dot dynamics}\label{sec:kineq}

\subsection{Outline}

The indirect measurement setup introduced in the previous section poses
several challenges for the theoretical treatment of the measurement
backaction. A central complication is that the environment of the qubit (the
SQD {\tmem{plus}} the electrodes) is {\tmem{not}} in a simple equilibrium state
since the detection is done by nonequilibrium transport. But even when
specializing to near-equilibrium conditions, one has to treat the SQD as a
strongly interacting quantum system with spin degeneracy. Both the
nonequilibrium conditions and the interactions in the SQD prevent a simple direct approach
where one averages over the environmental degrees of freedom, leaving only the
qubit degrees of freedom. Moreover, to obtain the sensor current, we need to
retain the sensor degrees of freedom as well. As we discuss in {\color{black}
Sec. \ref{sec:br}}, specifying the environmental state is the main difficulty
when trying to directly calculate the evolution of the qubit density operator
for an indirect detection setup.

Therefore, we integrate out \tmtextit{only the electrodes} to obtain the
density operator $\rho$ for the joint qubit-SQD system. The resulting
equation, which is of the form
\begin{eqnarray}
  \frac{d}{d t} \vecg{\rho} (t) & = & - i L \vecg{\rho} (t),  \label{eq:kineq0}
\end{eqnarray}
is given below [\Eq{eq:kineq}] and is the first main equation of this
work. Our main conclusion is that this provides a systematic description of
the \newer{measurement backaction}: in the weak tunneling, weak measurement limit $\Gamma / T \ll
\lambda / \Gamma \ll 1$, it does not get any simpler without making drastic
concessions. Yet, a mere reformulation of \Eq{eq:kineq} [see Eq.
(\ref{eq:kineq2})] already provides important insights into the measurement
backaction.

Still, we will also describe an attempt to eliminate the SQD degrees of
freedom in the high-temperature limit where corrections $\Gamma / T$ can be
dropped. This results in an effective Liouvillian $L_{\tmop{eff}}$ that
reproduces the outcome of \Eq{eq:kineq0} for the evolution of total
average isospin operator, ${\color{black} \vecg{\tau}} (t) = \left\langle
\op{{\color{black} \vecg{\tau}}} \right\rangle = {\color{black}
\vecg{\tau}}^0 (t) + {\color{black} \vecg{\tau}}^1 (t)$, in the long-time
limit,
\begin{eqnarray}
  \frac{d}{d t} {\color{black} \vecg{\tau}} (t) & = & - i L_{\tmop{eff}} 
  {\color{black} \vecg{\tau}} (t) \text{ \ \ } (t - t_0 \gg 1 / \Gamma),
  \label{eq:taueq}
\end{eqnarray}
\newer{with initial time $t_0$.} The preparations for this step provide interesting insights into Eq.
(\ref{eq:kineq0}). However, \Eq{eq:taueq} turns out to be invalid for
small times $t - t_0 \lesssim 1 / \Gamma$. The error made when still using Eq.
(\ref{eq:taueq}) to compute $\vecg{\tau} (t)$ starting from $t = t_0$ for $t
- t_0 \gg 1 / \Gamma$ can be compensated by a correction to the initial
condition ${\color{black} \vecg{\tau}} (t_0)$, a so-called initial slip.
This correction depends on the initial \tmtextit{qubit-sensor state} in an
essential way, preventing the sensor from being integrated out completely.
Before we discuss the details, let us first outline the further challenges
posed in deriving the above two equations.

The derivation of \Eq{eq:kineq0} has to include various effects: First, since we
incorporate the measurement backaction terms $\sim \Gamma \lambda / T$, we
must also include \tmtextit{next-to-leading order} tunnel processes $\sim
\Gamma^2 / T$ into the kinetic equations for the SQD-qubit evolution (see
{\Sec{sec:weakcoupling}}). \new{We have given} such a consistently expanded
kinetic equation in {\color{black} Ref. {\cite{Hell14a}}}.
However, there we employed an additional Markov approximation with
respect to the electrodes, which is valid
to obtain the stationary long-time limit studied in {\color{black} Ref.
{\cite{Hell14a}}}. Here, by contrast, we are interested in the
{\tmem{transient}} dynamics, where non-Markovian effects induced by the electrodes must be accounted for,
as we explain in {\Sec{sec:nonMarkovian}}. We include the
required leading non-Markovian correction perturbatively in the tunnel
\braggio{coupling $\Gamma$ along the lines of Refs.
{\cite{Braggio06,Splettstoesser06,Flindt08PRB,Flindt08PRL,Splettstoesser10,Karlewski14}}.}
We present and explain
the resulting time-local kinetic equations in Secs.
{\sec{sec:timelocal}}--{\sec{sec:nmdiscuss}}.

We next analyze in {\Sec{sec:projection}} how the qubit is
affected by the measurement within the resulting description. For this
purpose, we first solve in {\Sec{sec:zerocoupling}} the
kinetic equations for zero capacitive interaction $\lambda = 0$. An important
step is to identify a set of {\tmem{quasistationary modes}} that contain the
degrees of freedom of the qubit only, i.e., $\vecg{\tau} = \vecg{\tau}^0 +
\vecg{\tau}^1$. This identification remains valid also for nonzero capacitive
interaction $\lambda \neq 0$. The time scale $\sim 1 / \Omega$ for the
evolution of these modes -- connected with the slow qubit dynamics -- is well
separated from that for the evolution of the residual {\tmem{decaying modes}}.
Those are strongly damped on a short time $\sim 1 / \Gamma$ due to the fast
tunneling dynamics of the SQD. We then introduce new dynamical variables to analyze
the coupling between the quasistationary and decay modes in {\color{black}
Sec. \ref{sec:couplingmodes}} for nonzero capacitive interaction $\lambda$.
This will reveal the mitigation of the measurement backaction by the coherent
backaction, the first key result of the paper. Finally, we derive the evolution of
the quasistationary modes by effectively incorporating the impact of the decay
modes (see {\Sec{sec:effliouville}). 
\newer{Importantly, the resulting equations are not independent of the
detector evolution and even in the long-time limit an explicit
dependence on the initial overlap with the decay modes remains as we
will see in \Sec{sec:results}}

\subsection{Kinetic equation}

\subsubsection{Integrating out the electrodes}\label{sec:nonMarkovian}
Whenever the time evolution of an open system is considered, non-Markovian
features arise from the memory of the environment, that is, its correlation
functions decay within a nonzero correlation time $\tau_C$
{\cite{Breuer,Saptsov14}}. When integrating out the environment (here the
electrodes, \new{see} {\color{black} Fig. \ref{fig:model}}), the time evolution of the
reduced density operator $\rho (t)$ of the open system (here the SQD plus
qubit) is governed by a time-nonlocal kinetic equation:
\begin{eqnarray}
  \dot{\rho} (t) & = & - i L_{Q S} \rho (t) + \int_{t_0}^t d t' W (t -
  t') \rho (t') .  \label{eq:gme}
\end{eqnarray}
Here, $L_{Q S} \bullet = [H_Q + H_S + H_I, \bullet]$ is the internal
Liouvillian of the reduced system with ``$\bullet$'' denoting the operator the
Liouvillian acts on. Moreover, all effects of the environment are contained in
a kernel $W$ that we compute by a real-time diagrammatic approach
{\cite{Schoeller09a,Leijnse08a}}. If the initial value $\rho (t_0)$ is
specified, \Eq{eq:gme} can be used to compute $\rho (t)$ without
explicitly keeping track of the state of the electrodes. A key assumption enabling
such a closed description of $\rho (t)$ is that the reservoir is stationary,
i.e., $[H_R, \rho_R] = 0$, which is satisfied here because we assume the
electrodes to be a thermal equilibrium state $\rho_R$ (\new{see}, e.g.,
{\color{black} Ref. {\cite{Schoeller09a}}}).

When calculated to leading order in $\Gamma$, the kernel roughly decays as $W
(t - t') \sim \Gamma e^{- (t - t') / \tau_C}$ with correlation time $\tau_C
\sim 1 / T$ {\cite{Saptsov14}}. Within the Markov approximation with
respect to the electrodes, one replaces
$W (t - t') = W (i 0) \delta (t - t')$, where on the right-hand side $W (z) =
\int_0^{\infty} e^{i z t} W (t)$ denotes the Laplace-transformed kernel. This
yields a time-local kinetic equation $d \rho / d t \approx (- i L_{Q S} + W (i
0)) \rho (t)$ when inserted into \Eq{eq:gme}. In general, $(- i L_{Q S}
+ W (i 0)) \rho_{\text{st}} = 0$ determines the \tmtextit{exact} stationary
state, a fact which is often overlooked but easily shown
{\cite{Schoeller09a}}. Physically, this makes sense since a nearly constant
state cannot ``remember'' much. As $\rho (t)$ approaches the constant stationary
state $\rho_{\text{st}}$, non-Markovian corrections in \Eq{eq:expand}
become weaker [for fixed $t'$ the memory kernel $W (t - t')$ decays as $t$
increases].

To go beyond this Markovian approximation to obtain the transient dynamics, we
include the non-Markovian corrections induced by the electrodes perturbatively in $\Gamma / T$. To do so, we
insert the Taylor expansion for the reduced density operator,
\begin{eqnarray}
  \rho (t') & = & \rho (t) + \dot{\rho} (t) (t' - t) + \ldots, 
  \label{eq:expand}
\end{eqnarray}
recursively into \Eq{eq:gme}, as explained in Refs.
{\cite{Splettstoesser06,Splettstoesser10}}.
 As we argue in {\color{black} Appendix
\ref{app:nonmarkov}}, the derivatives $d^n \rho / d t^n$ are on the order of
$\Gamma^n$, and we estimate $(t' - t)^n \lesssim \tau_C^n \sim 1 / T^n$ within
the correlation time of the kernel. Thus, higher-order terms in the expansion
(\ref{eq:expand}) correspond to higher orders in the tunneling expansion in
$\Gamma / T$. Truncating the expansion after the leading-order memory
correction ($n = 1$), one can derive a time-local kinetic equation for $\rho
(t)$ as we show in {\color{black} Appendix \ref{app:nonmarkov}}.

 \braggio{The above treatment is closely
related to the techniques developed for full counting statistics
\cite{Braggio06,Flindt08PRL,Flindt08PRB} and to the recent study in
Ref. {\cite{Karlewski14}}.} \new{There is also a conceptual
connection to time-convolutionless master equations
{\cite{Bates69,Tokuyama76,Timm11}}: In the latter approach, the
full density operator evaluated at time $t'$ is obtained by evolving
the full density operator at time $t$ backwards in time before
integrating out the electrodes, resulting also
in an effectively time-local kinetic equation.}

\subsubsection{Kinetic equation}\label{sec:timelocal}

Including all terms of order $\Delta$, $\Gamma$, as well as $\Gamma^2 \text{/}
T$ and $\Delta \Gamma \text{/} T \nocomma$, where $\Delta \sim \Omega,
\lambda$, as well as the leading memory corrections we obtain the kinetic
equation expressed here in the representation (\ref{eq:rhorep}) of $\rho$ (no
time arguments written):

{\wideeq{\begin{eqnarray}
  \frac{d}{d t} \left(\begin{array}{c}
    p^0\\
    p^1\\
    {\color{black} \vecg{\tau}}^0\\
    {\color{black} \vecg{\tau}}^1
  \end{array}\right) & = & \left(\begin{array}{cccc}
    - 2 \gamma^0 & + \gamma^1 & + 2 c \vecg{\lambda} \cdot & + c \vecg{\lambda}
    \cdot\\
    + 2 \gamma^0 & - \gamma^1 & - 2 c \vecg{\lambda} \cdot & - c \vecg{\lambda}
    \cdot\\
    + 2 c \vecg{\lambda} & + c \vecg{\lambda} & - 2 \gamma^0 + \left(
    \vecg{\Omega} - \kappa {\color{black} \vecg{\lambda}} \right) \times & +
    \gamma^1 + \kappa \vecg{\lambda} / 2 \times\\
    - 2 c \vecg{\lambda} & - c \vecg{\lambda} & + 2 \gamma^0 + \kappa
    {\color{black} \vecg{\lambda}} \times & - \gamma^1 + \left( \vecg{\Omega}
    + {\color{black} \vecg{\lambda}} - \kappa {\color{black}
    \vecg{\lambda}} / 2 \right) \times
  \end{array}\right) \left(\begin{array}{c}
    p^0\\
    p^1\\
    {\color{black} \vecg{\tau}}^0\\
    {\color{black} \vecg{\tau}}^1
  \end{array}\right)  \label{eq:kineq}
\end{eqnarray}}

When computing the matrix product with the column vector in the above
equation, the dot ``$\cdot$'' (cross ``$\times$'') in the entries of the
matrix indicates that a three-dimensional scalar (vector) product is to be
formed with the corresponding entries of $\vecg{\rho}$. The above
equation is valid under the weak-coupling assumption $\Gamma / T \ll \lambda /
\Gamma \ll 1$ introduced in {\Sec{sec:weakcoupling}} such
that corrections of order $\Gamma^3 / T^2$, $\Gamma^2 \Delta / T^2$, and
$\Gamma \Delta^2 / T^2$ can be neglected.\\
The above kinetic equation is the first central equation of this paper. It goes
beyond a simple master equation by including all relevant coherences (see
{\Sec{sec:charge-specific-isospin}}) and extends the
kinetic equation of {\color{black} Ref. {\cite{Hell14a}}}, which is
Markovian with respect to the electrodes, to access the
transient dynamics by including the kernel frequency dependence.
 The kinetic equation (\ref{eq:kineq}) respects the
probability conservation, $\dot{p}^0 + \dot{p}^1 = 0$, and also the recently
found {\cite{Hell14a}} exact isospin sum rule (\ref{eq:taudotsumrule}),
$\dot{{\color{black} \vecg{\tau}}}^0 + \dot{\vecg{\tau}}^1 = \vecg{\Omega}
\times \vecg{\tau} + {\color{black} \vecg{\lambda}} \times \vecg{\tau}^1$. The
latter derives from the conservation of the total isospin, $\vecg{\tau} = \vecg{\tau}^0 + \vecg{\tau}^1$, when electrons tunnel from the electrodes into the SQD and vice
versa, a generic feature {\cite{Salmilehto12}} of indirect measurement models
of type (\ref{eq:H}). We next discuss the expressions and physical
significance of the four new coefficients $\gamma^0, \gamma^1$, $c$, and $\kappa$
occurring in \Eq{eq:kineq}\newer{; for the definition of $\vecg{\lambda}$ and
$\vecg{\Omega}$ see Eqs. \eq{eq:lambdadef} and \eq{eq:hqubit}, respectively.}
\subsubsection{Stochastic, dissipative, and coherent
backaction}\label{sec:review}
First, \Eq{eq:kineq} incorporates the SQD switching rates $\gamma^{0,1}
= \sum_r \gamma^{0,1}_r$ with
contributions from each junction $r = s, d$ reading
\begin{eqnarray}
  \gamma^{0, 1}_r & = & \sum_{r = s, d} \left[ \eta \Gamma_r f_r^{\pm} +
  \sum_{q = s, d} \tfrac{\Gamma_r \Gamma_q}{2 T} (f^{\pm}_r)' \phi_q \right.
  \nonumber\\
  &  & \left. \mp \sum_{q = s, d} \tfrac{\Gamma_r \Gamma_q}{2 T} \phi'_r (2
  f^+_q + f^-_q) \right] ,  \label{eq:dissrate}
\end{eqnarray}
where $0, 1$ corresponds to $\pm$. Let us first focus on the meaning of the
three different physical terms in \Eq{eq:dissrate}. The first term in
the first line of \Eq{eq:dissrate} is the sequential tunneling
contribution, whose dependence on the voltages is governed by the Fermi
functions $f^{\pm}_r = f^{\pm} ((\varepsilon - \mu_r) \text{/} T)$ for electrode $r
= s, d$ with $f^+ (x) = 1 / (e^x - 1)$ and $f^- (x) = 1 - f^+ (x)$. We comment
on the non-Markovian correction factor $\eta$ [\Eq{eq:alpha}] in
{\Sec{sec:nmdiscuss}}. The second term is a correction to
the sequential tunneling rate accounting for a renormalization of the level
position $\varepsilon$, incorporating the derivative of the Fermi function,
\begin{eqnarray}
  (f^{\pm}_r)' & = & \left. \frac{\partial f^{\pm} (x)}{\partial x} \right|_{x
  = (\varepsilon - \mu_r) / T}, 
\end{eqnarray}
and the renormalization function,
\begin{eqnarray}
  \phi_r & = & \phi ((\varepsilon - \mu_r) \text{/} T),  \label{eq:phir}
\end{eqnarray}
with
\begin{eqnarray}
  \phi (x) & = & \mathcal{P}  \int_{- \Lambda}^{+ \Lambda} \frac{d y}{\pi} 
  \frac{f^+ (y)}{x - y}  \label{eq:phi}\\
  & = & \frac{1}{\pi} \left[ - \tmop{Re} \text{ } \psi \left( \frac{1}{2} + i
  \frac{x}{2 \pi} \right) + \ln \left( \frac{\Lambda}{2 \pi} \right) \right] .
  \label{eq:phipsi}
\end{eqnarray}
Indeed, combining this second term with the first, $f^{\pm}_r (\varepsilon) +
(f^{\pm}_r)' (\varepsilon) (\delta / T) \approx f^{\pm}_r (\varepsilon +
\delta)$, one identifies the shift $\delta = \sum_q \Gamma_q \phi_q / 2$. The
function $\phi (x)$ is plotted in {\color{black} Fig. \ref{fig:phi}} and shows
a maximum at $x = 0$ with logarithmic tails. In \Eq{eq:phi},
$\mathcal{P}$ denotes the principal value of the integral with a cutoff
$\Lambda = W \text{/} T$, yielding the real part of the digamma function
$\psi$ with a logarithmic correction. The latter depends on the electrode
bandwidth $W$, which must be set to $W \sim U$, where $U$ is the large but
finite local Coulomb interaction energy of the SQD (we excluded the doubly
occupied state from the SQD Hilbert space).

The term in the second line of \Eq{eq:dissrate} relates to the
cotunneling processes through the SQD, which incorporates \newer{the derivative
of the renormalization function,}
\begin{eqnarray}
  \phi_r' & = & \left. \frac{\partial \phi (x)}{\partial x} \right|_{x =
  (\varepsilon - \mu_r) / T}, 
\end{eqnarray}
which is also plotted in {\color{black} Fig. \ref{fig:phi}}. The contribution
from each electrode $r$ changes its sign close to the resonance $\varepsilon =
\mu_r$ and takes its extremal values of $\phi'_r \approx \mp 0.143$ at
$\varepsilon - \mu_r \approx \pm 1.911 T$. While the terms in the first line of
\Eq{eq:dissrate} depend exponentially on the distance to the resonance
$| \varepsilon - \mu_r |$, the cotunneling term is only algebraically suppressed,
since {\footnote{Here we use $\tmop{Re} \text{ } \psi (1 / 2 + i x / 2 \pi)
\approx \ln |x|$ for $|x| \gg 1$.}}
\begin{eqnarray}
  \phi_r' & \approx & - \frac{1}{\pi}  \frac{T}{\varepsilon - \mu_r}
  \hspace{1em} \text{for $| \varepsilon - \mu_r | \gg T$} . 
  \label{eq:phirprime}
\end{eqnarray}
and $2 f^+_q + f^-_q = f^+_q + 1 \geqslant 1$ in \Eq{eq:dissrate}. When
these terms are added together, they result (for each electrode $r$) in a
temperature-broadened step function, which approaches its asymptotes
algebraically. Therefore, this must be accounted for when studying the
qubit-sensor dynamics at the onset of Coulomb blockade where typically the
readout is performed.
\begin{center}
  \Figure{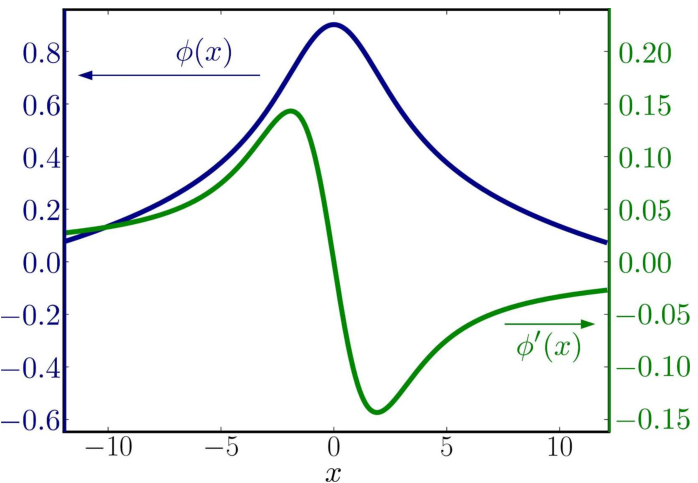}{Renormalization
  function $\phi (x)$, \Eq{eq:phi}, and its derivative $\phi' (x) = d
  \phi (x) / d x$. For illustration purposes we chose $\Lambda = 15$ to
  be rather small, noting that $\Lambda$ only shifts the $\phi$ vertically and
  drops out in $\phi'$ [see \Eq{eq:phipsi}].\label{fig:phi}}
\end{center}
All the above-mentioned tunneling processes contribute to a stochastic
switching of the SQD occupation $n$, which results via the capacitive
interaction $H_I = \hat{n}  \vecg{\lambda} \cdot \op{\vecg{\tau}} / 2$ in a
fluctuation of the effective field $\vec{B}_{\tmop{eff}}^n = {\color{black}
\vecg{\Omega}} + n {\color{black} \vecg{\lambda}}$ acting on the qubit as
explained in {\Sec{sec:charge-specific-isospin}} [see Eq.
(\ref{eq:beff})]. The importance of the capacitive interaction to produce this
{\tmem{stochastic}} contribution to the total measurement backaction becomes
apparent when rewriting the kinetic equation (\ref{eq:kineq}) in terms
of \newer{quasistationary and decaying}
dynamical variables (see {\Sec{sec:couplingmodes}}). It
causes the decoherence of the qubit already in lowest order as we show in
{\Sec{sec:comparison}}. \\
\new{Before we enter the detailed analysis of the next-to-leading order
corrections, let us right away indicate their importance for the
stochastic backaction on a more qualitative level. 
In Fig. \ref{fig:types}, we compare the evolution of the $x$-component $\tau_x(t)$ of the isospin, obtained by solving the kinetic equations \eq{eq:kineq}, when
higher-order terms are included (red) or neglected by hand (green).
The figure illustrates that noise from O($\Gamma^2$) terms
indeed contributes to the qubit decoherence as naively expected. On a
quantitative level, however, one would expect an algebraic suppression
of the measurement backaction with $\varepsilon$ based on \Eq{eq:dissrate} when entering
the Coulomb blockade regime. It will turn out in {\color{black} Sec.
\ref{sec:mitigation1}} that this expectation is incorrect, i.e., the
backaction is weaker than expected.\\
We further see from Fig. \ref{fig:types} that the
oscillation period of the qubit is notably changed due to next-to-leading order corrections. This is due to the mean field,
$\tilde{\vecg{\Omega}} = \vecg{\Omega} + \langle n \rangle
\vecg{\lambda}$, acting on the qubit in the presence of the
sensor QD. The average occupation $
\langle n \rangle = p^1_{\text{st}}$ on the sensor is significantly modified by
higher-order tunneling terms (see Sec. \ref{sec:meanfieldpicture}).}
\\

\begin{center}
  \Figure{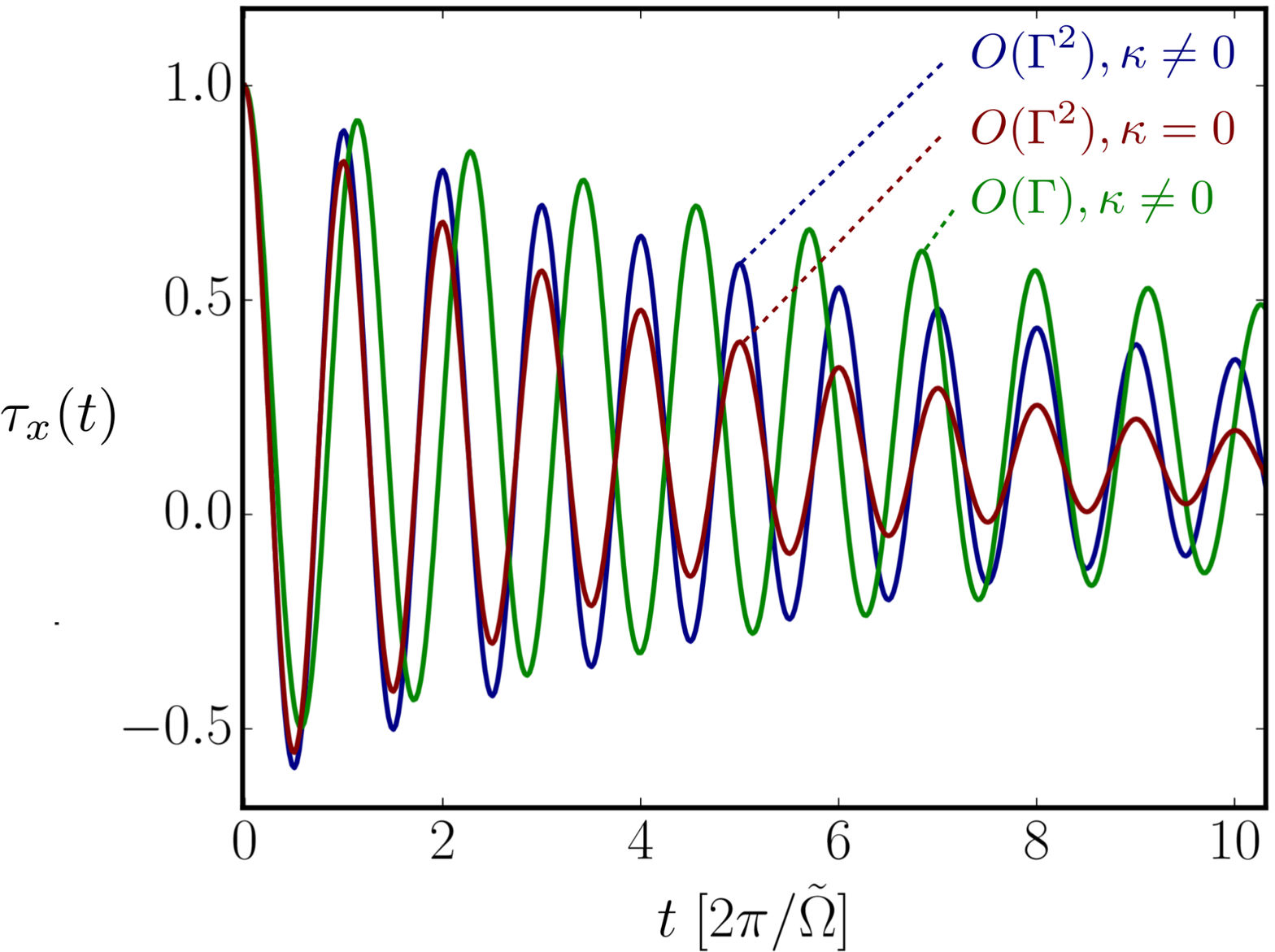}{\new{Modification of
 the sensor backaction due to the next-to-leading order $\Gamma^2$
 stochastic backaction and the coherent backaction. Shown is the $x$-component of the
 isospin, $\tau_x(t) = \tau_x^0(t) + \tau_x^1(t)$ as a function
 of time. The isospin is obtained by solving the kinetic
 equations \eq{eq:kineq} when all terms are taken into account (blue),
 when higher-order $\Gamma^2$ corrections are neglected (green), or when the
 coherent backaction $\propto \kappa$ is neglected (red). 
One can see that the isospin precession period is larger if next-to-leading-order
 contributions are neglected since the isospin experiences
 a different mean field as a consequence of the differing average occupation $\langle n \rangle$. The time is
 given for all three curves in the same unit,
 the inverse of the natural frequency for the full result, $\tilde{\Omega} =
 \sqrt{\Omega^2 + (\langle n \rangle \lambda)^2}$, which is the precession
 frequency including next-to-leading corrections.
 We insert here $\langle n \rangle = p^1_{\text{st}} = 2 \gamma^0 / \gamma$ from the
 stationary solution of the full kinetic equation \eq{eq:kineq} at
 $\lambda=0$ [see \Eq{eq:pst}].
The parameters are $\Gamma_s  = \Gamma_d  = 0.2 T$, $\lambda  = 0.02 T
 $, $\Omega = 0.001 T$,
 $V_b = 0$, $V_g=-3 T$, 
 $W=1000 T$.
Initially, the sensor is empty, $p^0(0) = 1-p^1(0) = 1$, 
and conditional upon this, the qubit isospin vector is prepared
perpendicular to the measurement vector, $\vecg{\tau}^0(0)=\vec{e}_x$ and $\vecg{\tau}^1(0) = \vec{0}$.} \label{fig:types}}
\end{center}

In addition to the stochastic backaction, there is also a {\tmem{dissipative}} backaction of the SQD on the
qubit: These terms are related to the \tmtextit{isospin-charge conversion}
rates $\sim c \vecg{\lambda}$ with coefficient
\begin{eqnarray}
  c & = & \sum_r \frac{\Gamma_r}{2 T} \nobracket (- f_r^+ \nobracket)' . 
  \label{eq:conversion}
\end{eqnarray}
This coupling appears in \new{two} ways. The isospins influence the SQD dynamics
(allowing for the readout) and \tmtextit{vice versa} the SQD occupation
probabilities directly influence the isospins (backaction). This dissipative
backaction drives the sensor QD and qubit into \new{a correlated state
in the stationary limit,
in contrast to the stochastic backaction, which only changes the
occupation probabilities of the qubit state. For the parameters chosen for
Fig. \ref{fig:types}, the dissipative backaction has a negligible impact
on the qubit evolution and therefore we show no comparison. The reason
for this suppression is that the SQD is already mildly Coulomb blockaded
for these parameters and
the dissipative backaction is exponentially
peaked around the resonance as Eq. (\ref{eq:conversion}) shows. 
The dissipative backaction therefore only
becomes relevant close to resonance.}\\
Finally, there is a third type of backaction: the tunneling gives rise to
isospin torque terms $\sim \kappa \vecg{\lambda}$, where
\begin{eqnarray}
  \kappa & = & \sum_r \frac{\Gamma_r}{T} \phi_r',  \label{eq:kappa}
\end{eqnarray}
incorporates the derivative of the renormalization function shown in
{\color{black} Fig. \ref{fig:phi}}. This signals the {\tmem{coherent}} nature
of this contribution to the backaction: it characterizes the {\tmem{response}}
of the sensor QD \tmtextit{level renormalization} to a change in the state of
the qubit. Similar to the cotunneling corrections, the coherent backaction
gains importance with the onset of Coulomb blockade {\cite{Hell14a}}.\\
\new{
The importance of coherent backaction for the qubit evolution stands out in 
Fig. \ref{fig:types}. We can see that neglecting the coherent
backaction (red) noticably (but artificially) enhances the qubit decoherence as compared to the
full solution (blue). This points to a
cancellation effect between coherent backaction and cotunneling noise
that we discuss in detail in Sec. \ref{sec:mitigation1}. We further
note that the coherent backaction has a negligible effect on the qubit
oscillation period [the period is the same for both the curves including
$O(\Gamma^2)$ corrections]. The coherent backaction can thus not be interpreted as a simple correction to
the qubit mean field \eq{eq:field}; it is the joint system of qubit \emph{and} sensor QD
that is renormalized and not just the qubit system. We finally emphasize
that even though Fig. \ref{fig:types} shows theoretical results when different
contributions of the kinetic equation are neglected, they cannot be
switched off individually in a real experiment - there they always appear together
and have to be taken into account altogether.
}

\subsubsection{Impact of non-Markovian corrections}\label{sec:nmdiscuss}

It remains to discuss the three ways in which non-Markovian corrections induced by the electrodes are
contained in \Eq{eq:kineq} and how the latter differs from the Markovian
kinetic equations (Eq. (18) of {\color{black} Ref. {\cite{Hell14a}}}). First,
the non-Markovian corrections modify the leading-order SQD tunneling rates [see \Eq{eq:dissrate}] just by introducing the prefactor
\begin{eqnarray}
  \eta & = & 1 + \sum_r \frac{\Gamma_r}{T} \phi'_r .  \label{eq:alpha}
\end{eqnarray}
Since the correction $\eta - 1$, the cotunneling broadening term [see Eq.
(\ref{eq:dissrate})], and the coherent backaction coefficient $\kappa$ [see \Eq{eq:kappa}], all depend on the same
\newer{factor} $\sum_r \Gamma_r / T \phi'_r$ with algebraic tails,
the non-Markovian effects should clearly be accounted for {\footnote{Note that we investigate the impact
of the coherent backaction by setting here and below ``by hand'' $\kappa = 0$ in our
results. Although one can express the non-Markovian correction factor as $\eta
= 1 + \kappa$, we do not set $\eta = 1$ since this would affect the
stochastic backaction and would not lead to the comparison we intend to
make.}}. The correction
factor $\eta - 1$ is an appreciable quantitative correction that yields a
contribution of $O (\Gamma^2 / T)$ to the switching rates
{\cite{Splettstoesser06}}. However, in contrast to the cotunneling and
coherent backaction, the correction $\eta - 1$ is multiplied with the
exponentially scaling SET contribution [see \Eq{eq:dissrate}] and
therefore it has no qualitative impact here.
 
By contrast, the second type of non-Markovian correction affects the coherent
backaction terms by qualitatively changing them in general relative to the
Markov approximation. The {\tmem{direction}} of the tunneling-induced isospin
torque terms changes: In {\color{black} Ref. {\cite{Hell14a}}}, we also found
a contribution to the coherent backaction $\propto \kappa \vecg{\Omega}$. These
terms are canceled out here up to $O (\Gamma \Omega \text{/} T) = O (\Gamma
\lambda \text{/} T)$. This is expected on physical grounds as the backaction
is mediated by the capacitive interaction $\vecg{\lambda}$ and therefore we
expect these terms to vanish when setting $\lambda = 0$. We emphasize that the
isospin torque terms $\propto \kappa \vecg{\Omega} \times \vecg{\tau}^n$ do not
affect the {\tmem{stationary}} state, which we studied in {\color{black} Ref.
{\cite{Hell14a}}}, and therefore all the conclusions drawn in {\color{black}
Ref. {\cite{Hell14a}}} remain valid. The third effect of the non-Markovian
corrections is a {\tmem{sign}} change of the isospin torque terms $\sim \kappa
\vecg{\lambda}$ in the last column of the matrix (\ref{eq:kineq}) as compared to
{\color{black} Ref. {\cite{Hell14a}}}.

Both the above modifications of the coherent backaction have important
physical consequences illustrated in {\color{black} Appendix
\ref{app:positivity}}: If one naively computes the transient dynamics of the
SQD-qubit system using the equations of {\color{black} Ref.
{\cite{Hell14a}}}, which neglect non-Markovian terms induced by the electrodes, one obtains exponentially increasing transient modes
leading to a violation of the positivity of the density operator. Moreover,
within the Markovian approximation the coherent backaction strongly enhances
the measurement backaction in the Coulomb-blockade regime for a large
parameter regime, while the coherent backaction suppresses the measurement
backaction for nearly all parameter values when non-Markovian corrections are
correctly accounted for (see {\color{black} Appendix \ref{app:positivity}}). This
clearly illustrates that non-Markovian corrections go hand in hand with
renormalization effects, which in an indirect measurement set up go hand in
hand with the cotunneling effects of the sensor rates. All of these are of vital
importance for describing the indirect measurement.



\subsection{Coupling of modes}\label{sec:projection}}

With the kinetic equations (\ref{eq:kineq}) now in hand we can proceed to
analyze the measurement backaction, but still \tmtextit{without} integrating
out the sensor. To achieve this goal, we make use of the separation of
different time scales in the coupled evolution of SQD and qubit in the
weak-coupling, weak-measurement limit $\Gamma / T \ll \Delta / \Gamma \ll 1$.
To identify these time scales, we first solve in {\color{black} Sec.
\ref{sec:zerocoupling}} the unperturbed problem of the decoupled SQD-qubit
system ($\lambda = 0$) as described by \Eq{eq:kineq}. This produces
eigenmodes which are well-separated in energy by $\Gamma \gg \Omega$ and
correspond to the wide-band limit for the \tmtextit{sensor quantum dot}. It turns
out that one needs to compute the evolution of only a part of the modes
-- referred to as the
quasistationary modes in the following -- to construct the evolution of the
total isospin $\vecg{\tau}$.\\ 
In {\Sec{sec:couplingmodes}},
we restore the coupling $\lambda$ and by simply writing the kinetic equation
\eq{eq:kineq} in the basis of these eigenmodes, we can immediately
extract $\lambda^2 / \Gamma$ as the relevant time scale for the qubit
decoherence time. However, this is not the full story of the backaction: there is a prefactor which strongly affects this time scale. We analytically identify a nontrivial
competition of the coherent backaction and the cotunneling-induced stochastic
backaction determining this prefactor. Finally, introducing an exact [relative to \Eq{eq:kineq}]
projection of the dynamics onto the quasistationary modes in {\color{black}
Sec. \ref{sec:effliouville}} we gain further insight, still without
integrating out the sensor. This projection incorporates the effect of the
coupling between the modes and provides the starting point for
\newer{deriving an effective equation for the isospin evolution}
in {\color{black} Sec.
\ref{sec:results}}.

\begin{center}
  \Figure{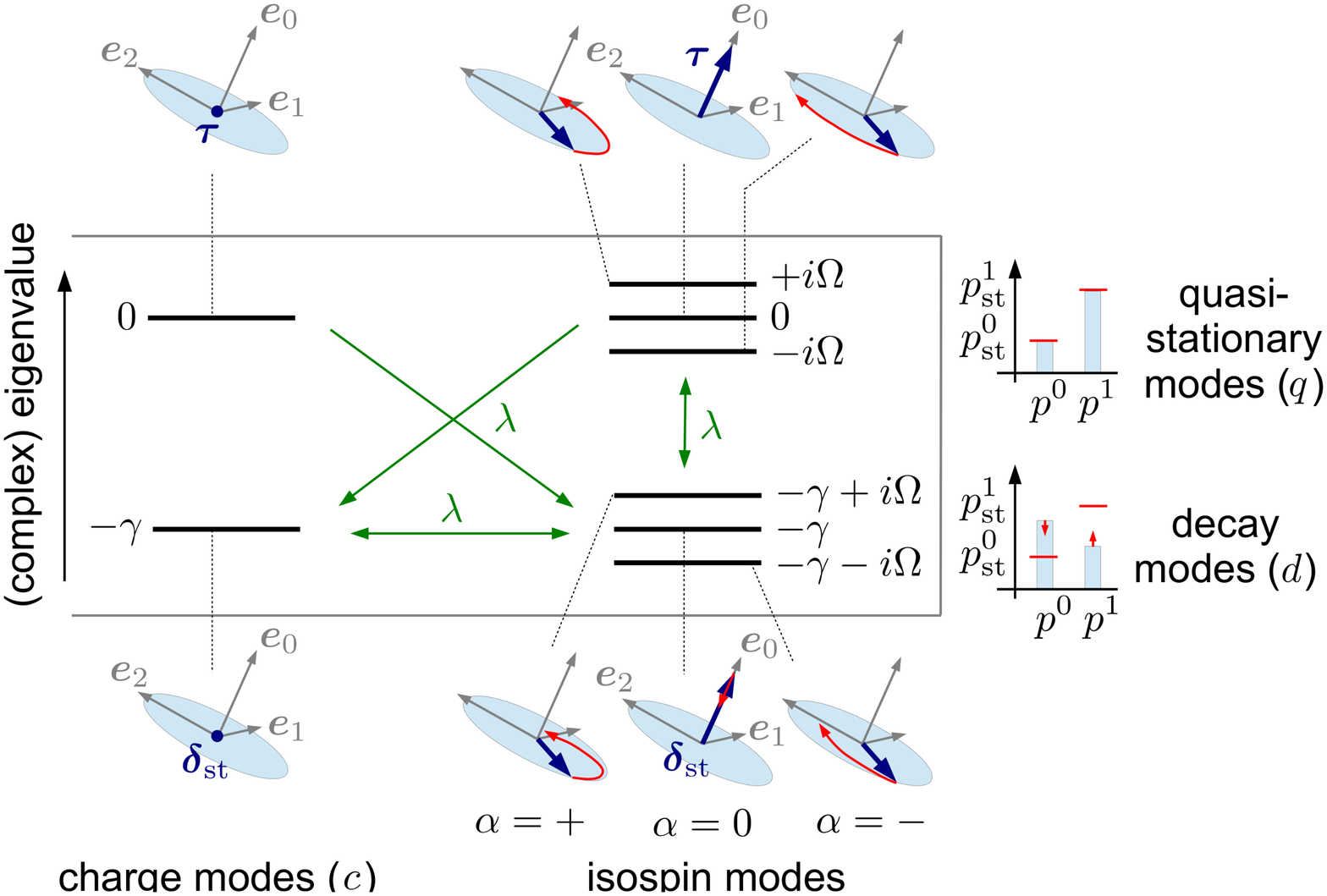}{Sketch
  of the complex eigenvalues and the associated dynamics of the joint qubit
  \tmtextit{plus} sensor QD system without readout ($\lambda = 0$). The four
  upper sketches depict the evolution of the total isospin $\vecg{\tau} =
  \vecg{\tau}^0 + \vecg{\tau}^1$ described by the coefficients of the four
  quasistationary modes in the expansion of $\vecg{\rho}$ [\new{see} Eq.
  (\ref{eq:rhogen})]. Moreover, the lower four sketches depict the evolution
  of the weighted difference $\vecg{\delta} = p^1_{\tmop{st}} \vecg{\tau}^0 -
  p^0_{\tmop{st}} \vecg{\tau}^1$ of the charge-specific isospins associated
  with the coefficients of the decaying isospin modes [\new{see} Eq.
  (\ref{eq:rhogen})]. The indices $\alpha = + 1, 0, - 1$ label the three
  different polarizations for the precessional motion. The unit vector
  $\vec{e}_0 = \hat{\vecg{\Omega}} \text{ }$is the ``bare'' qubit field and
  the vectors $\vec{e}_1$ and $\vec{e}_2$ can be chosen arbitrarily in the
  transverse plane. The two sketches on the right show the different dynamics
  of the occupation probabilities $p^0$ and $p^1$ for the stationary and the
  decaying charge mode, respectively. Note that $p^0$ and $p^1$ are not the
  coefficients in \Eq{eq:rhogen} but only their combinations $p^0 + p^1
  = 1$ and $\delta_{\text{st}} = p^1_{\tmop{st}} p^0 - p^0_{\tmop{st}} p^1$. If the
  readout is included ($\lambda \neq 0$), the modes become coupled. Since $L
  \neq L^{\dag}$, the transitions between the modes are not always possible in
  both directions as indicated by the arrows.\label{fig:modes}}
\end{center}

\subsubsection{Quasistationary and decay modes for $\lambda =
0$}\label{sec:zerocoupling}

We first solve the kinetic equation (\ref{eq:kineq}) for $\lambda = 0$ in
which case the dynamics of the occupation probabilities $p^0$ and $p^1$
decouples from the dynamics of the isospins ${\color{black} \vecg{\tau}}^0$
and ${\color{black} \vecg{\tau}}^1$ as shown by the ``unperturbed''
time-evolution generator $L_0 \assign L|_{\lambda = 0}$:
\begin{eqnarray}
  - i L_0 & = & \left(\begin{array}{cccc}
    - 2 \gamma^0 & + \gamma^1 & 0 & 0\\
    + 2 \gamma^0 & - \gamma^1 & 0 & 0\\
    0 & 0 & - 2 \gamma^0 + \vecg{\Omega} \times & + \gamma^1 \\
    0 & 0 & + 2 \gamma^0 & - \gamma^1 + \vecg{\Omega} \times
  \end{array}\right) . \nonumber\\
  &  & 
       \label{eq:L0}
\end{eqnarray}
This can be brought into diagonal form easily by noting that the cross product
operation is diagonal in the basis of the complex
unit vectors [\Eq{eq:omegacross}]
\begin{eqnarray}
  \vec{e}_0 & = & \hat{\vecg{\Omega}} \text{ \ } = \text{ \ } \vecg{\Omega}
  \text{/} \Omega,  \label{eq:e0}\\
  \vec{e}_{\pm} & = & \left( \vec{e}_1 \mp i \vec{e}_2 \right) / \sqrt{2} . 
  \label{eq:epm}
\end{eqnarray}
These are constructed from a right-handed orthonormal system $\left(
\vec{e}_0, \vec{e}_1, \vec{e}_2 \right)$ with $\vec{e}_0 =
\hat{\vecg{\Omega}}$ and an arbitrary choice of unit vectors $\vec{e}_1$ and
$\vec{e}_2$ in the plane perpendicular to $\vec{e}_0$. The complex unit
vectors satisfy the orthonormality and completeness relations ($\alpha = 0,
\pm$):
\begin{eqnarray}
  & \begin{array}{lllllll}
    \vec{e}_{\alpha}^{\dag} \cdot \vec{e}_{\alpha'} & = & \delta_{\alpha
    \alpha'}, &  & \sum_{\alpha}^{} \vec{e}_{\alpha}  \vec{e}_{\alpha}^{\dag}
    \cdot & = & \mathbbm{1},
  \end{array} &  \label{eq:eortho}
\end{eqnarray}
\newer{where here the dot $\cdot$ denotes the scalar product taken with the
object to its right.}
Writing $L_0$ in diagonal form, we find
\begin{eqnarray}
  - i L_0 & = & \sum_{p} \left[ i \Omega_p \vec{V}^q_p
  \tilde{\vec{V}}^{q \dag}_p + (i \Omega_p - \gamma) \vec{V}^d_p
  \tilde{\vec{V}}^{d \dag}_p \right],  \label{eq:l0}
\end{eqnarray}
with $p=c,0,\pm$, and eigenfrequencies
\begin{eqnarray}
  \Omega_c \text{ \ } = \text{ \ } \Omega_0 \text{ \ } = \text{ \ } 0, &  &
  \Omega_{\pm} \text{ \ } = \text{ \ } \pm \Omega, 
\end{eqnarray}
and $\tilde{\vec{V}}^{k \dag}_p$ and $\vec{V}^k_p$ are the left
and right eigenvectors, respectively, using dyadic notation. The indices $p =
c, 0, \pm$ and $k = q, d$ label in total eight different modes. Before we discuss
\newer{the physical meaning of} these modes, we first note the following general property: since $L_0$ is
diagonalizable (although $L_0^{\dag} \neq L_0$), all the left and right
eigenvectors are mutually biorthonormal,
\begin{eqnarray}
  \tilde{\vec{V}}^{k \dag}_p \cdot \vec{V}^{k'}_{p'} & = & \delta^{k k'}
  \delta_{p p'},  \label{eq:biortho}
\end{eqnarray}
and they satisfy the completeness relation
\begin{eqnarray}
  \sum_{k = q, d} \sum_{p = c, 0, \pm} \vec{V}^k_p \text{ }
  \tilde{\vec{V}}^{k \dag}_p \cdot & = & \mathbbm{1} . 
  \label{eq:complete}
\end{eqnarray}
This also follows explicitly by using \Eq{eq:eortho} and the expressions
below. This can be exploited to expand the state vector $\vecg{\rho} (t)$,
defined by \Eq{eq:rhorep}, as follows:
\begin{eqnarray}
  \vecg{\rho} & = & \text{ \ \ \ \ \ \ \ \ \ } (p^0 + p^1) \vec{V}^q_c +
  \sum_{\alpha}^{} \left( \tau^0_{\alpha} + \tau^1_{\alpha} \right)
  \vec{V}^q_{\alpha}  \label{eq:rhogen}\\
  &  & + (p^1_{\tmop{st}} p^0 - p^0_{\tmop{st}} p^1) \vec{V}^d_c +
  \sum_{\alpha} \left( p^1_{\tmop{st}} \tau^0_{\alpha} - p^0_{\tmop{st}}
  \tau^1_{\alpha} \right) \vec{V}^d_{\alpha} \nonumber\\
  & = & \text{ \ \ \ \ \ \ \ \ \ \ \ \ \ \ \ \ \ \ \ \ } \vec{V}^q_c +
  \sum_{\alpha} e^{i \Omega_{\alpha} t} v_{\alpha}^q (0)  \vec{V}^q_{\alpha} 
  \label{eq:xexpand}\\
  &  & \text{ \ \ \ \ } + e^{- \gamma t} v_c^d (0)  \vec{V}^d_c +
  \sum_{\alpha} e^{(i \Omega_{\alpha} - \gamma ) t} v_{\alpha}^d (0) 
  \vec{V}^d_{\alpha}, \nonumber
\end{eqnarray}
where $\tau^n_{\alpha} \assign \vec{e}_{\alpha}^{\dag}
\cdot \vecg{\tau}$ [\new{see} Eqs. (\ref{eq:e0}) and (\ref{eq:epm})] and the
coefficients are given by
\begin{eqnarray}
  v_{\alpha}^k (0) & = & \tilde{\vec{V}}^{k \dag}_{\alpha} \cdot
  \vecg{\rho} (0),  \label{eq:xcoeff}
\end{eqnarray}
taking the initial time $t_0=0$ here.
Importantly, equality (\ref{eq:rhogen}) is generally valid for any state
$\vecg{\rho} (t)$, while the second equality (\ref{eq:xexpand}) holds only if
$\vecg{\rho} (t) = e^{- i L_0 t}  \vecg{\rho} (0)$.

We now discuss the explicit form of the modes. The most fundamental one
is the \tmtextit{stationary charge mode} with the conjugated left eigenvector and the
right eigenvector
\begin{eqnarray}
  \tilde{\vec{V}}^q_c \text{ \ } = \text{ \ } \left(\begin{array}{c}
    1\\
    1\\
    0\\
    0
  \end{array}\right), &  & \vec{V}^q_c \text{ \ } = \text{ \ }
  \left(\begin{array}{c}
    p^0_{\tmop{st}}\\
    p^1_{\tmop{st}}\\
    0\\
    0
  \end{array}\right),  \label{eq:vqc}
\end{eqnarray}
expressed in the occupation probabilities of the SQD in the stationary limit
and for zero coupling $\lambda = 0$,
\begin{eqnarray}
    p^0_{\tmop{st}} \ \ = \ \  \frac{\gamma^1}{\gamma}, & &
    p^1_{\tmop{st}} \ \ = \ \  \frac{2 \gamma^0}{\gamma},  \label{eq:pst}
\end{eqnarray}
introducing the often recurring rate combination
\begin{eqnarray}
  \gamma & \assign & 2 \gamma^0 + \gamma^1 .  \label{eq:gamma}
\end{eqnarray}
The right zero eigenvector corresponds to a physical state, a valid density
operator, which is factorizable, $\rho_c^q = \rho_{S, \tmop{st}} \otimes
\tfrac{1}{2} \op{\mathbbm{1}}$. In this state,  the SQD is stationary,
$\rho_{S, \tmop{st}} = \sum_n p_{\tmop{st}}^n  \hat{P}^n$, and the qubit
is in
the completely mixed state with zero Bloch vector ($\vecg{\tau}^0 =
\vecg{\tau}^1 = 0$, c.f. Fig. \ref{fig:modes}, upper left). Any valid solution of the $\lambda = 0$ kinetic equation
$\dot{\rho} (t) = - i L_0 \rho (t)$ always involves this stationary charge
mode superposed with other modes. \new{These additional modes contain
the isospin precession as we explain in the next paragraph.} As one can see from expansion
(\ref{eq:rhogen}), the coefficient $v_c^q
(t) = 1$ for all $t$ and irrespective of the initial condition because the
corresponding left zero eigenvector is just the trace operation,
guaranteeing that $\rho (t)$ has unit trace for all times $t$:
\begin{eqnarray}
  \tmop{tr} (\rho (t)) \text{ \ = \ } \sum_n p^n & = & \tilde{\vec{V}}^{q
  \dag}_c \cdot \vecg{\rho} (t) \text{ \ = \ } 1. 
\end{eqnarray}
The remaining seven ``modes'' have zero trace (\new{see} \Eq{eq:biortho} with
$k = q$, $p = c$) and therefore cannot represent proper density operators on
their own. These modes cannot be excited alone: They always appear in
combination with the stationary charge mode. In this respect these modes
differ from modes encountered in, e.g., pure-state unitary evolution problems.

There are three more \tmtextit{quasistationary modes} ($k = q$), for which the
SQD remains in the stationary state $\rho_{S, \tmop{st}}$ but the qubit state
is not completely mixed, i.e., the isospin $\vecg{\tau}$ is polarized
(see upper right of Fig. \ref{fig:modes}). The
related conjugated left and right eigenvectors, respectively, read:
\begin{eqnarray}
  \tilde{\vec{V}}^q_{\alpha} \text{ \ } = \text{ \ }
  \left(\begin{array}{c}
    0\\
    0\\
    \vec{e}_{\alpha}\\
    \vec{e}_{\alpha}
  \end{array}\right), &  & \vec{V}^q_{\alpha} \text{ \ } = \text{ \ }
  \left(\begin{array}{c}
    0\\
    0\\
    p^0_{\tmop{st}} \vec{e}_{\alpha}\\
    p^1_{\tmop{st}} \vec{e}_{\alpha}
  \end{array}\right) .  \label{eq:vqa}
\end{eqnarray}
The expansion (\ref{eq:rhogen}) shows that the coefficients of these
quasistationary isospin modes are connected with the total isospin $\vecg{\tau}
= \vecg{\tau}^0 + \vecg{\tau}^1$. If the mode $\alpha = 0$ is excited, the total
isospin $\vecg{\tau}$ points along the qubit axis $\vec{e}_0$ and does not
precess. If the other two modes $\alpha = + (-)$ are excited, the total
isospin $\vecg{\tau}$ precesses (counter)clockwise in the plane perpendicular
to $\vec{e}_0$ with frequency $\Omega$. \new{Thus, if the isospin was
nonzero initially, expansion (\ref{eq:rhogen}) involves at
least one of these three modes in addition to
the quasistationary charge mode. In the case $\lambda = 0$, the former are
not damped and the magnitude of the total isospin remains unchanged, reproducing exactly the free unitary evolution of the qubit.}

In addition to these four (quasi)stationary modes, there are four more
\tmtextit{decay modes} ($k = d$) that are exponentially damped in time. As
{\color{black} Fig. \ref{fig:modes}} illustrates, the eigenvalues of these
modes are well-separated from the quasistationary modes in the complex plane
since $\Omega \ll \gamma \sim \Gamma$ as seen by inserting Eq.
(\ref{eq:dissrate}) into \Eq{eq:gamma} and noting $\Gamma / T \ll 1$.
The conjugated left and right eigenvector of the charge decay mode read,
respectively:
\begin{eqnarray}
  \tilde{\vec{V}}^d_c \text{ \ } = \text{ \ } \left(\begin{array}{c}
    + p^1_{\tmop{st}}\\
    - p^0_{\tmop{st}}\\
    0\\
    0
  \end{array}\right), &  & \vec{V}^d_c \text{ \ } = \text{ \ }
  \left(\begin{array}{c}
    + 1\\
    - 1\\
    0\\
    0
  \end{array}\right) .  \label{eq:vdc}
\end{eqnarray}
If only this mode is excited in addition to the fundamental stationary charge
mode, the SQD state deviates from the stationary state, i.e., the coefficient
$\delta_{\text{st}} \assign p^1_{\tmop{st}} p^0 - p^0_{\tmop{st}} p^1 \neq 0$ in Eq.
(\ref{eq:rhogen}), while the qubit remains in the completely mixed state
($\vecg{\tau}^0 = \vecg{\tau}^1 = 0$). Clearly, such deviations from the
stationary SQD state decay on the short time scale $1 / \gamma$ set by the SQD
tunneling dynamics (\new{see} {\color{black} Fig. \ref{fig:modes}}). Finally, there
are three decaying modes with conjugated left and right eigenvectors,
respectively, $(\alpha = 0, \pm)$ in which the isospins are polarized:
\begin{eqnarray}
  \tilde{\vec{V}}^d_{\alpha} \text{ \ } = \text{ \ }
  \left(\begin{array}{c}
    0\\
    0\\
    + p^1_{\tmop{st}}  \vec{e}_{\alpha} \\
    - p^0_{\tmop{st}}  \vec{e}_{\alpha} 
  \end{array}\right), &  & \vec{V}^d_{\alpha} \text{ \ } = \text{ \ }
  \left(\begin{array}{c}
    0\\
    0\\
    + \vec{e}_{\alpha} \\
    - \vec{e}_{\alpha}
  \end{array}\right) .  \label{eq:vda}
\end{eqnarray}
The coefficients of these modes characterize the decay of the weighted
{\tmem{difference}} $\vecg{\delta}_{\tmop{st}} = p^1_{\tmop{st}} \vecg{\tau}^0 -
p^0_{\tmop{st}} \vecg{\tau}^1$ of the charge-specific isospins [\new{see} Eq.
(\ref{eq:rhogen})]. The weighted difference $\vecg{\delta}_{\tmop{st}}$ and the
sum $\vecg{\tau} = \vecg{\tau}^0 + \vecg{\tau}^1$ are linearly independent and
together uniquely determine $\vecg{\tau}^0$ and $\vecg{\tau}^1$. Note that
$\vecg{\delta}_{\tmop{st}}$ involves the stationary occupation probabilities in
contrast to $\vecg{\delta}$ defined by \Eq{eq:faccond}. Similar to the total isospin for the
quasistationary modes, the difference $\vecg{\delta}_{\tmop{st}}$ can point
along $\vec{e}_0 = \hat{\vecg{\Omega}}$ without any precessional motion for
$\alpha = 0$ or it can precess (counter)clockwise in the plane perpendicular
to $\vec{e}_0$ for $\alpha = + (-)$ (c.f. lower right of Fig. \ref{fig:modes}).\\
We see now that for $\lambda =
0$, the total isospin $\vecg{\tau}$ decouples from the motion of all
other degrees of freedom and particularly also from $\vecg{\delta}_{\text{st}}$. In
contrast to $\vecg{\tau}$, the difference $\vecg{\delta}_{\text{st}}$ is strongly
susceptible to the tunneling dynamics of the SQD, i.e., the switching between
charge states $n = 0$ and $n = 1$. This generates the noncollinearities of
$\vecg{\tau}^0$ and $\vecg{\tau}^1$, which are responsible for the qubit
decoherence for nonzero capacitive coupling [\new{see}
discussion in {\Sec{sec:charge-specific-isospin}}, Eq.
(\ref{eq:puritydecay}) ff.]. 
What is important for the following is that that the expansion
(\ref{eq:rhogen}) carries over to the case of \tmtextit{nonzero} coupling
$\lambda \neq 0$; even in this case, it suffices to compute the evolution of
the quasistationary modes to obtain the evolution of the total
isospin \footnote{\newer{Note that the dynamics of the
quasistationary modes does not contain information about the response of
the sensor QD to the qubit since the coefficient of the stationary
charge mode $\vec{V}^q_c$ trivially equals 1. To describe the sensor
response, one must compute the dynamics of $\delta_{\text{st}}=p^1_{\text{st}} p^0
-p^0_{\text{st}} p^1$, \new{see} Eq. (\ref{eq:rhoqrhod}) (while the isospin
degrees of freedom may be projected out for that purpose).}}.
This observation provides the
starting point of the subsequent discussion where we first investigate how the
quasistationary modes -- ``containing'' the qubit dynamics -- are coupled to
the decay modes and then even explicitly eliminate the decay modes (except for
their initial state).

\subsubsection{Coupling of quasistationary and decay
modes}\label{sec:couplingmodes}

We now turn the capacitive interaction back on, $\lambda \neq 0$, and
investigate what we can say about the evolution using the above discussion of the
eigenmodes of the Liouvillian $L_0$. We can identify new dynamical variables
that characterize the evolution in the quasistationary and decay subspace,
respectively. Based on \Eq{eq:rhogen}, we introduce
\begin{eqnarray}
  \vec{X}^q \text{ = } \left(\begin{array}{c}
    p^0 + p^1\\
    \vecg{\tau}^0 + \vecg{\tau}^1
  \end{array}\right), & \vec{X}^d \text{ = } \left(\begin{array}{c}
    p^1_{\tmop{st}} p^0 - p^0_{\tmop{st}} p^1\\
    p^1_{\tmop{st}} \vecg{\tau}^0 - p^0_{\tmop{st}} \vecg{\tau}^1
  \end{array}\right),  & \label{eq:rhoqrhod}
\end{eqnarray}
and rewrite the kinetic equation as
\begin{eqnarray}
  \frac{d}{d t} \left(\begin{array}{c}
    \vec{X}^q\\
    \vec{X}^d
  \end{array}\right) & = & - i \left(\begin{array}{cc}
    L_0^{q q} + \Lambda^{q q} & \Lambda^{q d}\\
    \Lambda^{d q} & L_0^{d d} + \Lambda^{d d}.
  \end{array}\right) \left(\begin{array}{c}
    \vec{X}^q\\
    \vec{X}^d
  \end{array}\right). \nonumber\\
  &  &  \label{eq:kineq2}
\end{eqnarray}
This is the second main equation of our study.
The $\Lambda^{k k'}$ blocks are given and discussed below except for
$\Lambda^{d d}$, which is not needed here [it is given by \Eq{eq:ldd} in {\color{black} Appendix
\ref{app:liouville}}]. First, the action of the unperturbed Liouvillian on
these variables reads trivially
\begin{eqnarray}
  - i L_0^{q q} & = & \left(\begin{array}{cc}
    0 & 0\\
    0 & \vecg{\Omega} \times
  \end{array}\right),  \label{eq:l0qqmain}\\
  - i L_0^{d d} & = & \left(\begin{array}{cc}
    - \gamma & 0\\
    0 & - \gamma \mathbbm{1}_{} + \vecg{\Omega} \times
  \end{array}\right), 
\end{eqnarray}
and $L_0^{q d} = L_0^{d q}=0$ for the $\lambda = 0$ solution. The
perturbation $\Lambda$ has two effects. It first introduces a direct action on the
quasistationary variables:
\begin{eqnarray}
  - i \Lambda^{q q} & = & \left(\begin{array}{cc}
    0 & 0\\
    0 & p^1_{\tmop{st}} \vecg{\lambda} \times
  \end{array}\right) .  \label{eq:transqq}
\end{eqnarray}
This produces the mean-field backaction, which amounts to a tilting of the
internal qubit field $\vecg{\Omega}$ as anticipated in {\color{black} Sec.
\ref{sec:meanfield}}: adding $\Lambda^{q q}$ to $L^{q q}_0$ gives the
effective qubit field
\begin{equation}
  \begin{array}{lll}
    \tilde{\vecg{\Omega}} & = & \vecg{\Omega} + p^1_{\tmop{st}} 
    \vecg{\lambda} .
  \end{array} \label{eq:mf}
\end{equation}
The term $\Lambda^{q q}$, i.e., the term \emph{linear} in $\lambda$ thus does not lead to dissipative dynamics of the
isospin contrary to what one might naively expect. The isospin
decoherence is at least quadratic in $\lambda$ (see below). We also note that the effective qubit field
$\tilde{\vecg{\Omega}}$ is modified by $O (\Gamma^2 / T)$ processes
(cotunneling broadening, level shift), which affect the stationary occupation
probability $p_{\tmop{st}}^1$ [\new{see} \Eq{eq:pst}]. Moreover, if one is
close to resonance, the probability $p^1_{\tmop{st}}$ may be a
sizable fraction of 1, which it approaches in the Coulomb blockade regime for
$\varepsilon \ll \mu_s, \mu_d$. Since we allow for $\lambda \sim \Omega$, this
implies that the correction $p^1_{\tmop{st}}  \vecg{\lambda}$ leads
to a large change in the qubit frequency and the direction of the qubit axis for a
large range of parameters.

The second effect of the perturbation is due to the coupling of the
quasistationary variables to the decaying variables due to the off-diagonal
blocks $\Lambda^{q d}$ and $\Lambda^{d q}$. As a consequence, the decaying
variables cannot be just ignored after a time $\sim 1 / \gamma$ since they are
permanently excited by virtual transitions from the quasistationary modes into
the decay modes and back. These virtual processes are responsible for the
qubit relaxation and decoherence; the multiplet of quasistationary eigenvalues
in \Eq{eq:l0} acquires an imaginary part $\Omega_{\alpha} \rightarrow
\tilde{\Omega}_{\alpha} + i \gamma_{\alpha}$ for $\alpha = 0, \pm$, which
induces a shrinking of the isospin Bloch vector. From \Eq{eq:kineq2}
and {\color{black} Fig. \ref{fig:modes}}, we expect that $1 / T_1, 1 / T_2 \sim
\Lambda^{q d} \Gamma^{- 1} \Lambda^{d q} \propto \lambda^2 / \Gamma$, i.e., if
the couplings $\Lambda^{q d}$ and/or $\Lambda^{d q}$ \ are small, then so will
the backaction be. Before we make this more precise [\new{see} Eq.
(\ref{eq:leffzmain})], we investigate the detailed form of these couplings,
which contains the first main result of the paper.

\subsubsection{Mitigation of cotunneling noise by coherent
backaction}\label{sec:mitigation1}

The transition matrix from the quasistationary modes into the decay modes
reads [\new{see} {\color{black} Appendix \ref{app:leffcalc}}]
\begin{eqnarray}
  \Lambda^{d q} & = & \left(\begin{array}{cc}
    0 & (1 + p^0_{\tmop{st}}) c \vecg{\lambda} \cdot\\
    (1 + p^0_{\tmop{st}}) c \vecg{\lambda} & - r \vecg{\lambda} \times
  \end{array}\right),  \label{eq:transdq}
\end{eqnarray}
with the transition factor
\begin{eqnarray}
  r & = & p^0_{\tmop{st}} p^1_{\tmop{st}} - \kappa \left( \frac{1}{2}
  p^1_{\tmop{st}} - p^0_{\tmop{st}} \right),  \label{eq:r}
\end{eqnarray}
while the transition matrix back into quasistationary modes is given by
\begin{eqnarray}
  \Lambda^{q d} & = & \left(\begin{array}{cc}
    0 & 0\\
    0 & - \vecg{\lambda} \times
  \end{array}\right) .  \label{eq:transqd}
\end{eqnarray}
We first note that $\Lambda$, like $L_0$, is non-Hermitian, $\Lambda^{\dag}
\neq \Lambda$, since the qubit-sensor evolution is nonunitary due the
tunneling. As a consequence, the two transition matrices are markedly
different: while transitions from the decay modes to quasistationary modes
($\Lambda^{q d}$) do {\tmem{not}} depend on the parameters of the SQD (level
position, bias voltage, and tunneling rates $\Gamma$), transitions from the
quasistationary modes ($\Lambda^{d q}$) exhibit a strong dependence on the
sensor QD parameters that we discuss below. That $\Lambda^{q d}$ is entirely
induced by the readout interaction $H_I$ can be shown to be a consequence of
the probability conservation together with the conservation of the isospin
during tunneling processes. The latter is specific to the indirect measurement
setup.

With \Eq{eq:transdq} in hand, we can now precisely pinpoint what we
mean by \emph{stochastic backaction}: the diagonal component of $\Lambda^{d q}$ can be
split into a first term, $p^0_{\tmop{st}} p^1_{\tmop{st}} \vecg{\lambda}$,
associated with the stochastic backaction, and a second term, $- \kappa
(p^1_{\tmop{st}} / 2 - p^0_{\tmop{st}}) \vecg{\lambda}$, associated with the
coherent backaction as signaled by the factor $\kappa$. 
\newer{The combination $p^0_{\tmop{st}} p^1_{\tmop{st}}$ appears as a simple consequence of the charge
fluctuations of the SQD, which are characterized by $\langle n^2 \rangle -\langle n \rangle^2
= (1 -\langle n \rangle) \langle n \rangle 
= p^0_{\tmop{st}} p^1_{\tmop{st}}$ for a two-level system 
(\new{see} Sec. \sec{sec:semiclassical}).
The rates $\gamma^{0,1}$ determining $p^0_{\tmop{st}} p^1_{\tmop{st}}$ 
incorporate both the effect of the SET tunneling as well as that of next-to-leading order
corrections.
 The stochastic term is} multiplied with $\vecg{\lambda}$ because to act back on
the qubit, the ``tunneling noise'' has to act together with the internal
interaction $H_I = \hat{n}  \vecg{\lambda} \cdot \op{\vecg{\tau}} / 2$ to evoke
a $\lambda$-induced transition mediated by $\Lambda^{d q}$.

The most striking finding is that $\Lambda^{d q}$ is strongly suppressed when
tuning the SQD towards Coulomb blockade. To see this, we first note that Eq.
(\ref{eq:transdq}) incorporates the isospin-to-charge conversion rates
(\ref{eq:conversion}), $c \sim \sum_r \frac{\Gamma_r}{2 T} \nobracket (- f_r^+
\nobracket)'$, which depend on the derivative of the Fermi function. These
rates are thus {\tmem{exponentially}} suppressed in the Coulomb blockade
regime. We next inspect the diagonal element of $\Lambda^{d q}$. It is useful
to first consider the expansion of the transition factor $r$ to zeroth order
in $\Gamma / T$:
\begin{eqnarray}
  r & = & \frac{2 \Gamma^+ \Gamma^-}{(2 \Gamma^+ + \Gamma^-)^2} + O \left(
  \frac{\Gamma}{T} \right),  \label{eq:rlowest}
\end{eqnarray}
with $\Gamma^{\pm} \assign \sum_r \Gamma_r f_r^{\pm}$ and $2 \Gamma^+ +
\Gamma^- > \Gamma \assign \sum_r \Gamma_r$. The terms in Eq.
(\ref{eq:rlowest}) derive only from the stochastic part $p^0_{\tmop{st}}
p^1_{\tmop{st}}$ in \Eq{eq:r}. Thus, in the single-electron tunneling
approximation, the factor $r$ is exponentially suppressed with gate voltage
since either $p^0_{\tmop{st}}$ or $p^1_{\tmop{st}}$ becomes exponentially
small when going off-resonance. One would expect that this exponential
dependence is removed by including cotunneling corrections and the coherent
backaction, which scale algebraically with $\varepsilon - \mu_r$ \newer{and
start to dominate over the} single-electron tunneling rates as one
moves into the Coulomb blockade regime. In our calculation, we include these
terms as well, but {\tmem{still}} obtain an exponential suppression of the
transition factor. Indeed, expanding \Eq{eq:r} to the next
order in $\Gamma/T$, we find
\begin{eqnarray}
  r & = & \frac{1}{(2 \Gamma^+ + \Gamma^-)^2} \left[ 2 \Gamma^+ \Gamma^-
  \left( 1 + \frac{\sum_r \Gamma_r \phi'_r}{T} \right) \right. \nonumber\\
  &  & + \left. (\Gamma^- - 2 \Gamma^+) (\Gamma^- + \Gamma^+)
  \frac{(\Gamma^+)'  \sum_r \Gamma_r \phi_r}{2 \Gamma^+ + \Gamma^-} \right]
  \nonumber\\
  &  & + O \left( \frac{\Gamma^2}{T^2} \right),  \label{eq:rexp}
\end{eqnarray}
with 
$\Gamma' = d \Gamma^+ (\varepsilon) / d \varepsilon = \sum_r \Gamma_r
(f^+_r)^{\prime} / T $. 
Clearly, $\Gamma^+$, $\Gamma^-$, and $\Gamma'$ are determined
by the Fermi functions and their derivatives. Thus, transitions from the
quasistationary modes into the decay modes become exponentially small when
tuning the SQD into the Coulomb blockade regime. What this implies is that any
deviation from the exponential suppression must be due to even higher order
tunneling contributions (i.e., beyond cotunneling) and thus must be a
higher power law \newer{$\sim 1/(\varepsilon - \mu_r)^n$ with $n \geq
2$}. 
The experimentally important conclusion that we can draw from this
is that the sensor can be switched off better with the gate voltage than
naively expected [\new{see} \Eq{eq:dissrate}].

What has happened in \Eq{eq:rexp} is that the coherent backaction $\kappa \sim \sum_r \Gamma_r
\phi'_r / T \propto 1 / (\varepsilon - \mu_r)$, which depends algebraically on
$\varepsilon$, has completely canceled out the algebraically scaling
cotunneling corrections to the stochastic backaction in the first term of Eq.
(\ref{eq:r}).
Hence, in \Eq{eq:transdq}, the coherent backaction term
($\Gamma \lambda / T$) counteracts the \tmtextit{change} in the stochastic
backaction term due to a change in the sensor QD rates by the cotunneling
($\lambda / \Gamma \times \Gamma^2 / T \sim \Gamma \lambda / T$). This can be
seen by explicitly comparing \Eq{eq:rlowest} to \Eq{eq:rexp} \new{and is clearly
visible in \Fig{fig:types}.}
\newer{We emphasize that for this cancellation also non-Markovian
effects induced by the electrodes are important, which modify the
coherent backaction (see Sec. \sec{sec:nmdiscuss}). Without these, the
transition factor $r$ exhibits a different dependence on the level
position $\varepsilon$ that can lead to a violation of positivity of the
qubit-SQD density operator as we discuss further in
Appendix \sec{sec:impnm}.}
\new{Moreover, the cancellation does not imply that the backaction is
precisely the same as when only accounting for the
lowest-order $\Gamma$ approximation (see Fig. \ref{fig:types}); the renormalization of the level position and
non-Markovian contributions, both scaling
exponentially, still modify the backaction.}

By formulating the problem in \Eq{eq:kineq2} in terms of the $\lambda = 0$ eigenmodes one most
clearly sees how the cotunneling and coherent backaction, formally terms of
different order, conspire to effectively cancel out. Note also that the dissipative
backaction (through $c$) appears on its own. This highlights the importance of keeping track
of all three types of backaction that are revealed only \emph{after} integrating out
the electrodes coupled to the sensor QD to obtain our central
\Eq{eq:kineq}. 
\new{It should be noted that the dissipative backaction couples the quasistationary modes to the decay modes
and therefore is not relevant for the leading-order $\lambda^2/\Gamma$
dephasing times as Appendix \ref{app:leffcompute} shows; see also the discussions after Eq. (\ref{eq:conversion}) and after
 Eq. (\ref{eq:oneovert2telegraph}) below.
Rather, the dissipative backaction must be kept to be able to calculate the response of the dissipative
sensor current of which it represents the flip side, as explained in the introduction.}
As
emphasized in {\Sec{sec:review}}, we were careful
throughout our analysis to include all terms which depend on the function
$\phi'_r$ with algebraic tails that could possibly cancel out. In
{\color{black} Appendix \ref{app:validity}}, we further discuss the cancellation in
view of our weak-measurement, weak-coupling assumption (see also {\color{black}
Sec. \ref{sec:weakcoupling}}).\\
An important conclusion, which we draw in {\color{black} Sec.
\ref{sec:comparison}}, is that this cancellation of cotunneling noise and coherent backaction cannot be understood within simple classical fluctuator model. Although this approach could, in principle, be extended
to account for the cotunneling-induced noise by modifying the switching rates,
it seems not possible to include the coherent
backaction. Moreover, other approaches that aim at directly calculating the
qubit Bloch vector $\vecg{\tau}$ must make an assumption about the qubit
environment, in particular the sensor QD. Here one is liable to miss the above
cancellation as we also discuss in {\Sec{sec:whynotbr}}.\\
It is furthermore interesting to observe that this cancellation appears even
though the coherent-backaction induced torque terms in the kinetic equations
(\ref{eq:kineq}) scale with $\lambda$, while the cotunneling corrections do
not. However, to affect the qubit, the ``cotunneling noise'' has to act
together with the internal interaction $H_I \sim \hat{n}  \vecg{\lambda} \cdot
\op{\vecg{\tau}}$ to evoke a $\lambda$-induced transition mediated by
$\Lambda^{d q}$. This is why they both affect the measurement backaction to
first order in $\lambda$. Note also that the transition factor $r$ is not only
independent of the SQD-qubit interaction $\vecg{\lambda}$ (it appears as a
factor in $\Lambda^{d q}$) but also of the internal qubit field
$\vecg{\Omega}$. This means that the relative importance of coherent backaction
over the stochastic backaction cannot be altered by measuring weaker or
stronger (i.e., by changing $\lambda$);
in the weak measurement limit, these effects physically come together
and should be calculated together 
{\footnote{The coherent
backaction does not only affect the transitions into the decaying subspace but
also affects the evolution in the decay space by modifying $\Lambda^{d d}$ in
\Eq{eq:kineq2}. However, these terms yields only a small correction as
compared to the large free evolution term $| L_0^{q q} | \sim \gamma$.}}.
Other experimental parameters alter this competition and their effect is studied in the next section.\\
In short, the delicate interplay of the qubit \tmtextit{plus} sensor renormalization
$\propto \Gamma$ and sensor cotunneling rates $\propto \Gamma^2$ in indirect
capacitive detection may be rationalized as follows. By keeping track
of the sensor-quit coherences (since both are quantum systems), we find that
coherent effects counteract decoherence, which is not really that unexpected.
This may in fact present a key difference of a sensor with quantized levels
from a single-electron transistor with a continuous spectrum. A comparison of
both types of detectors regarding the importance of renormalization effects is
therefore an interesting future task. Finally, we emphasize that for the above
cancellation the modeling of the qubit as a charge qubit is not relevant as
long as the isospin couples to the charge of the SQD.

\subsection{Experimental control over backaction strength}

Since \Eq{eq:rexp} is a key result of
this paper, characterizing -- together with $c$ -- the strength of the backaction
[beyond the mean-field effect in Eq. (\ref{eq:mf})], we now investigate its
dependence on experimental parameters in some detail.

\subsubsection{Gate-voltage dependence of transition factor $r$\label{sec:gatevoltagedep}.} In
{\color{black} Fig. \ref{fig:rates}}(a), we plot the dependence of the
transition factor $r$ on gate voltage. The figure shows two curves, one
including the coherent backaction (blue, $\kappa \neq 0$) and the other
excluding the coherent backaction (green, $\kappa = 0$). Clearly, the factor
$r$ is the largest in the sequential tunneling regime of the SQD ($| V_g |
\lesssim V_b / 2$). Here, transitions from the quasistationary modes into the
decay modes are induced by the fast succession of tunneling electrons, which
impose a strong noise on the qubit. The coherent backaction and cotunneling
are negligible in this regime.
This drastically changes when tuning the SQD into the Coulomb blockade regime
($| V_g | \gtrsim V_b / 2$). The full transition factor $r$ is actually
exponentially suppressed with gate voltage [linear on the scale of
{\color{black} Fig. \ref{fig:rates}}(a)]. By contrast, the gate-voltage
dependence is markedly nonexponential when neglecting the coherent backaction,
characteristic of cotunneling noise, \new{see also} the discussion of Eq.
(\ref{eq:rexp}).\\
Experimentally, we expect the cancellation to be reflected in the voltage
dependencies of the qubit relaxation and dephasing rates provided the sensor
can be made the dominant environment (which should be the case for a good
qubit, for which noise from manipulation ``channels'' can be switched off).
Measuring the qubit decoherence rates could clearly distinguish between an
algebraic and exponential dependence in an experiment. Here, we expect the
measurement-induced decay rates to scale exponentially into the Coulomb
blockade regime until higher-order tunneling processes at least of order
$\Gamma^3 / T^2$ become relevant. They can lead to \new{a} crossover to an
algebraic scaling $\propto 1 / | \varepsilon - \mu_r |^n$ with $n > 1$ deep in
the Coulomb blockade regime, \new{see} {\color{black} Appendix \ref{app:validity}}. In
any case, our numerical examples illustrate that a QD detector can be {\tmem{switched off}}
more efficiently with a gate voltage than naively expected.
\begin{center}
  \Figure{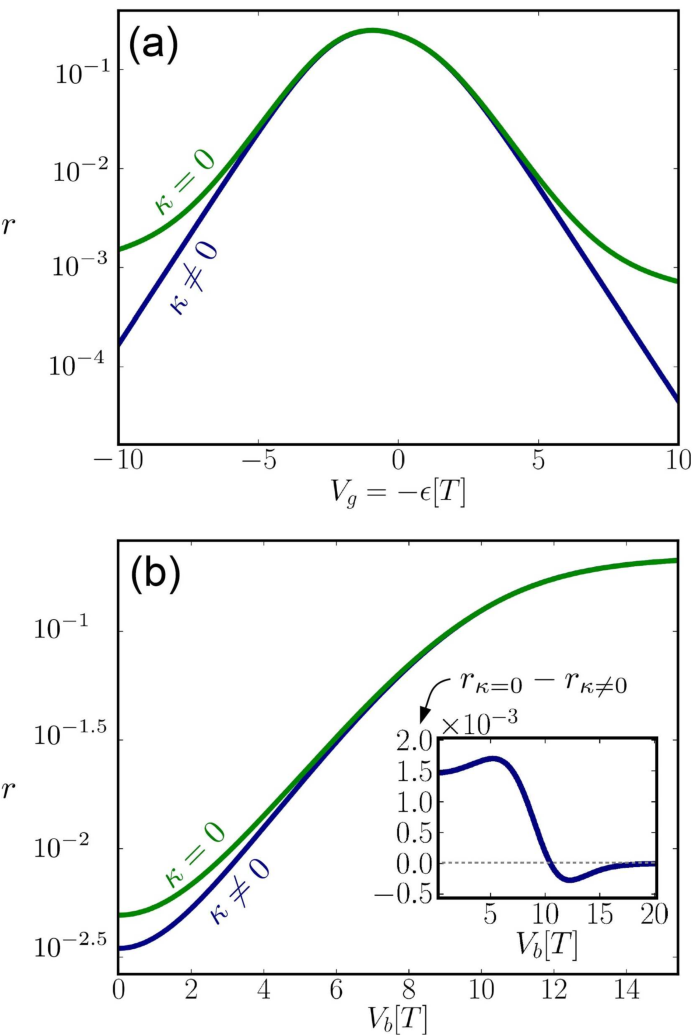}{Voltage
  dependence of the transition factor $r$, \Eq{eq:r}, which determines
  the measurement-induced backaction of the sensor QD on the qubit (e.g.,
  decoherence rates). We include the coherent backaction in the blue curves
  ($\kappa \neq 0$), while we exclude it by hand for the green curves ($\kappa
  = 0$). We show the dependence of $r$ in (a) on gate voltage $V_g$ for bias
  voltage $V_b = 2.5 T$ and in (b) on bias voltage $V_b$ for gate voltage $V_g
  = 5 T$. In all plots we use $\Gamma_s = \Gamma_d = \bar{\Gamma} = 0.02 T$
  and bandwidth $W = 1000 T$. The inset in (b) shows the difference of the two
  curves for $\kappa = 0$ and $\kappa \neq 0$ and illustrates that the
  coherent backaction can also enhance the backaction for larger bias
  voltages. Note that $r$ is independent of $\lambda$ and $\Omega$
 [choosing,
  e.g., $\lambda \sim \Omega \sim 0.1 \bar{\Gamma}$ would be a parameter
  combination consistent with the conditions for the kinetic equations
  (\ref{eq:kineq}) to be valid: $\bar{\Gamma} / T \ll \lambda / \Gamma \ll
  1$].\label{fig:rates}}
\end{center}
\subsubsection{Bias-voltage dependence of $r$\label{sec:biasdep}}We
  continue with the
discussion of the bias dependence of the transition factor $r$, which we
show in {\color{black} Fig. \ref{fig:rates}}(b). For small bias voltages, the
SQD is Coulomb-blockaded and the coherent backaction strongly suppresses $r$.
When the bias is increased, sequential tunneling sets in and when the level
position $\varepsilon \approx V_b / 2$ is resonant with the electrochemical
potential of the drain, the transition factor saturates. Here, the correction
due to the coherent backaction actually becomes {\tmem{positive}}, as shown in
the inset in {\color{black} Fig. \ref{fig:rates}}(b). Yet, one should note
that in the sequential tunneling regime the coherent backaction has only a
small impact. For even larger bias voltages, the correction from the coherent
backaction drops to zero. This is in accordance with the general finding
that renormalization effects can be neglected in the large-bias limit
{\cite{Gurvitz08,Hell14a}} because $\kappa = \sum_r \Gamma_r \phi'_r / T \sim
\sum_r \Gamma_r / (\varepsilon - \mu_r) \propto 1 / V_b$ is suppressed [\new{see}
\Eq{eq:phirprime}].

\subsubsection{Tunnel coupling dependence of transition factor transition factor $r$}We finally discuss the impact of
the coherent backaction when changing the tunnel couplings and their
asymmetries. Since the coherent backaction is linear in $\kappa \sim \Gamma /
T$ [\new{see} \Eq{eq:kappa}], increasing the average tunnel coupling
$\bar{\Gamma} = (\Gamma_s + \Gamma_d) / 2$ and lowering the temperature both
increase renormalization effects in a trivial way (within the
limit $\Gamma/T \ll \Delta/\Gamma \ll 1$). By contrast, the {\tmem{asymmetry}} of the tunneling rates in the
generic experimental situation, quantified
by
\begin{eqnarray}
  g & : = & \frac{\Gamma_s - \Gamma_d}{\Gamma_s + \Gamma_d}, 
\end{eqnarray}
may have a nontrivial effect.
Controlling the asymmetry has also been suggested {\cite{Gurvitz05}} as an
experimental strategy for optimizing sensor efficiency in the limit $\Gamma_d / \Gamma_s >
1$. In the stationary limit, we previously found {\cite{Hell14a}} that for
$\Gamma_d / \Gamma_s < 1$ the impact of the coherent backaction is strongly
enhanced.

The effect of asymmetries on the transition factor $r$ [\Eq{eq:r}] strongly depends on the chosen bias and gate
voltages. To illustrate this point, we plot in {\color{black} Fig.
\ref{fig:asym}}(a) the relative change in $r$,
\begin{eqnarray}
  K & = & \frac{r_{\kappa = 0} - r_{\kappa \neq 0}}{r_{\kappa \neq 0}}, 
\end{eqnarray}
due to the coherent backaction as a function of gate voltage $V_g$ for the
different values of $g$ as indicated. When introducing a nonzero junction
asymmetry $g \neq 0$, the exponential suppression with gate voltage effected
by the coherent backaction, found in {\color{black} Fig. \ref{fig:rates}}(a),
is simply rigidly shifted horizontally without changing its shape
considerably. This can be understood from the fact that the maximum of $r$ as
a function of $V_g$ is shifted by an asymmetry $g$ due to a basic effect of
Coulomb interaction
{\footnote{It is known that at finite temperature the
condition for a resonance is \tmtextit{not} $\varepsilon = \mu_r$, but there
is an offset linear in $T$ with a coefficient that grows with junction
asymmetry. This is related to Coulomb interaction effects on the QD and
appears already in $O (\Gamma)$ {\cite{Bonet02}}. This causes the maximum of
the decoherence rates to lie at
nonzero $V_g$ (see {\color{black} Fig. \ref{fig:rates}}) and also causes the shift with changing junction asymmetry $g$
in {\color{black} Fig. \ref{fig:asym}}.}}. 
In the vicinity of this maximum,
leading-order processes dominate and cotunneling and the coherent backaction
effects can be ignored. One then finds the rule that for large asymmetries
($|g| \gg 1$), the SQD level position $\varepsilon = - V_g$ effectively lies
closer to resonance with the electrochemical potential of the more strongly
coupled electrode. Now, for positive bias $V_b$, as in {\color{black} Fig.
\ref{fig:asym}}(a), the electrochemical potential $\mu_s = + V_b \text{/} 2$
($\mu_d = - V_b \text{/} 2$) of the source (drain) is resonant with the QD
level for negative (positive) values of $V_g$. Thus, when the coupling to the
source is larger (smaller) than that to the drain, the maximal transition
factor $r$ is achieved for negative (positive) $V_g$ as {\color{black} Fig.
\ref{fig:asym}}(a) demonstrates.
\begin{center}
  \Figure{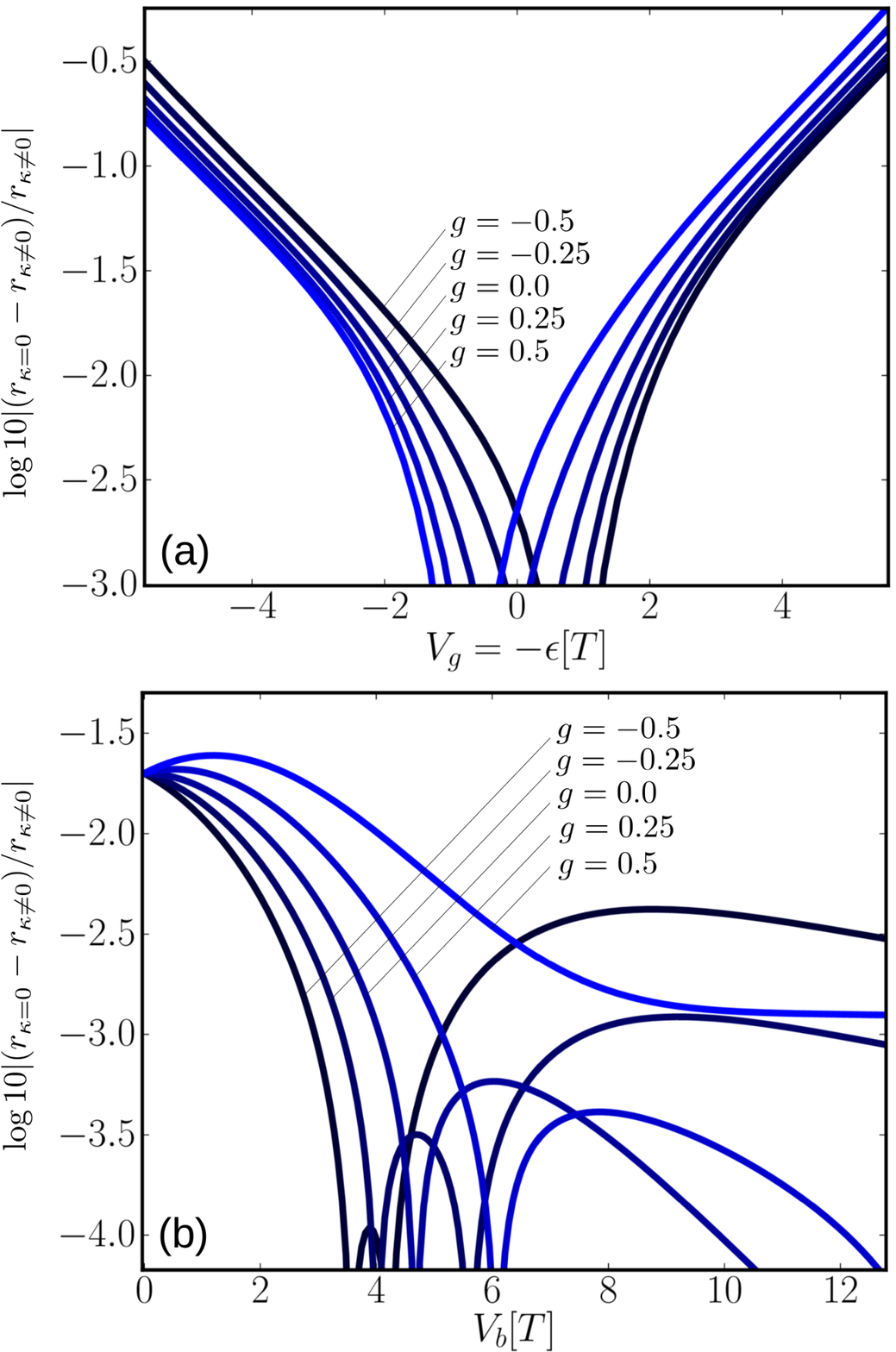}{Relative
  change $K$ in the transition factor $r$, \Eq{eq:r}, by the coherent
  backaction as a function of (a) the gate voltage $V_g$ for $V_b = 2.5 T$ and (b)
  of the bias voltage $V_b$ for $V_g = 1.5 T$. The different curves correspond
  to the indicated values of the tunneling asymmetry $g = (\Gamma_s -
  \Gamma_d) \text{/} (\Gamma_s + \Gamma_d)$ for fixed average tunneling rate
  $\bar{\Gamma} = (\Gamma_s + \Gamma_d) \text{/} 2 = 0.02 T$ and bandwidth $W
  = 1000 T$.\label{fig:asym}}
\end{center}
By contrast, {\color{black} Fig. \ref{fig:asym}}(b) shows that the impact of
the asymmetry $g$ on $K$ as function of the bias voltage is more complicated
close to resonance. In general, $K$ decreases with $V_b$, as {\color{black}
Fig. \ref{fig:rates}}(b) already \new{showed} for $g = 0$. In the limit $V_b
\rightarrow \infty$, one expects $\kappa \sim 1 / V_b \rightarrow 0$ and
therefore $K \rightarrow 0$. Moreover, we find $K > 0$ for small bias
[generally valid] and for large bias [specific to the parameters chosen in
{\color{black} Fig. \ref{fig:asym}}(b)]. Thus, both for large and small bias
the coherent backaction suppresses the measurement backaction. However, for
intermediate bias voltages, $K$ shows strong drops as {\color{black} Fig.
\ref{fig:asym}}(b) illustrates. This suppression appears since $\kappa$ and
$\nobracket (p^1_{\tmop{st}} / 2 - p^0_{\tmop{st}}) \nobracket$ change their
sign in the vicinity to resonance at $\varepsilon = \mu_r$ and therefore
become both very small. This suppresses $K$ and for a small intermediate bias
regime $K$ can become {\tmem{negative}} as pointed out already in
{\Sec{sec:biasdep}}. Figure {\color{black}
\ref{fig:asym}}(b) reveals that the position and even the existence of these
drops and sign changes depends crucially on the asymmetry $g$ and the gate and
bias voltage polarity. In particular, for $V_g > 0$, as assumed in
{\color{black} Fig. \ref{fig:asym}}(b), the source cannot become resonant with
the level for positive bias. Accordingly, the drop is absent for $g = 0.5$ in
{\color{black} Fig. \ref{fig:asym}}(b), i.e., when the source is coupled more
strongly to the SQD than the drain.

In summary, the coherent backaction starts to cancel the cotunneling-induced
stochastic backaction on the flanks of the SQD resonances where one enters
Coulomb blockade  -- the optimal spot for sensing -- with corrections depending on the junction asymmetry, see
{\color{black} Fig. \ref{fig:asym}}. Junction asymmetries commonly found in
experiments strongly affect the relative importance of the coherent backaction
terms and thus also the qubit decay rates.

\subsection{Effective evolution of quasistationary
modes}\label{sec:effliouville}

Since the qubit isospin is contained in the quasistationary degrees of freedom
[\new{see} \Eq{eq:rhoqrhod}], it is of interest to find a description only
for their evolution. Importantly, we do this without making further
approximations, i.e., we reproduce exactly the result for $\vec{X}^q (t)$ as
one does when solving \Eq{eq:kineq} for $\vecg{\rho}$.
For all practical purposes, one should solve the latter equation [or equivalently \Eq{eq:kineq2}]:
Eqs. \eq{eq:xqt} and \eq{eq:xqefft} below merely serve to highlight some properties of this solution,
in particular, its non-Markovian nature
\newer{and the impracticability for really getting rid of the sensor dynamics}.
It also serves as a basis for \Sec{sec:results}.

Equation (\ref{eq:kineq2}) is a coupled set of differential equations for
$\vec{X^{}}^q (t)$ and $\vec{X^{}}^d (t)$.
The equations can be solved most
conveniently by Laplace transformation, defined by $f (z) = \int_0^{\infty} d
t e^{i z t} f (t)$ for any (well-behaved) time-dependent function $f (t)$. This yields
\begin{eqnarray}
  \vec{X^{}}^q (z) & = &  \frac{i}{z - L^{q q}_{\tmop{eff}} (z)}
  \vec{X^{}}^q_{\tmop{eff}} (z),  \label{eq:xqz}
\end{eqnarray}
with a Laplace-frequency dependent effective Liouvillian
\begin{eqnarray}
  L^{q q}_{\tmop{eff}} (z) & = & L^{q q}_0 + \Lambda^{q q} + \Lambda^{q d}
  \frac{1}{z - L_0^{d d} - \Lambda^{d d}} \Lambda^{d q},  \label{eq:leffzmain}
\end{eqnarray}
and matrices given by Eqs. (\ref{eq:l0qqmain}) -- (\ref{eq:transqq}),
(\ref{eq:transdq}), (\ref{eq:transqd}), and (\ref{eq:ldd}). The
frequency-dependent initial condition is given by
\begin{eqnarray}
  \vec{X}^q_{\tmop{eff}} (z, 0) & = & \vec{X}^q (0) + \Lambda^{q d} \frac{1}{z
  - L_0^{d d} - \Lambda^{d d}} \vec{X}^d (0),  \label{eq:zinitial}
\end{eqnarray}
where we take the initial time to be $t = 0$. The intermediate steps of the
derivation of Eqs. (\ref{eq:xqz}) -- (\ref{eq:zinitial}) are given in
{\color{black} Appendix \ref{app:liouville}}. There we show how to reproduce the
above result in the general framework of a projection approach following
that of
Nakajima and Zwanzig {\cite{Fickbook,Breuer,Nakajima58,Zwanzig60}}. The
projection approach simply separates the dynamics in the complementary
subspaces spanned by the quasistationary and decaying modes. Importantly, this
projection technique treats all the different types of backaction we discussed
so far on the same footing, which allows us to go beyond the approaches of
{\color{black} Refs. {\cite{Emary08}}} and {\cite{Makhlin00}}.

Transforming the solution (\ref{eq:xqz}) back to time space and exploiting the
convolution theorem yields [\new{see} {\color{black} Appendix
\ref{app:liouville}}]
\begin{eqnarray}
  \vec{X^{}}^q (t) & = & \int_0^t d t' \Pi^{q q}_{\tmop{eff}} (t - t') 
  \vec{X^{}}^q_{\tmop{eff}} (t'),  \label{eq:xqt}\\
  \vec{X}^q_{\tmop{eff}} (t') & = & \vec{X}^q (0) \delta (t' - 0) + \Pi^{q
  d}_{\tmop{eff}} (t') \vec{X}^d (0),  \label{eq:xqefft}
\end{eqnarray}
where $\delta(t'- 0)$ indicates a $\delta$-function with infinitesimal shift and
\begin{eqnarray}
  \Pi^{q q}_{\tmop{eff}} (t) & = & \int_{- \infty}^{\infty} \frac{d z}{2 \pi}
  e^{- i z t}  \frac{i}{z - L^{q q}_{\tmop{eff}} (z)}, \\
  \Pi^{q d}_{\tmop{eff}} (t) & = & \int_{- \infty}^{\infty} \frac{d z}{2 \pi}
  e^{- i z t} \Lambda^{q d} \frac{1}{z - L_0^{d d} - \Lambda^{d d}} . 
  \label{eq:slipqd}
\end{eqnarray}
Equations \eq{eq:xqt} and \eq{eq:xqefft} are the third set of main equations.
We stress that they are generally valid in the sense
that they do not involve any approximations beyond those needed for the validity of the kinetic
equations (\ref{eq:kineq}): Eq. (\ref{eq:xqt}) exactly reproduces
$\vec{X^{}}^q (t)$ as obtained from the solution of \Eq{eq:kineq2}. The
expression for $\vec{X^{}}^q (t)$ simplifies drastically for times $t \gg 1 /
\gamma$ if one accounts only for the leading-order contributions in $\Delta /
\gamma$. Such an expansion is valid only in the weak-measurement limit $\Delta
/ \gamma$ and complies with our kinetic equations in in the high-temperature
limit $\Gamma / T \ll \Delta / \Gamma$ as we discuss further below in
{\Sec{sec:results}}. Before doing so, let us first note a
couple of general properties of the above solution.

\emph{Non-Markovian dynamics and time scales.} The effective Liouvillian (\ref{eq:leffzmain}) confirms
the discussion in {\Sec{sec:couplingmodes}}: the
unperturbed evolution in the quasistationary subspace is perturbed directly by
$\Lambda^{q q}$, which can be absorbed into a redefinition $\bar{L}_0^{q q} =
L^{q q}_0 + \Lambda^{q q}$ that just leads to the mean-field tilting of the
qubit axis.

Moreover, the third term in \Eq{eq:leffzmain} gives an explicit
expression for the indirect and \newer{in general non-Markovian} perturbation of the quasistationary evolution by
virtual transitions via the decay modes. This interpretation follows most
clearly from an alternative derivation in time space given in {\color{black}
Appendix \ref{app:derivationtimespace}} \newer{and illustrates that the time
delay between these transitions --- giving rise to non-Markovian effects ---
converts into the frequency dependence
of this expression.}
This term entails the effect of the
stochastically fluctuating deviations from the mean field (see {\color{black}
Sec. \ref{sec:charge-specific-isospin}}) and separates it from the mean-field
effect contained in $\Lambda^{q q}$. Since the denominator in this
expression does not grow exponentially (it rather tends to be constant because
$| L_0^{q q} | \sim \gamma$), the voltage dependence of this term is largely
determined by that of $\Lambda^{d q}$, which we discussed above.

The third term in \Eq{eq:leffzmain} has, in general, several effects:
the eigenvalues of $\bar{L}^{q q}_0$ in general acquire (i) a real part
leading to an additional shift of the qubit frequency, and (ii) an imaginary
part, which corresponds to the relaxation and dephasing rates of the qubit. To
extract both effects, one has to inverse the Laplace-transformed function
(\ref{eq:xqz}), which leads to an integral that should be computed by the
applying the residue theorem. The residues are determined by the zeros of the
denominator satisfying $z_p = L^{q q}_{\tmop{eff}} (z_p)$ and determine the
time scales of the qubit evolution. (iii) Moreover, the third term may not commute
with $\bar{L}^{q q}_0$. This induces transitions between the unperturbed
eigenstates and leads to a rotation of the qubit eigenbasis. (iv) Finally, since
the third term is not Hermitian due to the decoherence that it induces (in contrast
to $\bar{L}^{q q}_0$), the qubit eigenaxes may not be mutually orthogonal
any more. This renders the circular precession induced by $\bar{L}^{q q}_0$
slightly elliptical as we illustrate below in {\color{black} Sec.
\ref{sec:results}}.

The effective Liouvillian (\ref{eq:leffzmain}) also allows for a comparison
with earlier results. We note that our approach is conceptually quite similar
to that discussed in {\color{black} Ref. {\cite{Emary08}}}. However,
{\color{black} Ref. {\cite{Emary08}}} employs the clearly stated additional
assumption that the effect of the electrodes is only to modify the SQD dynamics
(assumed to obey a Lindblad equation) without affecting the (effective)
coupling to the qubit. In our formulation, this would mean that the coupling
to between the quasistationary and decay modes is mediated only by the
interaction $H_I = \hat{n}  \vecg{\lambda} \cdot \op{\vecg{\tau}} \text{/} 2$,
i.e., only the stochastic backaction. This means that in the approach of
{\color{black} Ref. {\cite{Emary08}}} both the dissipative and coherent
backaction are neglected.
Within this approximation one can thus not calculate the experimentally
measurable signal current; it must be consistently set to zero.
(The signal current was not calculated nor of interest in {\color{black} Ref. {\cite{Emary08}}}).
Moreover, {\color{black} Ref. {\cite{Emary08}}} focuses only on the time
scales and \new{does not}
consider initial-slip effects that we next turn to.

\emph{Initial slip.} Equations (\ref{eq:zinitial}) and (\ref{eq:xqefft})
show that the decay modes affect the quasistationary modes not only through
the effective Liouvillian but also through the effective initial state \cite{Haake83,Geigenmueller83,Haake85,Gaspard99,Flindt08PRL}: the
latter is {\tmem{not}} just given by the initial quasistationary variables
contained in $\vec{X}^q (0)$. In addition, there is a term in
(\ref{eq:xqefft}) that accounts for a initial contribution from the decaying
subspace, $\vec{X}^d (0)$, followed by a time integral over transitions into
the quasistationary subspace. This leads to an {\tmem{initial slip}} that
affects the quasistationary modes.
Like the time evolution \eq{eq:xqt}, the slip \eq{eq:xqefft} of the initial state has a a time-nonlocal
expression in terms of the initial state of the decay mode $\vec{X}^d(0)$ [in the Laplace transform \eq{eq:zinitial} the corresponding second slip term has frequency dependence].

An experimentally relevant question is how to eliminate or minimize the initial slip
since it can induce errors even for a perfectly prepared initial qubit state.
In {\color{black} Appendix \ref{app:slip}}, we
show that initial qubit-SQD states $\rho (0)$ that exhibit no initial slip form a
subset of measure zero in the set of all valid initial density operators $\rho (0)$.
In general, a sufficient condition for zero slip is that $\vec{X}^d (0) =
\vec{0}$, i.e., using \Eq{eq:rhoqrhod}:
\begin{eqnarray}
  \delta_{\tmop{st}} (0) \text{ \ = \ } p^1_{\tmop{st}} p^0 (0) -
  p^0_{\tmop{st}} p^1 (0) & = & 0,  \label{eq:suff1}\\
  \vecg{\delta}_{\tmop{st}} (0) \text{ \ = \ } p^1_{\tmop{st}} \vecg{\tau}^0 (0)
  - p^0_{\tmop{st}} \vecg{\tau}^1 (0) & = & \vec{0} .  \label{eq:suff2}
\end{eqnarray}
To find all initial states with zero slip, one has to compute the kernel of
matrix $\Lambda^{q d} (z - L_0^{d d} - \Lambda^{d d})^{- 1}$. Assuming the
frequency $z$ does not hit a pole of the denominator (for example when
considering the Markov approximation for $z = 0$), the inverse $(z - L_0^{d d}
- \Lambda^{d d})^{- 1}$ exists and has full rank. Thus, to determine the
dimension of the kernel, it suffices to determine the rank of $\Lambda^{q d}$,
which is 2 since
\begin{eqnarray}
  \Lambda^{q d} \left(\begin{array}{c}
    x\\
    y \op{\vecg{\lambda}}
  \end{array}\right) & = & \left(\begin{array}{cc}
    0 & 0\\
    0 & - \vecg{\lambda} \times
  \end{array}\right) \left(\begin{array}{c}
    x\\
    y \op{\vecg{\lambda}}
  \end{array}\right) \text{ \ = \ } \left(\begin{array}{c}
    0\\
    \vec{0}
  \end{array}\right)
\end{eqnarray}
for any $x, y \in \mathbbm{R}$ and using \Eq{eq:transqd}. Thus, the
matrix $\Lambda^{q d} (z - L_0^{d d} - \Lambda^{d d})^{- 1}$ has a kernel of
dimension 2, which means that the set of initial states has zero measure since
$\vec{X}^d$ can be taken from a four-dimensional set.

One way to eliminate the initial slip is to switch off the
capacitive interaction before $t = 0$, i.e., $\lambda(t)=0$ for
$-1/\gamma \ll t <0$. Then the detector can establish a stationary state and the
initial SQD-qubit state factorizes: $\vec{X}^d(0)=\vec{0}$ [\new{see} the $\lambda = 0$ solution (\ref{eq:xexpand}]
and see also {\color{black} Appendix \ref{app:slip}}). By contrast, if one
switches off the current through the sensor, $\Gamma \rightarrow 0$,
before $t=0$, then one starts with a sensor in
a definite charge state $p^1 (0) = 0$ or $1$, which is highly
nonstationary for typical operation parameters of the SQD (tuned close
to resonance for high sensitivity, one finds usually $p_{\tmop{st}}^0 \sim p^1_{\tmop{st}}$).
The backaction-induced initial slip
thus leads to an essential difference between two ways of switching off a
sensor, which should be taken into account in designing detection
protocols.

\newer{The magnitude of the slip in general depends on frequencies only for $z
\gtrsim L_0^{d d} + \Lambda^{d d} \sim \gamma$. This means that the initial
value $\vec{X}^d (0)$ influences $\vec{X}^q (t)$ for times $t \sim 1 / \gamma$
[through the integral (\ref{eq:xqefft})].} In Sec.
\ref{sec:slip}, we investigate the slip magnitude in more
detail in a high-temperature limit to leading order in $\lambda / \gamma$;
importantly, we find that even in this simple case there is a slip effect of
order $\lambda / \gamma$ which affects the overall qubit dynamics by, e.g.,
phase-shifting the solution in a way depending on the sensor initial state.

From the above we can generally conclude within the regime of validity
of the kinetic equation (\ref{eq:kineq}) that due to the backaction-induced
initial slip \tmtextit{additional errors} are generated; since the zero-slip
states are sets of measure zero, it is clear that preparation errors of the
\tmtextit{qubit-sensor state} will invariably lead to an initial slip.
Models of such possible errors are actually relevant for a different branch of
quantum information.
In quantum-error correction, one deals with decoherence from the
environment in a phenomenological way by introducing additional bit-flip
errors \cite{NielsenBook}. This requires assumptions to be made about the
type and statistics of the different possible errors. 
Our work thus provides in this context a possible scenario
how such errors may arise and how they can be modeled, after, e.g., a
measurement has been performed. This will become more concrete in the
next section, where we discuss simplified equations, which have,
however, only a limited range of applicability.


\section{High-temperature qubit dynamics}\label{sec:results}

The kinetic equation, either in representation (\ref{eq:kineq}) or
(\ref{eq:kineq2}), form central results of this paper. They fully suffice to
compute the transient dynamics of the charge-specific isospins $\vecg{\tau}^0
(t)$ and $\vecg{\tau}^1 (t)$. From this result, the total isospin $\vecg{\tau}
(t) = \vecg{\tau}^0 (t) + \vecg{\tau}^1 (t)$, i.e., the reduced qubit state can
be constructed. However, there are several reasons to attempt to obtain a
closed description in terms of $\vecg{\tau} (t)$ only.

First, from \Eq{eq:kineq}, it is not directly clear on which time scales
$\vecg{\tau} (t)$ evolves or decays. Second, an effective qubit description
plays an important role, for example, in quantum error correction. It is an interesting question how far Eq.
(\ref{eq:kineq}) can actually be reduced to a closed equation for the reduced
density operator for the qubit alone. One indication that this requires
additional assumptions is that (\ref{eq:kineq}) and (\ref{eq:kineq2}) can be
solved only if the full initial state vector [either in form Eq.
(\ref{eq:rhorep}) or \Eq{eq:rhoqrhod}] is specified and not just the
sum $\vecg{\tau}^0 (0) + \vecg{\tau}^1 (0)$. We have already seen that the
initial values of the other degrees of freedom produce an initial slip [see Eq.
(\ref{eq:zinitial})] and will see below that even in lowest nonvanishing order
this slip cannot be avoided. The third consideration is related to this and
concerns the minimization of backaction in quantum-information processing: One
would like to know the effective qubit eigenmodes, e.g., to construct initial
states that are least sensitive to backaction by setting experimental parameters.

To investigate all this further systematically, an effective theory for the qubit evolution
in a simple limit is useful. In this section, we consider the regime where the coherent
backaction and $O (\Gamma / T)$ corrections to the stochastic backaction can both be neglected.
We stress that this is for the purpose of illustration mostly since the latter effects are relevant under typical experimental conditions.
This simplification allows us to perform a expansion of the effective Liouvillian
(\ref{eq:leffzmain}) and the initial slip (\ref{eq:zinitial}) to leading order
in $\lambda / \gamma$ and we investigate the resulting transient qubit
evolution here in some detail. In {\Sec{sec:gatevoltagedep}}, we
give tangible analytical expressions for the relaxation and dephasing rates as
well as for the qubit precession frequency. In {\color{black} Sec.
\ref{sec:meanfield}}, we assess the accuracy of the approximate theory by
comparing with the numerical solution of the full kinetic equation
(\ref{eq:kineq}). We further discuss the slip of the initial condition of the
qubit isospin in {\Sec{sec:slip}} which relates to a
``kick'' that the qubit experiences during the relaxation time of the sensor QD.
Finally, we show in {\Sec{sec:elliptical}} that the
measurement backaction forces the isospin to precess about a tilted axis in an
elliptical way. The eccentricity is connected to oscillations in the decay of
the purity of the qubit state. \newer{This illustrates concretely that our
density-operator approach goes beyond standard master-equation
approaches as we discuss in Sec. \sec{sec:comparison}.}

\newer{\subsection{Effective Liouvillian, initial slip, and mode vectors}\label{sec:leffexpand}}
The kinetic equation [\Eq{eq:kineq} or \Eq{eq:kineq2}] was
derived using the weak-coupling, weak-measurement limit, $\Gamma / T \ll
\Delta / \Gamma \ll 1$ where $\Delta \sim \lambda, \Omega$. This prevents one
from just expanding in $\lambda / \Gamma$ since that would imply taking
$\lambda / \Gamma \ll \Gamma / T$. However, if we consistently neglect the
corrections $\Gamma / T$ (the cotunneling corrections to the stochastic backaction as
well as dissipative and coherent backaction), then we can take $\lambda /
\Gamma \rightarrow 0$ and expand in this parameter. We refer to this as the
high-temperature limit (since it is the large temperature that allows one to take the
infinitely weak-measurement limit). It should be noted that in this
approximation the current through the sensor QD is zero, i.e., at this level
of the theory one is not accounting for the actual backaction effects due to
the current \tmtextit{measurement} (rates $\sim \Gamma \lambda / T$),
but only for the leading effect of the tunnel coupling. Below, the stationary
occupations $p^{0, 1} = \gamma^{1, 0} / \gamma$ are given by their leading
order expressions (SET rates) $\gamma^{0, 1} = \sum_{r = s, d} \Gamma_r
f^{\pm}_r$ [see \Eq{eq:dissrate}] and $\gamma = 2 \gamma^0 + \gamma^1$.\\
We thus simplify the isospin evolution obtained from the effective Liouvillian
(\ref{eq:leffzmain}) and for concreteness assume from hereon 
\begin{eqnarray}
  \vecg{\Omega} & = & \Omega \vec{e}_x,
\end{eqnarray}
perpendicular to the capacitive interaction vector
$\vecg{\lambda} = \lambda \vec{e}_z$. This means that
if we ignored the mean-field tilting $\vecg{\Omega} \rightarrow
\tilde{\vecg{\Omega}}$ (which we do not), the qubit would oscillate in the measurement basis.\\
As we explain in detail in {\color{black} Appendix \ref{app:liouville}}, the
isospin evolution contained in Eqs. (\ref{eq:leffzmain}) and
(\ref{eq:zinitial}) can be simplified by performing a Markov
approximation,
i.e., by replacing $z = 0$. This Markov approximation with respect to
memory induced by the sensor QD (after integrating out the electrodes) is valid in the
weak-measurement limit $\lambda / \Gamma \ll 1$ (see {\color{black} Appendix
\ref{app:nm}}). We note that in the high-temperature limit also non-Markovian
corrections due to tunneling processes ($\Gamma$) are consistently neglected
as next-to leading order $\Gamma^2 / T$ corrections are not accounted for.
With $z = 0$, the Laplace-transform inverse of Eq. (\ref{eq:xqz}) can be easily performed.
Expanding the denominator in powers of $\Delta / \Gamma$ and extracting
the isospin from $\vec{X}^q$, we find
\begin{eqnarray}
  {\color{black} \vecg{\tau}} (t) & = & e^{- i L_{\tmop{eff}} t} 
  {\color{black} \vecg{\tau}}_{\tmop{eff}} (0) + O (\Delta^2 / \gamma^2) . 
  \label{eq:taueq2}
\end{eqnarray}
In this approximation the stationary state ${\color{black} \vecg{\tau}}
(\infty) = \vec{0}$. The effective Liouvillian in \Eq{eq:taueq2} reads
in diagonal form:
\begin{eqnarray}
  - i L_{\text{eff}} & = & \sum_{\alpha = 0, \pm} (i \tilde{\Omega}_{\alpha} -
  \gamma_{\alpha}) \vec{e}_{\tmop{eff}, \alpha} 
  \tilde{\vec{e}}^{\dag}_{\tmop{eff}, \alpha},  \label{eq:leff}
\end{eqnarray}
where the eigenvalues, the left and right eigenvectors, and the effective initial
state $\vecg{\tau}_{\tmop{eff}} (0)$ are specified below by Eqs.
(\ref{eq:taueff0})--(\ref{eq:e2eff}). Equation (\ref{eq:taueq2}) is valid for
times $1 / \gamma \ll t \ll \gamma^2 / \Delta^3$ \ as {\color{black} Appendix
\ref{app:liouville}} shows and also our numerical checks below confirm. The lower limit
indicates that we consider the wide-band limit
{\footnote{The kinetic
equations for sensor QD plus qubit are considered here in the wide-band limit with respect
to the bandwidth $W$ of the electrodes to which the sensor is
coupled.}} 
with
respect to the \tmtextit{sensor QD} band-width $\gamma$ (by setting $\gamma \gg
z \rightarrow i 0$ above), whereas for times $t \gg \gamma / \Delta^2$,
corrections of $O (\Delta^3 / \gamma^2)$ to the effective Liouvillian
(\ref{eq:leff}) accumulate and the error made for ${\color{black}
\vecg{\tau}} (t)$ may become sizable.\\

\newer{\subsubsection{Initial slip}}
The effective initial condition appearing in Eq.
(\ref{eq:taueq2}) reads (see {\color{black} Appendix \ref{app:liouville}}):
\begin{eqnarray}
  {\color{black} \vecg{\tau}}_{\tmop{eff}} (0) & = & \vecg{\tau} (0) -
  \frac{1}{\gamma}  {\color{black} \vecg{\lambda}} \times (p^1_{\tmop{st}}
  {\color{black} \vecg{\tau}}^0 (0) - p^0_{\tmop{st}} {\color{black}
  \vecg{\tau}}^1 (0)) .  \label{eq:taueff0}
\end{eqnarray}
This shows that even in this simple limit the qubit description is still not
closed. Although the entire sensor variables (electrodes plus sensor QD) have been
eliminated, the initial condition does not depend only on $\vecg{\tau} (0)$
(see {\color{black} Appendix \ref{app:slip}}). Instead, it additionally requires
the specification of the component of the initial qubit-sensor state $\rho
(0)$ in the decaying subspace. Thus, \tmtextit{both} initial charge-specific
isospins $\vecg{\tau}^n (0)$ are needed to compute $\vecg{\tau} (t)$.
In contrast to the general case discussed in {\color{black}
\Sec{sec:effliouville}}, \Eq{eq:taueff0} only relies on
$\vecg{\delta}_{\tmop{st}} (0) = p^1_{\tmop{st}} {\color{black}
\vecg{\tau}}^0 (0) - p^0_{\tmop{st}} {\color{black} \vecg{\tau}}^1 (0)$
and does not involve $\delta_{\tmop{st}} (0) = p^1_{\tmop{st}} p^0 (0) -
p^0_{\tmop{st}} p^1 (0)$. The reason is that the denominator in Eq.
(\ref{eq:slipqd}) is approximated here by $ - \gamma$ and $\Lambda^{q d}$
does not act on the charge sector [see \Eq{eq:transqd}].}\\ 
As mentioned
in the general discussion of the initial slip, \Eq{eq:xqefft}, initial
qubit-sensor states $\rho (0)$ with zero initial slip form a zero-measure
subset of all possible initial states. As a result, the initial slip adds to
preparation errors, an error that depends on the sensor dynamical state.
In the present simple limit, the
initial slip is a time-local expression, \Eq{eq:taueff0}, in
contrast to the general case, \Eq{eq:xqefft}. If the measurement is not
weak any more, i.e., if $\lambda \sim \gamma$, one can expect the dynamics
of the charge-specific
isospins to become important on the entire time scale of the qubit decay
\newer{and not just through an effective slip of the initial condition.}
Still, even in the weak-measurement limit studied here, \Eq{eq:taueff0}
shows explicitly that the slip is of non-negligible order $\lambda / \gamma$.\\

\newer{\subsubsection{Qubit time scales}}The simple formulas (\ref{eq:taueq2})-(\ref{eq:taueff0})
 for the relevant precession, relaxation, and dephasing time scales form the third central set of equations of the paper. We now discuss their contents. The eigenvalues in \Eq{eq:leff} contain the effective qubit
frequencies
\begin{eqnarray}
  \tilde{\Omega}_{\alpha} & = & \alpha \tilde{\Omega}, \hspace{1em} \alpha =
  0, \pm,  \label{eq:omegaalpha}
\end{eqnarray}
which up to $O (\Delta^3 / \gamma^2)$ read
\begin{eqnarray}
  \tilde{\Omega} \text{ \ } = \text{ \ } \left| \tilde{\vecg{\Omega}}
  \right| & = & \sqrt{\Omega^2 + (p^1_{\tmop{st}} \lambda)^2} . 
  \label{eq:omegatilde}
\end{eqnarray}
Equation (\ref{eq:omegatilde}) is precisely the length of the mean isospin
field announced earlier in \Eq{eq:field}, but now with $\langle n
\rangle = p^1_{\tmop{st}}$,
\begin{eqnarray}
  & \begin{array}{lllll}
    \begin{array}{l}
      \tilde{\vecg{\Omega}}
    \end{array} & = & \vecg{\Omega} + p^1_{\tmop{st}} \vecg{\lambda} & = &
    \tilde{\Omega}  \vec{e}_{\parallel},
  \end{array} &  \label{eq:omegatildevec}
\end{eqnarray}
whose unit direction vector is relevant for the following:
\begin{eqnarray}
  \vec{e}_{\parallel} & = & \left( \Omega \op{{\color{black} \vecg{\Omega}}}
  + p^1_{\tmop{st}} \lambda \op{{\color{black} \vecg{\lambda}}} \right) /
  \tilde{\Omega} .  \label{eq:epara}
\end{eqnarray}
Moreover, we will need the perpendicular unit vector,
\begin{eqnarray}
  \vec{e}_{\perp} & = & \left( \Omega \hat{\vecg{\lambda}} -
  p_{\tmop{st}}^1 \lambda \hat{\vecg{\Omega}} \right) / \tilde{\Omega}, 
  \label{eq:eperp}
\end{eqnarray}
lying in the plane spanned by $\vecg{\Omega}$ and $\vecg{\lambda}$. Our
calculation thus confirms the intuitive picture explained in {\color{black}
Sec. \ref{sec:charge-specific-isospin}}: The mean-field effect of the average
SQD charge $ \langle \hat{n} \rangle = p^1_{\tmop{st}}$ is just to tilt
the qubit axis (see also below in {\color{black} Sec.
\ref{sec:meanfieldpicture}}) and does not rely on the tunneling-induced
fluctuations $\sim \delta \hat{n}$, see \Eq{eq:beff}. Tunneling
influences the mean sensor charge only
indirectly as noted in the discussion of \Eq{eq:field}.

Compared to the first term of \Eq{eq:l0}, the eigenvalues of the quasistationary modes
have acquired small dissipative parts [see \Eq{eq:leff}]:
\begin{eqnarray}
    \gamma_0 \ \  = \ \ \frac{1}{T_1}, & &\gamma_{\pm} \ \  = \ \ \frac{1}{T_2} .
   \label{eq:gammaalpha}
\end{eqnarray}
Here, the relaxation rate is given up to $O (\Delta^4 / \gamma^3)$ by
\begin{eqnarray}
  \frac{1}{T_1} & = & r \frac{\left( \vecg{\lambda} \cdot \vec{e}_{\perp}
  \right)^2}{\gamma} = r \left( \frac{\Omega}{\tilde{\Omega}} \right)^2 
  \frac{\lambda^2}{\gamma},  \label{eq:t1}
\end{eqnarray}
which is quadratic in the component of the measurement vector $\vecg{\lambda}$
 {\tmem{perpendicular}} to the average isospin field $\tilde{\vecg{\Omega}}$
with the transition factor r given up to zeroth order in $\Gamma/T$ as given by \Eq{eq:rlowest}.
The dephasing rate $1
/ T_2$ is expressed compactly in terms of the pure dephasing rate $1
/ T_{\phi} = 1/T_2 - 1/(2 T_1)$ {\cite{Chirolli08}} up to order $O (\Delta^4 / \gamma^3)$ as
\begin{eqnarray}
  \frac{1}{T_{\phi}} & = & r \frac{\left( \vecg{\lambda} \cdot \vec{e}_{\parallel} \right)^2}{\gamma}
  = r \left( \frac{\lambda p^1_{\tmop{st}}}{\tilde{\Omega}} \right)^2 
  \frac{\lambda^2}{\gamma},  \label{eq:tphi}
\end{eqnarray}
which is quadratic in the projection of $\vecg{\lambda}$ on the unit vector
$\vec{e}_{\parallel} = \tilde{\vecg{\Omega}} / \tilde{\Omega}$
{\tmem{along}} the average isospin field. In the following, we refer to both
relaxation and dephasing as {\tmem{decoherence}} because both drive the qubit
into a mixed state.

Note that in both decay rates, the transition factor $r$ appears, which links to the discussion of the previous sections. If
higher-order $\Delta / \gamma$ terms are included into the relaxation and
dephasing rate, additional terms appear that depend on the dissipative
backaction terms $\sim c$ stemming from the off-diagonal elements in Eq.
(\ref{eq:transqd}).

It is easy to see that the relaxation and pure dephasing time $T_1$ and
$T_{\phi}$ are positive {\footnote{The transition factor satisfies $r > 0$ even
when cotunneling and coherent backaction terms are included, see Eq.
(\ref{eq:rexp}).}} , since the transition factor is given to lowest order
by $r \approx 2 \Gamma_+ \Gamma_- / (2 \Gamma_+ + \Gamma_-)^2 > 0$ [see Eq.
(\ref{eq:rlowest})].
Since $T_{\phi} > 0$, the ratio $T_2 / 2 T_1$ further
satisfies the relation {\cite{Chirolli08}}:
\begin{eqnarray}
    \frac{T_2}{2 T_1} & = & \frac{1}{1 + 2 (p^1_{\tmop{st}} \lambda /
    \Omega)^2} \ \  < \ \  1.
\end{eqnarray}
Equations (\ref{eq:t1}) and (\ref{eq:tphi}) again confirm our intuitive
expectation from {\Sec{sec:charge-specific-isospin}} that
only the fluctuating part of the SQD charge $\sim \vecg{\lambda} \delta n$
(involving here virtual transitions into the decay modes) is responsible for
the qubit decoherence. The energy scale $\lambda^2 / \gamma$ for the
decoherence rates exhibits the expected quadratic scaling with the weak
coupling $\lambda$ and inverse scaling with the large detector band width
$\gamma$, as discussed further below in {\color{black} Sec.
\ref{sec:semiclassical}}. The decay time is thus slow compared to the time
scale of the relaxation $\sim 1 \text{/} \gamma$ of the sensor QD and that of
the intrinsic evolution of the SQD-qubit system $\sim 1 \text{/} \Delta$. We
emphasize that this picture and our approach hold only in the limit $\Delta
\ll \Gamma$: it breaks down if the tunneling becomes to strong relative to
either the measurement $\lambda$ or the qubit internal field
$\Omega$.
{\footnote{Note that the decay rates would diverge in the limit
$\lambda \text{/} \Gamma \rightarrow \infty$.}}

One should note that the corrections to the decoherence rates
$\gamma_{\alpha}$ [see Eqs. \eq{eq:t1} and (\ref{eq:tphi})] are of $O (\Delta^4
/ \gamma^3)$, while the corrections to the qubit frequency
(\ref{eq:omegaalpha}) already appear in lower $O (\Delta^3 / \gamma^2)$. The
reason is that these quantities behave differently under a simultaneous
reversal of the orientation of $\vecg{\lambda}$ and $\vecg{\Omega}$ as one can
see from a simple physical argument: mapping $\vecg{\lambda} \rightarrow -
\vecg{\lambda}$ and $\vecg{\Omega} \rightarrow - \vecg{\Omega}$ corresponds to
spatially mirroring the detection setup about the vertical axis in
{\color{black} Fig. \ref{fig:model}}. This clearly inverts the sense of the
precessional motion, i.e., one has $\Omega_{\alpha} \left( - \vecg{\lambda}, -
\vecg{\Omega} \right) = - \Omega_{\alpha} \left( \vecg{\lambda}, \vecg{\Omega}
\right)$ but the qubit decay cannot depend on mirroring the setup, i.e., we
have $\gamma_{\alpha} \left( - \vecg{\lambda}, - \vecg{\Omega} \right) =
\gamma_{\alpha} \left( \vecg{\lambda}, \vecg{\Omega} \right)$. This implies that
corrections to $\gamma_{\alpha}$ must be of even order in $\Delta$ and therefore at
least of fourth order in $\Delta$, in agreement with our calculation.

\newer{\subsubsection{Mode vectors}}To complete the specification of the effective
Liouvillian (\ref{eq:leff}), we give the explicit formulas for the unit
vectors $\vec{e}_{\tmop{eff}, \alpha}$. Expressed in $\vec{e}_{\parallel}$ and
$\vec{e}_{\perp}$ given by \Eq{eq:epara} and (\ref{eq:eperp}),
respectively, they read:
\begin{eqnarray}
  \vec{e}_{\tmop{eff},0} & \propto & \vec{e}_{\parallel} + r
  \frac{\lambda^2}{\gamma \tilde{\Omega}}  \frac{\Omega \lambda
  p^1_{\tmop{st}}}{\tilde{\Omega}^2}  \vec{e}_{\parallel} \times
  \vec{e}_{\perp},  \label{eq:e0eff}\\
  \vec{e}_{\tmop{eff},1} & \propto & \vec{e}_{\perp} + r
  \frac{\lambda^2}{\gamma \tilde{\Omega}}  \frac{\Omega}{\tilde{\Omega}^2} 
  \frac{\tilde{\Omega}^2 - \Omega^2 / 2}{\Omega + \tilde{\Omega}} 
  \vec{e}_{\parallel} \times \vec{e}_{\perp},  \label{eq:e1eff}\\
  \vec{e}_{\tmop{eff},2} & \propto & \vec{e}_{\parallel} \times
  \vec{e}_{\perp}  \label{eq:e2eff}\\
  &  & + r \frac{\lambda^2}{\gamma \tilde{\Omega}} 
  \frac{\Omega}{\tilde{\Omega}^2} \left( \lambda p^1_{\tmop{st}}
  \vec{e}_{\parallel} + \frac{\Omega^2 - \tilde{\Omega}^2 + \Omega
  \tilde{\Omega} / 2}{\Omega + \tilde{\Omega}} \vec{e}_{\perp} \right),
  \nonumber
\end{eqnarray}
where $\propto$ indicates that we suppress the normalization constants. As
before [see Eqs. (\ref{eq:e0}) and (\ref{eq:epm})], we define
$\vec{e}_{\tmop{eff}, \pm} = \left( \vec{e}_{\tmop{eff}, 1} \mp i
\vec{e}_{\tmop{eff}, 2} \right) / \sqrt{2}$ and we note that
$\vec{e}_{\parallel} \times \vec{e}_{\perp} = \hat{\vecg{\Omega}} \times
\op{\vecg{\lambda}}$ [see Eqs. (\ref{eq:epara}) and (\ref{eq:eperp})].

In stark contrast to the unperturbed case, the real unit vectors
$\vec{e}_{\tmop{eff}, 0}$, $\vec{e}_{\tmop{eff}, 1}$, and
$\vec{e}_{\tmop{eff}, 2}$ form a real {\tmem{non}}orthogonal basis. This
implies that to decompose a vector in the basis $\left\{ \vec{e}_{\tmop{eff},
\alpha} \right\}$, one needs to take the scalar product with the dual basis
denoted $\left\{ \tilde{\vec{e}}_{\tmop{eff}, \alpha} \right\}$, see
{\color{black} Fig. \ref{fig:modevectors}}. The dual basis vectors
$\tilde{\vec{e}}_{\tmop{eff}, \alpha}$ are non-unit vectors orthogonal to
the plane spanned by $\vec{e}_{\tmop{eff}, \beta}$ and $\vec{e}_{\tmop{eff},
\gamma}$,
\begin{eqnarray}
  \tilde{\vec{e}}_{\tmop{eff}, \alpha} & = & \frac{\vec{e}_{\tmop{eff},
  \beta} \times \vec{e}_{\tmop{eff}, \gamma}}{\left| \vec{e}_{\tmop{eff},
  \alpha} \cdot \left( \vec{e}_{\tmop{eff}, \beta} \times \vec{e}_{\tmop{eff},
  \gamma} \right) \right|},  \label{eq:tildedef}
\end{eqnarray}
where $(\alpha, \beta, \gamma)$ is a cyclic permutation of $(0, 1, 2)$. We refer to
this nonorthogonality of the eigenvectors in the following simply as a
distortion of the isospin modes. 
\begin{center}
  \Figure{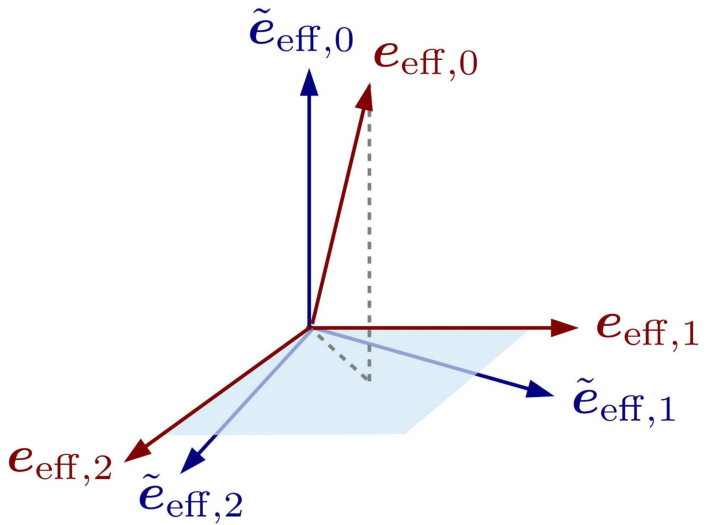}{Distortion
  of isospin mode vectors: (a) Three-dimensional sketch of the effective qubit
  axes $\vec{e}_{\tmop{eff}, \alpha}$ [$\alpha = 0, \pm$, Eqs.
  (\ref{eq:e0eff}) -- (\ref{eq:e2eff})] and its dual basis
  $\tilde{\vec{e}}_{\tmop{eff}, \alpha}$. Note that $\vec{e}_{
  \tmop{eff}, 0}$ has a nonzero projection on the plane spanned by $\vec{e}_{
  \tmop{eff}, 1}$ and $\vec{e}_{\tmop{eff}, 2}$, while $\tilde{\vec{e}}_{
  \tmop{eff}, 0}$ is orthogonal to this plane. Likewise, a vector in the
  $\vec{e}_{\tmop{eff}, 1}$-$\vec{e}_{\tmop{eff}, 2}$ plane has a nonzero
  component along $\vec{e}_{\tmop{eff}, 0}$.\label{fig:modevectors}}
\end{center}
The distortion of the mode vectors scales with the small ratio of the
magnitude of the decoherence rates $\lambda^2 / \gamma$ relative
to the effective qubit frequency $\tilde{\Omega} \sim \Delta$ and is furthermore suppressed by the transition factor
$r$ when going off-resonance [see \Eq{eq:r}]. Since we allow for
$\lambda \sim \Omega$, the remaining factors in Eqs.
(\ref{eq:e0eff})--(\ref{eq:e2eff}) can be of order 1.
This distortion also has tangible physical
consequences: \Eq{eq:e0eff} shows that the precessional motion of the
qubit isospin $\vecg{\tau}$ becomes tilted. It is also not circular any more,
but becomes slightly elliptical instead as we investigate in more detail in
{\Sec{sec:elliptical}}.

\newer{\subsubsection{Other approaches}}At this point, it is instructive to compare with
some other approaches.
\newer{At first sight, the mode distortion may appear peculiar and one
may wonder why it does not show up in other approaches.
 In fact, the mode distortion disappears in the limit $\lambda/\Omega \rightarrow 0$, i.e., when the capacitive coupling becomes smaller than
\emph{all} other energy scales. In this limit, the modes become
orthogonal with $\vec{e}_0 = \hat{\vecg{\Omega}}$. This diminishes also the
mean-field effect. This limit is equivalent to the frequently
made secular
approximation that Davies has shown to be exact in this strict
weak-coupling limit \cite{Davies74,Davies75}.  Here, however, we
consider the more general situation that the capacitive coupling $\lambda$ can be of the same magnitude as
the internal qubit energy scale $\Omega$. The
secular approximation is not applicable in this case as we showed
previously \cite{Hell14a} in accordance with other works
\cite{Salmilehto12}. This is furthermore signalled by the observation
that the secular approximation conflicts with the isospin conservation when electrons tunnel between the
electrodes and the sensor QD. The strength of the mode distortion
thus reflects the importance of nonsecular corrections (coherences).}\\
We next compare to Ref. {\cite{Makhlin00}}, whose approach is similar to
ours 
\footnote{We stress that {\color{black} Ref.
{\cite{Makhlin00}}} is based on a kinetic equation to leading order in $\Gamma$,
i.e., cotunneling and renormalization effects are excluded from the start. Thus, the authors do not address the same questions as we do here.}.
This study applies a
two-step procedure to derive a closed, effective description of the qubit
dynamics, starting from a generalized master equation for qubit plus SQD, as
we do. To treat the limit $\tilde{\Omega} \gg \lambda^2 / \gamma$, it is
additionally assumed that $\left| \vecg{\lambda} \cdot \vec{e}_{\perp} \right|
\ll \left| \vecg{\lambda} \cdot \vec{e}_{\parallel} \right|$, i.e.,
perpendicular fluctuations of $\vec{B}_{\tmop{eff}}$ along $\vec{e}_{\perp}$
are small as compared to longitudinal fluctuations along the mean field
$\tilde{\vecg{\Omega}}$. Perpendicular fluctuations are therefore treated
there perturbatively. In this way coherences between isospin states quantized
along $\vec{e}_{\parallel}$ are only effectively included. By contrast, we
allow for comparable fluctuations in the direction of both
$\vec{e}_{\parallel}$ and $\vec{e}_{\perp}$, which is a more general case.\\

\subsection{Accuracy of effective isospin dynamics}\label{sec:meanfield}

Before we illustrate \newer{the effects of the initial slip and the mode-vector
distortion}, we discuss the accuracy of our Liouville-space perturbation
theory as compared to the full solution of the kinetic equations
(\ref{eq:kineq}) in the high-temperature limit and
up to times times $t \sim \gamma / \lambda^2 \gg 1 / \gamma$.

\subsubsection{Undistorted, mean-field qubit
modes}\label{sec:meanfieldpicture}

As a starting point for this discussion, we first \newer{discuss our
result in view of the} rough mean-field
picture of the detector backaction. It is valid in zeroth order in
$\lambda$ and $\Omega$ relative to $\Gamma$, where we neglect all relaxation
and decoherence rates $\lambda^2 / \gamma \ll \lambda$. Expanding the expressions (\ref{eq:e0eff}) -- (\ref{eq:e2eff}) for the mode vectors to the
corresponding zeroth order yields an orthonormal basis,
\begin{eqnarray}
  \vec{e}_{\tmop{eff}, 0} \approx \vec{e}_{\parallel}, & \vec{e}_{\tmop{eff},
  1} \approx \vec{e}_{\perp}, & \text{and \ } \vec{e}_{\tmop{eff}, 2} \approx
  \vec{e}_{\parallel} \times \vec{e}_{\perp},  \label{eq:eundist}
\end{eqnarray}
with $\vec{e}_{\parallel}$ and \ $\vec{e}_{\perp}$ given by (\ref{eq:epara})
and (\ref{eq:eperp}), respectively. Thus, in this approximation the dual basis
$\left\{ \tilde{\vec{e}}_{\tmop{eff}, \alpha} \right\}$ coincides with
$\left\{ \vec{e}_{\tmop{eff}, \alpha} \right\}$. In analogy to the unperturbed
case discussed in {\Sec{sec:zerocoupling}}, the isospin
just precesses circularly about $\vec{e}_{\parallel}$; however, the precession
frequency -- $\tilde{\Omega}$ instead of $\Omega$ -- and the precession axis
-- along $\vec{e}_{\parallel} = \tilde{\vecg{\Omega}} / \tilde{\Omega}$
instead of $\op{\vecg{\Omega}}$ -- are different. This rough mean-field picture
of the measurement is therefore to {\tmem{tilt}} the ``bare'' isospin field to
the mean isospin field (\ref{eq:omegatildevec}) by an angle [see Eq.
(\ref{eq:field})]:
\begin{eqnarray}
  \tan \theta & = & \frac{p^1_{\tmop{st}} \lambda}{\Omega}.
  \label{eq:tantheta}
\end{eqnarray}
In our concrete charge-qubit model, this means that the capacitive readout
simply detunes the charge qubit due the gating effect of the sensor QD with
mean charge $p^1_{\tmop{st}}$. Here the mean charge is identified with the
ensemble averaged charge, see the related discussion in {\color{black} Sec.
\ref{sec:charge-specific-isospin}}. Since we only require $\Delta
\ll \gamma$ but impose no constraint on the ratio $\lambda / \Omega$, this
angle can be large. In {\color{black} Fig. \ref{fig:compana}}, we illustrate
this effect by showing the evolution of the three isospin components in the
basis $( \vec{e}_x, \vec{e}_y, \vec{e}_z) = 
(\op{\vecg{\Omega}}, \op{\vecg{\lambda}} \times \op{\vecg{\Omega}},
\op{\vecg{\lambda}})$ on a long time scale $t \gg 1 \text{/} \gamma$
when the isospin initially points into the direction of $\op{\vecg{\Omega}}$,
i.e., perpendicular to $\op{\vecg{\lambda}}$. If the coupling $\lambda$ was
switched off, we would expect the isospin not to precess at all and to remain
stable along $\op{\vecg{\Omega}}$. By contrast, the oscillations of all
components in {\color{black} Fig. \ref{fig:compana}} clearly demonstrates
that the isospin revolves about a very different axis, roughly pointing in the
direction of $\vecg{\Omega} + \vecg{\lambda}$ in line with Eq.
(\ref{eq:tantheta}) for the parameters employed here.
\begin{center}
  \Figure{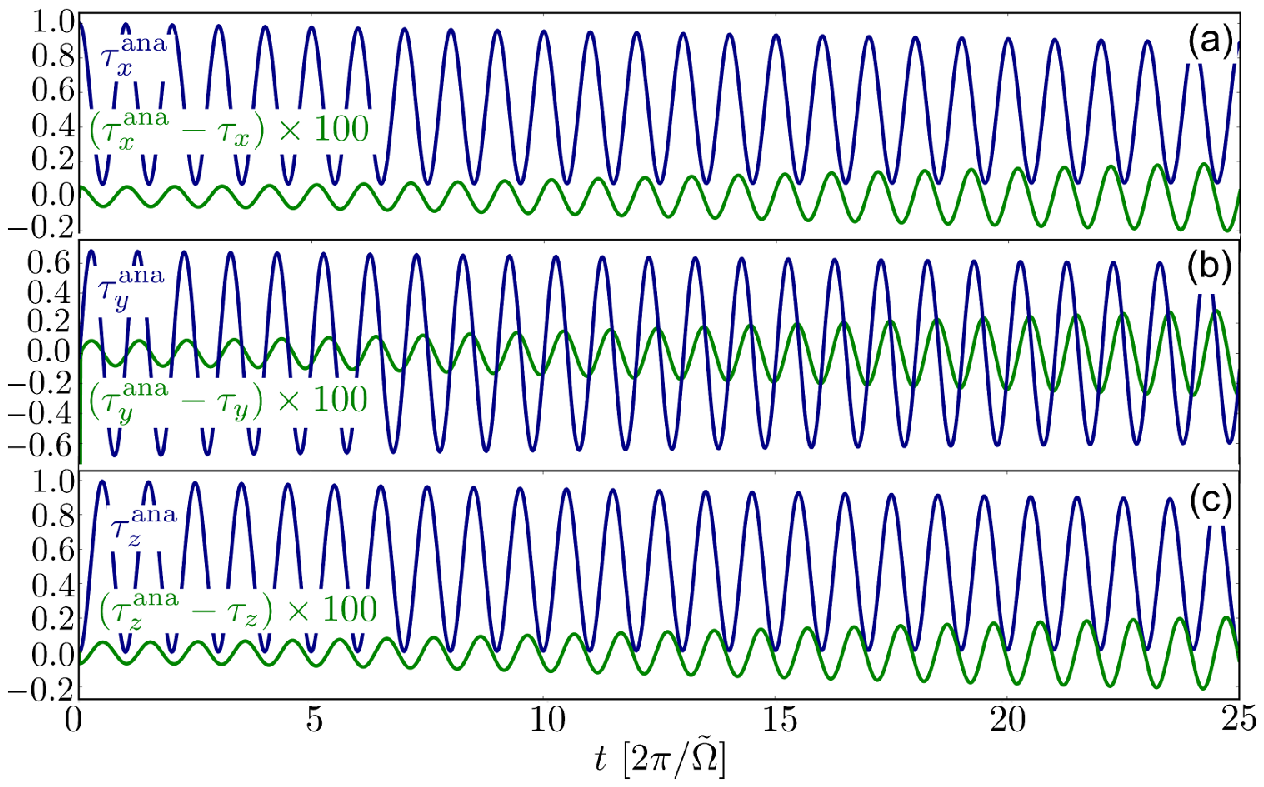}{Time
  evolution of the isospin vector components (a) $\tau_x$, (b) $\tau_y$, and
  (c) $\tau_z$ for times $t \gg 1 / \gamma$. The blue curves show the analytic
  solution $\vecg{\tau}^{\text{ana}}$ given by \Eq{eq:taueq2}, whereas
  the green curves show \tmtextit{100 times} the error with the $\vecg{\tau}$
  obtained from the solution of the full kinetic equations (\ref{eq:kineq}).
  The coordinate system is chosen as $( \vec{e}_x, \vec{e}_y, \vec{e}_z
  ) = ( \op{\vecg{\Omega}}, \op{\vecg{\lambda}} \times
  \op{\vecg{\Omega}}, \op{\vecg{\lambda}} )$. The initial state of sensor
  \tmtextit{plus} qubit is $p^0 (0) = 0$, $\vecg{\tau}^0 (0) = \vec{0}$, $p^1
  (0) = 1$, and ${\color{black} \vecg{\tau}}^1 (0) = \op{\vecg{\Omega}}$. The
  remaining parameters are $\Gamma_s = \Gamma_d = \bar{\Gamma} = 10^{- 3} T$,
  $\lambda = \Omega = 0.1 \bar{\Gamma} = 10^{- 4} T$, $V_b = V_g = 2.5 T$, and
  $W = 1000 T$. For these parameters, we find $\gamma \approx 3.75 \cdot 10^{- 3} T$
  and $\tilde{\Omega} \approx 1.37 \cdot 10^{- 4} T$ and therefore $1 / \gamma
  \approx (2 \pi / \tilde{\Omega}) / 200 \ll t \ll \gamma^2 / \Delta^3 \approx
  300 (2 \pi / \tilde{\Omega})$ is well fulfilled for the times shown
  above.\label{fig:compana}}
\end{center}
\subsubsection{Accuracy of weak-measurement expansion}

The green curves in {\color{black} Fig. \ref{fig:compana}} \ depict the
difference between the isospin evolution obtained by solving the kinetic
equations (\ref{eq:kineq}) and the evolution computed from Eq.
(\ref{eq:taueq2}) based on our perturbation theory, indicating that both agree
well (plotted is the error multiplied by 100). The remaining deviation is
mostly due to a small phase shift between the full and perturbative solution
that accumulates in time. The origin lies in ignored corrections of order of
$\sim \Delta^3 \text{/} \gamma^2$ to the effective qubit frequency
(\ref{eq:omegaalpha}). In {\color{black} Fig. \ref{fig:compana}}, this
accumulates after the shown time $\Delta t \sim 25 \cdot (2 \pi /
\tilde{\Omega})$ to a phase difference $\Delta \varphi \lesssim 25 \cdot
(\lambda / 4 \bar{\Gamma})^2 \sim 0.01$, which is just visible. However, we
emphasize that the accuracy of the {\tmem{decay rates}} is higher as discussed
below \Eq{eq:r} and therefore the exponentially decaying
{\tmem{envelope}} of the isospin evolution agrees with much larger accuracy up
to longer times.

The rough mean-field picture introduced above complies with the physical
picture developed in prior works
{\cite{Galperin04,Makhlin00,Chirolli08,Ithier05}}. It also forms the basis of
a simple classical understanding of the qubit decoherence in fluctuator
models, to which we compare our results in {\color{black} Sec.
\ref{sec:semiclassical}}. However, there are important corrections to this
picture even in the high-temperature limit, which we next discuss.

\subsection{Effect of initial slip}\label{sec:slip}

A first illustration of the corrections to the mean-field
picture is the effect of the slippage of the initial condition, Eq.
(\ref{eq:taueff0}). To illustrate this effect, we compare in {\color{black}
Fig. \ref{fig:initial}} the solution for the isospin for two different initial
states. We start from a factorizable initial state $\rho_{Q S} = \rho_Q
\otimes \rho_S$ with a fixed total isospin $\vecg{\tau} (0) =
\op{\vecg{\Omega}}$ along the ``bare'' internal field [determining $\rho_Q = \left( \op{\mathbbm{1}}_Q +
\vecg{\tau} (0) \cdot \op{\vecg{\tau}} \right) / 2$] while changing the initial
condition for the SQD through the sensor charge equal to $p^1 (0)$ [determining $\rho_S = (1 - p^1
(0)) \op{P}^0 + p^1 (0) \op{P}^1$]. We show the outcome for
$\tau_y^{\tmop{ana}}$ for the two cases of an initially empty SQD ($p^1 (0) =
0$) and a SQD hosting an electron ($p^1 (0) = 1$). Figure {\color{black}
\ref{fig:initial}}(a) exhibits a phase shift between the two isospin evolutions
that persists over an entire qubit cycle (and in fact for all future times,
which are not shown here).

The approximate analytical solution (\ref{eq:taueq2}) is valid over the
entire time scale shown in Fig. \ref{fig:initial}(a) except
for very small times $t < 1 / \gamma$. In {\color{black} Fig.
\ref{fig:initial}}(b), we illustrate how the isospin computed from the full
kinetic equations (\ref{eq:kineq}) (red) approaches the approximate analytical
solution (\ref{eq:taueq2}) (blue): all curves for the full solution (red)
start from the same value for $\tau_y (0) = 0$ but immediately develop
differently depending on the initial SQD charge $p^1 (0)$. On a time scale
$\sim 1 / \gamma$, they approach the approximate analytical solutions
$\tau_y^{\tmop{ana}} (0)$ (blue), which are offset by the initial slip
(\ref{eq:taueff0}). Figure {\color{black}
\ref{fig:initial}}(b) confirms that precisely due to this slip the analytic
solution accurately approximates the full numerical one for times $t \gg 1 /
\gamma$. The latter time scale is expected since the approximate curve relies
on the SQD wide-band limit. (The analytic solution may even be unphysical \
for times $t \lesssim 1 / \gamma$: In certain cases, including {\color{black}
Fig. \ref{fig:initial}}, one may find $| {\color{black}
\vecg{\tau}}_{\tmop{eff}} (0) | > 1$.)

If one, however, neglects the initial slip (\ref{eq:taueff0}) one obtains a
curve similar to the blue one in {\color{black} Fig. \ref{fig:initial}} but
with zero vertical offset for time $t = 0$. This clearly leads to a
nonnegligible deviation from the full solution for initial conditions with
nonstationary sensor. After a time $t \sim 1 / \gamma$, the qubit phase
is advanced by $\Omega / \gamma$ which can be of the same order as the phase
angle $\sim \lambda / \gamma$ of the initial slip (\ref{eq:taueff0})
(depending on the relation $\lambda, \Omega$, within the restriction $\lambda,
\Omega \ll \Gamma$). We stress that this slip leads to a cumulative
effect: Even at long times $t \gg 1 / \gamma$ the approximate solution without slip remains offset relative to the full solution.

Altogether, this shows clearly that one cannot get rid of the detector
completely -- even though we describe the qubit state using only the Bloch
vector $\vecg{\tau} (t)$. \cut{Importantly, the slip is not negligible even for weak
interaction $\lambda$, giving corrections of order $\lambda / \gamma$.} It is
difficult to eliminate the slip by a choice of the initial state of the coupled qubit-sensor system
as mentioned (see {\Sec{sec:effliouville}}) and further discussed in {\color{black}
Appendix \ref{app:slip}}.
Importantly, one should note that it may not be possible to remove the initial-slip backaction by any qubit preparation (i.e., just its  \emph{reduced} density operator $\vecg{\tau}$).
For quantum error correction, it is thus important to model the failure
of the preparation not only of the qubit dynamical state, but also the
dynamical state of the
sensor QD and their mutual correlations.
%
\begin{center}
  \Figure{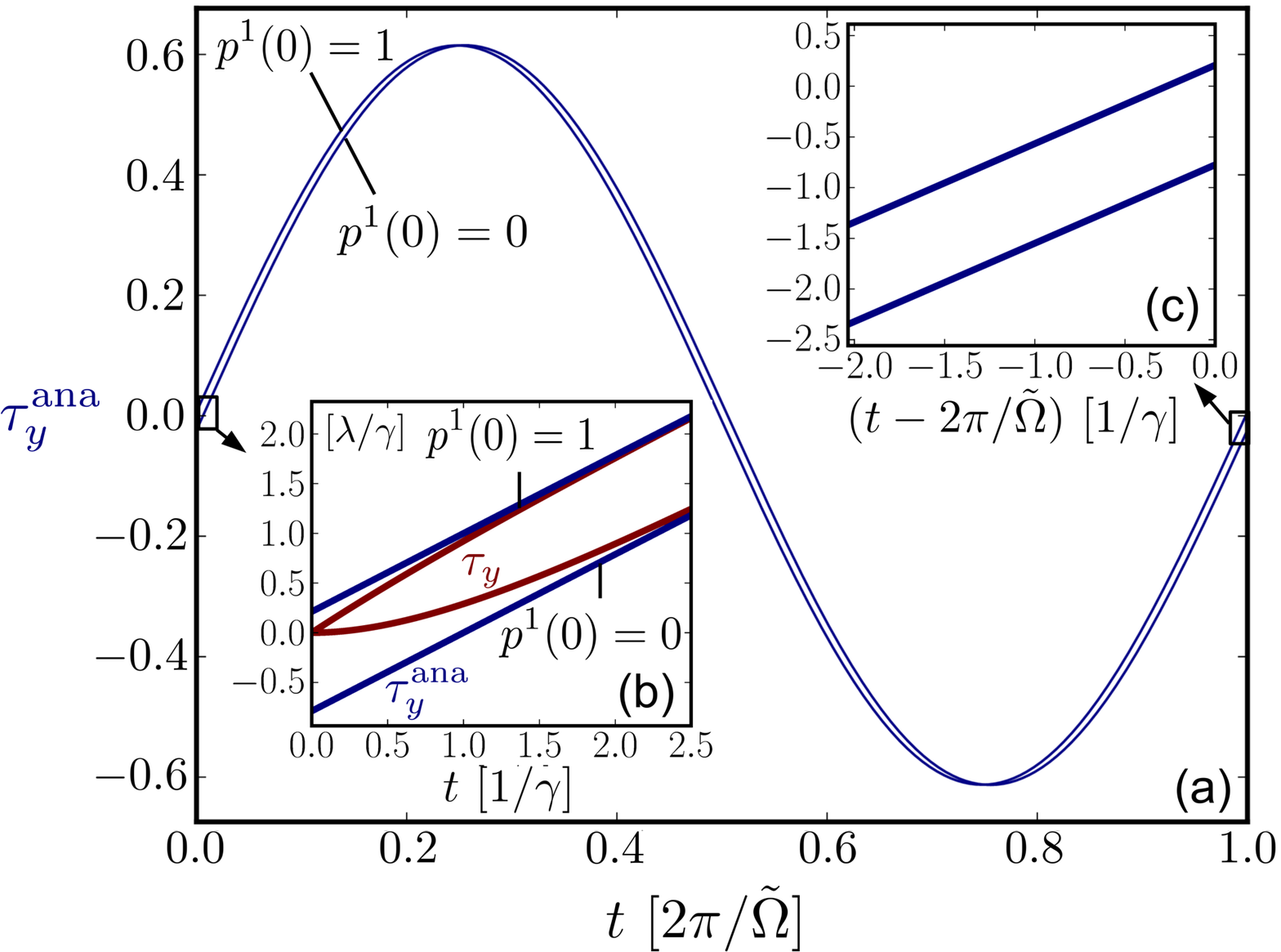}{Effect
  of initial slip: time evolution of the isospin component
  $\tau_y^{\tmop{ana}}$ computed from the analytical expression
  (\ref{eq:taueq2}). (a) Component $\tau_y^{\tmop{ana}}$ for a full qubit
  cycle. We assume a factorizable initial state of SQD \tmtextit{plus} qubit
  factorizes by taking $\vecg{\tau}^n (0) = p^n (0) \op{\vecg{\Omega}}$ for $n =
  0, 1$. Both solutions show a phase shift with respect to each other that does not
  die out, i.e., it persists even over many qubit cycles. In inset (b), we
  compare $\tau_y^{\tmop{ana}}$ (blue) with the component $\tau_y$ computed
  from the full kinetic equations (\ref{eq:kineq}) (red). The high-temperature
  approximation $\tau_y^{\tmop{ana}}$ approaches the full solution $\tau_y$ on
  the time scale $\sim 1 / \gamma$ during which the SQD approaches
  stationarity. The inset (c) shows $\tau_y^{\tmop{ana}}$ after one qubit
  cycle for the two initial conditions and illustrates that the offset of both
  curves has not changed appreciably as compared to the evolution close to $t
  = 0$. The coordinate system is chosen as $( \vec{e}_x, \vec{e}_y,
  \vec{e}_z) = ( \op{\vecg{\Omega}}, \op{\vecg{\lambda}} \times
  \op{\vecg{\Omega}}, \op{\vecg{\lambda}})$. The parameters are $\Gamma_s
  = \Gamma_d = \bar{\Gamma} = 10^{- 3} T$, $\lambda = \Omega = 0.1
  \bar{\Gamma} = 10^{- 4} T$, $V_b = 3 T$, $V_g = T$, $W = 1000 T$, resulting
  in $\gamma = 2 \gamma^0 + \gamma^1 \approx 3.30 \cdot 10^{- 3} T$ [see Eq.
  (\ref{eq:dissrate})] and $\tilde{\Omega} \approx 1.27 \cdot 10^{- 4} T$ [see Eq.
  (\ref{eq:omegatilde})].\label{fig:initial}}
\end{center}
\subsection{Distortion of isospin mode vectors}\label{sec:elliptical}

The rough mean-field picture also breaks down when accounting for the
backaction effect on the \tmtextit{isospin modes vectors}: the eigenvectors
${\color{black} \mathbf{e}}_{\tmop{eff}, \alpha}$ are modified from Eq.
(\ref{eq:eundist}) to Eqs. (\ref{eq:e0eff})--(\ref{eq:e2eff}) when taking into
account the finite decoherence rate $\lambda^2 / \gamma$. This leads to both a
tilting of the qubit axis and elliptical isospin precession.

\subsubsection{Tilting of precession axis}

As a first consequence, the effective precession axis $\vec{e}_{\tmop{eff},
0}$ acquires an additional tilting beyond the mean-field effect. This
manifests as a nonzero component of $\vec{e}_{\tmop{eff}, 0}$ along 
$\vec{e}_{\parallel} \times \vec{e}_{\perp} = \op{{\color{black}
\vecg{\Omega}}} \times \op{{\color{black} \vecg{\lambda}}}$, which is
perpendicular to both the intrinsic qubit precession axis $\op{\vecg{\Omega}}$
and the measurement vector $\op{\vecg{\lambda}}$ \newer{and therefore
perpendicular to $\op{\vecg{\Omega}}$.} By virtue of Eq.
(\ref{eq:e0eff}) this rotates the qubit axis relative to $\vec{e}_{\parallel}$
by an angle
\begin{eqnarray}
  \chi & \approx & r \frac{\lambda^2}{\gamma \tilde{\Omega}}  \frac{\Omega
  \lambda p^1_{\tmop{st}}}{\tilde{\Omega}^2} 
\end{eqnarray}
plus higher-order corrections. This tilt becomes noticeable close to
resonance, where detection is performed,
as we illustrate in {\color{black} Fig. \ref{fig:axis}}(a), where we
plot the projection of ${\color{black} \vecg{\tau}} (t)$ onto the mean-field
axis $\vec{e}_{\parallel} = \tilde{{\color{black} \vecg{\Omega}}} /
\tilde{\Omega}$. In addition to an exponential decay with the relaxation rate
$1 / T_1$, this component acquires an additional oscillatory component for a
general initial state. This simply indicates that we are looking at the
component along a vector that is not the zero eigenmode of the qubit: the
backaction additionally tilts the relaxation mode vector from $\vec{e}_{\parallel}
\rightarrow \vec{e}_{0, \tmop{eff}}$ through the virtual-transition
terms $\propto \lambda^2/\gamma$.
\begin{center}
  \Figure{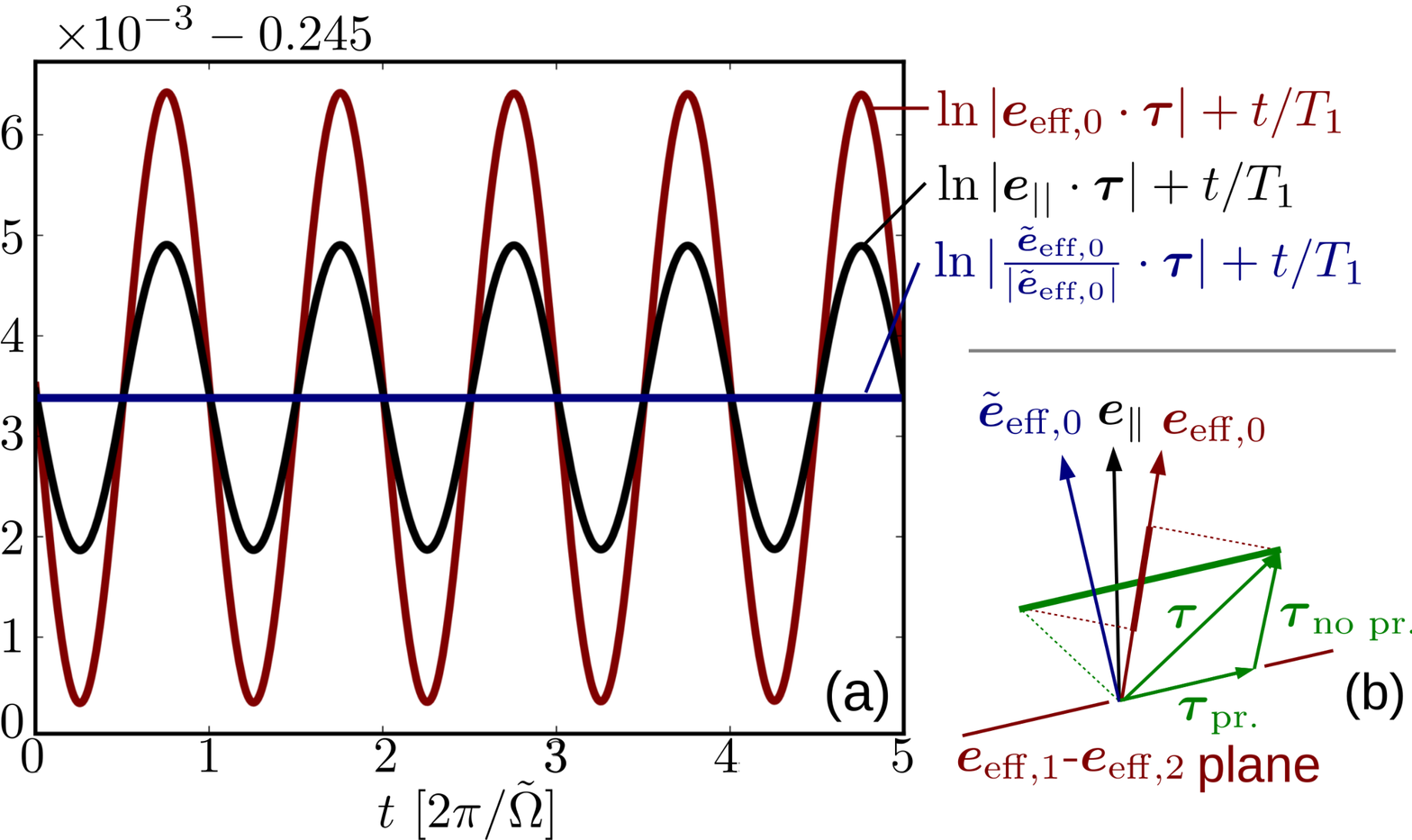}{Distortion
  of isospin modes: (a) Comparison of the components of the high-temperature
  approximation (\ref{eq:taueq2}) for the isospin ${\color{black}
  \vecg{\tau}} (t)$ along the mean-field qubit axis, $\vec{e}_{\parallel} =
  \tilde{{\color{black} \vecg{\Omega}}} \text{/} \tilde{\Omega}$,
  (black), along the tilted relaxation mode vector $\vec{e}_{\tmop{eff}, 0}$
  (red), and along its dual $\tilde{\vec{e}}_{\tmop{eff}, 0}$, (blue) as
  a function of time. Plotted are $\ln \left( \vec{e}_{\parallel} \cdot
  {\color{black} \vecg{\tau}} \right) + t \text{/} T_1$ (black), $\ln
  \left( \vec{e}_{\tmop{eff}, 0} \cdot {\color{black} \vecg{\tau}} \right) +
  t \text{/} T_1$ (red), and $\ln \left( \left( \tilde{\vec{e}}_{\tmop{eff}, 0} / \left| \tilde{\vec{e}}_{\tmop{eff}, 0} \right| \right)
  \cdot {\color{black} \vecg{\tau}} \right) + t \text{/} T_1$ (blue),
  respectively. The initial state is given by $p^1 (0) = 1 - p^0 (0) = 0$,
  $\vecg{\tau} (0) = \vecg{\tau}^0 (0) = \op{\vecg{\Omega}}$ and all other
  parameters as in {\color{black} Fig. \ref{fig:initial}}.  \newer{Note that
 the decay of the precessional component is hardly visible here because
 $T_2 \approx T_1$ for the parameters chosen here and because of the
 short time window shown in this figure (see Fig. \ref{fig:compana})}
  (b) Sketch of the
  orientation of $\vec{e}_{\parallel}$, $\vec{e}_{\tmop{eff}, 1}$, and
  $\tilde{\vec{e}}_{\tmop{eff}, 1}$ relative to the $\vec{e}_{\tmop{eff}, 1}$-$\vec{e}_{\tmop{eff}, 2}$ plane. The thick green line indicates
  the precession of the isospin in a plane parallel $\vec{e}_{\tmop{eff}, 1}$-$\vec{e}_{\tmop{eff}, 2}$ plane and shifted along $\vec{e}_{\tmop{eff}, 0}$. The precessing part of the total isospin leads to an
  oscillation along the components along $\vec{e}_{\tmop{eff}, 0}$ (as
  indicated) and also along $\vec{e}_{| |}$. Only the component along
  $\tilde{\vec{e}}_{\tmop{eff}, 0}$ does not oscillate.\label{fig:axis}}
\end{center}
However, the component of ${\color{black} \vecg{\tau}} (t)$
along the tilted relaxation mode vector $\vec{e}_{\tmop{eff}, 0}$ shows \emph{also}
an oscillation (superposed on an exponentially decaying
contribution) as the red curve in {\color{black} Fig. \ref{fig:elliptical}}(a)
illustrates. This is an effect of the mode distortion: the plane spanned by
the unit vectors $\vec{e}_{\tmop{eff}, 1}$ and $\vec{e}_{\tmop{eff}, 2}$ is
not anymore orthogonal to $\vec{e}_{\tmop{eff}, 0}$. Since the precessing
part of the isospin lies in this plane, the projection of the isospin
$\vecg{\tau} (t)$ on $\vec{e}_{\tmop{eff}, 0}$ becomes oscillatory as
sketched in {\color{black} Fig. \ref{fig:axis}}(b). These
oscillations are therefore damped with the \emph{dephasing} rate $1 / T_2$, which differs from
the relaxation rate $1 / T_1$ that sets the time scale for the exponential
decay of the nonoscillatory contribution. Thus, the nonorthogonality of the
isospin mode vectors mixes relaxation and dephasing in a nontrivial
way.

The only component of $\vecg{\tau} (t)$ that does not oscillate
for an {\tmem{arbitrary}} initial state is the one along
$\tilde{\vec{e}}_{\tmop{eff}, 0}$ as the blue curve in {\color{black} Fig.
\ref{fig:axis}}(a) shows. The reason is that the dual vector
$\tilde{\vec{e}}_{\tmop{eff}, 0} \propto \vec{e}_{\tmop{eff}, 1} \times
\vec{e}_{\tmop{eff}, 2}$ is normal to the $\vec{e}_{
\tmop{eff},1}$ - $\vec{e}_{\tmop{eff}, 2}$ plane [see definition
(\ref{eq:tildedef})]. This normal component is also independent of the dephasing
time: it simply decays exponentially with the relaxation rate $1 \text{/}
T_1$.

However, the dual vector $\tilde{\vec{e}}_{\tmop{eff}, 0}$
should not be confused with the zero mode $\vec{e}_{\tmop{eff},
0}$: If we initially have ${\color{black} \vecg{\tau}}_{\tmop{eff}} (0)
\propto \tilde{\vec{e}}_{\tmop{eff}, 0}$, then the isospin still has a
precessional component since $\tilde{\vec{e}}_{\tmop{eff}, 0}$ is not the
relaxation mode vector. If one aims to prepare the qubit in a state whose
Bloch vector \tmtextit{direction} is stable under the time evolution that
includes the measurement backaction, one should take
\begin{eqnarray}
  {\color{black} \vecg{\tau}}_{\tmop{eff}} (0) & = & F \vec{e}_{0,\tmop{eff}},  \label{eq:leastspiralinit}
\end{eqnarray}
where $F$ is suitable real constant. \footnote{Note that ${\color{black}
\vecg{\tau}}_{\tmop{eff}} (0)$ differs from the initial isospin $\vecg{\tau}
(0)$ due to the initial slip (\ref{eq:taueff0}) and therefore $F$ is not
necessarily limited to the range $[- 1, + 1]$.} For the initial state
(\ref{eq:leastspiralinit}), we indeed find pure exponential decay to the
stationary state ${\color{black} \vecg{\tau}} (\infty) = \vec{0}$:
\begin{eqnarray}
  {\color{black} \vecg{\tau}} (t) & = & F e^{- t \text{/} T_1} 
  \vec{e}_{\tmop{eff}, 0} .  \label{eq:leastspiral}
\end{eqnarray}
The above illustrates that the notion of ``exciting a qubit mode'' has to be
treated with care due to both backaction-induced initial slip and due to mode
vector distortion. Due to the initial slip discussed above, one has to prepare
the qubit-sensor state [see \Eq{eq:xqefft}, \Eq{eq:taueff0}, and
{\color{black} Appendix \ref{app:slip}}] very carefully in order to achieve the initial
condition (\ref{eq:leastspiralinit}).

\subsubsection{Elliptical precession}

The second qualitative consequence of the distortion of the qubit modes due to
the finite decoherence rate concerns the precessional motion in the
$\vec{e}_{\tmop{eff}, 1}$ -- $\vec{e}_{\tmop{eff}, 2}$ plane with normal
$\tilde{\vec{e}}_{\tmop{eff}, 0}$. The trajectory of the isospin vector
${\color{black} \vecg{\tau}} (t)$ in this plane is changed from a circle to an
ellipse. We illustrate this in {\color{black} Fig. \ref{fig:elliptical}}(a) for
an effective initial isospin ${\color{black} \vecg{\tau}}_{\tmop{eff}} (0) =
F \vec{e}_{\tmop{eff}, 1}$ lying in this plane (again for a suitable real
constant $F$). Applying \Eq{eq:taueq2}, the evolution of the isospin
can be expressed as:
\begin{eqnarray}
  {\color{black} \vecg{\tau}} (t) & = & e^{- t \text{/} T_2} F \left[ \cos
  (\tilde{\Omega} t)  \vec{e}_{\tmop{eff}, 1} + \sin (\tilde{\Omega} t) 
  \vec{e}_{\tmop{eff}, 2} \right] \\
  & = & e^{- t \text{/} T_2} F \left[ \sqrt{1 - \epsilon / 2} \cos \left(
  \tilde{\Omega} t + \frac{\pi}{4} \right) \vec{v}_1 \right. \nonumber\\
  &  & \text{ \ \ \ \ \ \ \ \ \ } + \sqrt{1 + \epsilon / 2} \sin \left.
  \left( \tilde{\Omega} t + \frac{\pi}{4} \right) \vec{v}_2 \right] . 
  \label{eq:elliptical}
\end{eqnarray}
In the rewritten form, the part in the bracket describes an elliptical motion
with linear eccentricity
\begin{eqnarray}
  \epsilon \ = \ 2 \vec{e}_{\tmop{eff}, 1} \cdot \vec{e}_{\tmop{eff}, 2}
  &= &\frac{1}{\tilde{\Omega} T_1} \text{ \ } = \text{ \ }
  r \frac{\lambda^2}{\gamma \tilde{\Omega}^{}} 
  \frac{\Omega^2}{\tilde{\Omega}^2}, 
\end{eqnarray}
which is maximal near the resonance of the sensor QD [due to $r$, see \Eq{eq:r}] and
proportional to the scale $\lambda^2 / \gamma$ for the decoherence rate
relative to the qubit frequency $\tilde{\Omega}$. The precession plane is
spanned by two orthonormal vectors,
\begin{eqnarray}
  \vec{v}_{1, 2} & = & \frac{\vec{e}_{\tmop{eff}, 1} \pm \vec{e}_{\tmop{eff},
  2}}{\sqrt{2 \pm \epsilon}},  \label{eq:v12}
\end{eqnarray}
which are at the same time \newer{unit vectors} along the principal axes of the precession ellipse
as sketched in {\color{black}
Fig. \ref{fig:elliptical}}(b).
\begin{center}
  \Figure{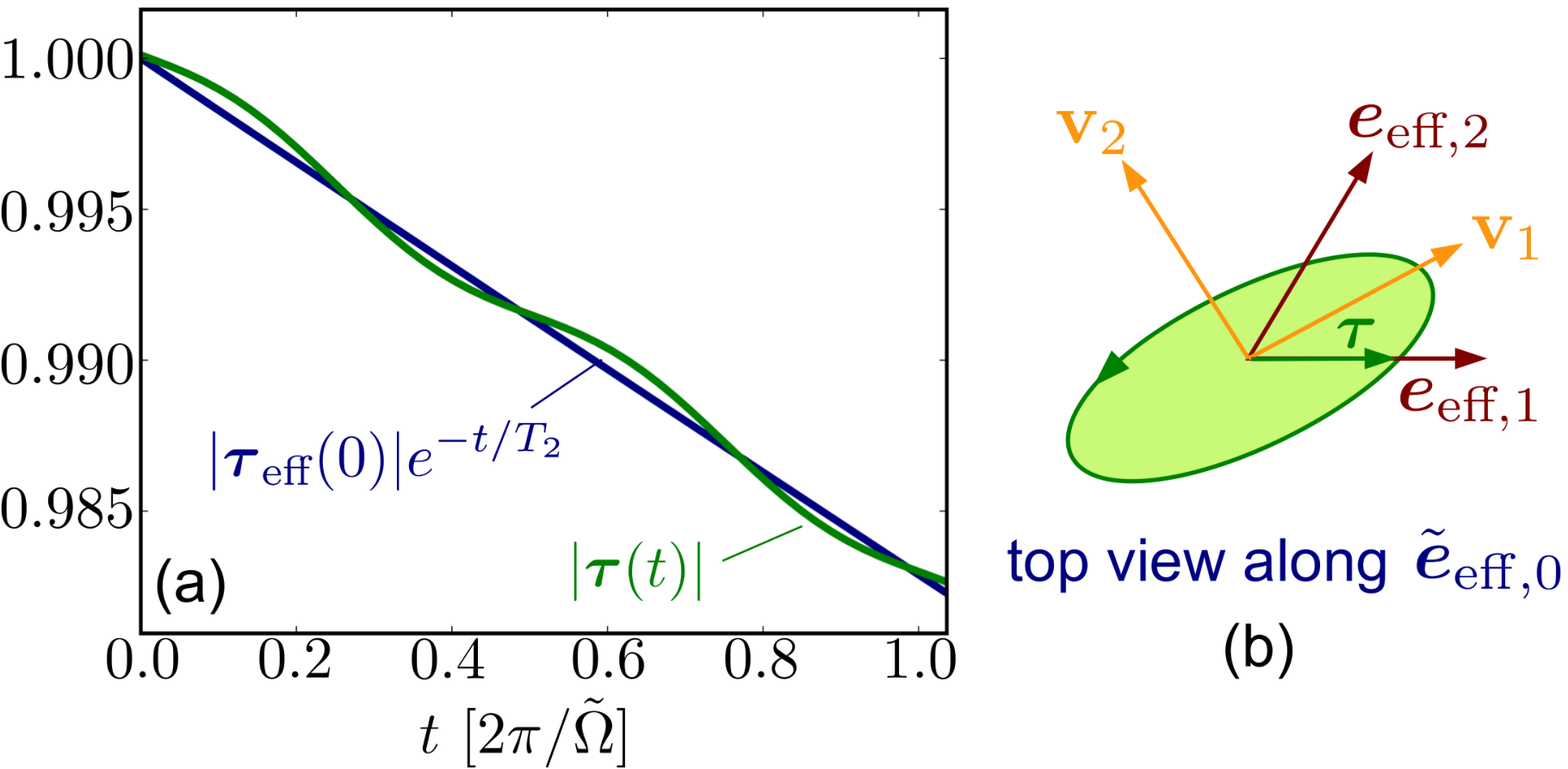}{Elliptical
  precession. (a) Magnitude of the high-temperature approximation of the
  isospin $| {\color{black} \vecg{\tau}} (t) |$ (green). We take the initial
  condition $p^1 (0) = 1 - p^0 (0) = 0$, $\vecg{\tau}_{\tmop{eff}} (0) =
  \vec{e}_{\tmop{eff}, 1}$ here such that ${\color{black} \vecg{\tau}} (t)$
  evolves in the $\vec{e}_{\tmop{eff}, 1}$ - $\vec{e}_{\tmop{eff}, 2}$
 plane. \newer{From Eqs. \eq{eq:e1eff} and \eq{eq:e2eff}, it is easy to see that
 $\vec{e}_{\tmop{eff}, 1}$ has a much larger component along
 $\op{\vecg{\lambda}}$ than $\vec{e}_{\tmop{eff}, 2}$, leading to a
 stronger decay initially (see explanation in the text).}
  The oscillating deviations of the isospin magnitude from the exponential
  dephasing, $| {\color{black} \vecg{\tau}}_{\tmop{eff}} (0) | e^{- t /
  T_2}$ (blue), reveal the elliptical precession. All other parameters are as
  in {\color{black} Fig. \ref{fig:initial}}. Although the effects are
 weak in our controlled perturbative calculations, they indicate
 qualitatively new features that can be expected to grow for stronger
 readout couplings. (b) Two-dimensional sketch of the ellipse with
  principal axis $\vec{v}_1$ and $\vec{v}_2$ [see \Eq{eq:v12}]
  described by the tip of the isospin for initial condition as in (a). The
  exponential shrinking with rate $1 / T_2$ is not indicated for
  simplicity.\label{fig:elliptical} }
\end{center}
\newer{Equation \eq{eq:elliptical} shows
that the magnitudes
of both principal axes shrink exponentially with rate $1 / T_2$. The magnitude of the isospin
[green curve in {\color{black} Fig. \ref{fig:elliptical}}(a)] thus oscillates
around a pure exponential decay with that rate [blue curve in
{\color{black} Fig. \ref{fig:elliptical}}(a)]. The oscillations are a
signature of the
elliptical distortion of the precessional motion.}

\newer{If there were no mode distortion, the above picture would be true
also for
slipped initial states
out of the $\vec{e}_{1,\tmop{eff}}$
-- $\vec{e}_{\tmop{eff}, 2}$ plane. However, due to the mode distortion
arbitrary initial states must again considered with care}
because the projection of $\vecg{\tau} (t)$ onto the $\vec{e}_{\tmop{eff}, 1}$
- $\vec{e}_{\tmop{eff}, 2}$ precession plane has an additional contribution
arising from the component of $\vecg{\tau}$ along $\vec{e}_{\tmop{eff}, 0}$
[see {\color{black} Fig. \ref{fig:axis}}(b)]: this causes the center of the
ellipse to be shifted away from origin. This shift of the center decays exponentially
towards the origin with the relaxation rate $1 / T_1$, which is again
different from the decay rate of the precession, which is damped with the
dephasing rate $1 / T_2$.

\emph{Relation to state purity.}
The discussed elliptical isospin motion reflects the exponentially damped but oscillatory
decrease of the qubit-state purity due to the readout by the sensor QD. 
This can be roughly understood as follows. The dephasing
is the strongest when the isospin and the measurement vector
$\op{\vecg{\lambda}}$ are perpendicular to each other because this corresponds
to the charge qubit electron being delocalized between the two sites. Here,
small fluctuations in the detuning, induced by the stochastic switching of the
SQD, then introduce a strong dephasing. By contrast, no dephasing appears when
the isospin and the measurement vector $\op{\vecg{\lambda}}$ are collinear,
i.e., when the charge qubit electron is localized in one of the QDs. This
interpretation is consonant with the situation in {\color{black} Fig.
\ref{fig:elliptical}}(a), which shows the magnitude of the qubit Bloch vector
as a function of time. The qubit Bloch vector has initially a large overlap
with $\op{\vecg{\lambda}}$, leading to a suppressed
decay in the first quarter of a precession period $2 \pi /
\tilde{\Omega}$. The oscillatory reduction of the qubit purity was anticipated also in our discussion of the exact relation
(\ref{eq:puritydecay}),
\begin{eqnarray}
  \frac{d}{d t} [| {\color{black} \vecg{\tau}} (t) |^2] & = & - 2
  {\color{black} \vecg{\lambda}} \cdot [{\color{black} \vecg{\tau}}^0 (t)
  \times {\color{black} \vecg{\tau}}^1 (t)].  \label{eq:puritydecay2}
\end{eqnarray}
If $\vecg{\tau} (t)$ precesses, its charge-specific components
$\vecg{\tau}^n (t)$ also precess and their components along $\vecg{\lambda}$
change in time, resulting in a nonexponential purity decay {\footnote{We note that applying \Eq{eq:puritydecay2}
requires a calculation of the charge-specific isospins from the full
kinetic equations (\ref{eq:kineq}). Using only the projections quasistationary
subspace, $\vec{X}^q (t)  = \left(
  1 \ \vecg{\tau} (t) \right)^{\dag}$, see \Eq{eq:xq}, would just yield zero when
inserted into \Eq{eq:puritydecay2}.}}.

Finally, we note that these oscillations of the purity decay are {\tmem{not}} an effect of
non-Markovian corrections [see Eqs. \eq{eq:xqt} and \eq{eq:xqefft}] induced by the sensor QD onto the qubit by
capacitive interaction $\lambda$: they may even be reproduced by modeling the
SQD charge as a classical fluctuator (see, e.g., Refs.
{\cite{Abel08,Neuenhahn09}}). In fact, to obtain Eqs. (\ref{eq:taueq2}) and
(\ref{eq:taueff0}), we employ a Markovian approximation with respect to the sensor on the qubit (by setting $z = 0$, see discussion in
{\color{black} Appendix \ref{app:nm}}). The oscillations of the qubit decay should
also not be mistaken for coherence revivals, i.e., an increase in the purity
of the qubit state. Figure {\color{black} \ref{fig:elliptical}}(a) clearly
shows that the magnitude of the qubit Bloch vector decreases for all times,
i.e., the information is permanently transferred to the environment during the
measurement process. However, the {\tmem{rate}} of information loss is
nonmonotonic, which simpler approaches might not predict (see, e.g.,  \Cite{Clerk10}). This shows most
clearly that our perturbation theory goes beyond the simple ``mean-field''
detector picture discussed in {\Sec{sec:meanfield}}.


\section{Comparison to other approaches}\label{sec:comparison}

In this final Section we compare our results for the measurement backaction
with the results of prior works. One of the central results that we obtained
-- \new{the strong} suppression of the measurement backaction when
tuning the sensor into the Coulomb blockade regime -- is surprising: One
expects to underestimate the measurement backaction by a too simplified
treatment which ignores cotunneling noise. We accounted for this cotunneling
noise contribution (``broadening'' contribution to the rates) but our result
(\ref{eq:rexp}) revealed that it is canceled out by the coherent backaction, a
renormalization effect. Such a cancellation should of course be viewed very
critically and we have carefully traced its origins. It also raises the
question, which physical assumptions and approximations may have caused it not
to be noticed before. This is discussed here.

We first show in {\Sec{sec:semiclassical}} that
semiclassical approaches can only reproduce the contributions to the decay
times induced by the {\tmem{stochastic}} backaction, but they fail to account
for corrections from the coherent backaction. This also quite easily happens
within a standard weak-coupling Bloch-Redfield approach aimed at finding a
qubit-only description as we discuss in {\Sec{sec:br}}.
Both these approaches are correct only under the assumption of the
weak-coupling, high-temperature limit, which may experimentally be violated
and does not allow the sensor QD current to be calculated. Finally, we show
that our findings are in accordance with exact quantum treatments of models of
decoherence due to noninteracting two-level fluctuators in equilibrium,
insofar a comparison is possible.

\subsection{Semiclassical stochastic approaches}\label{sec:semiclassical}

A popular way to study the decoherence a qubit suffers from the coupling to
its environment is a semiclassical {\tmem{qubit-fluctuator model}}
{\cite{Galperin04,Ithier05,Schriefl06,Chirolli08,Bergli09}}. In this approach,
the qubit is considered subjected to noise generated by a randomly switching
two-level system. The basic idea is to replace the occupation number $\op{n}$
of the SQD in the interaction Hamiltonian, $H_I = \op{n} {\color{black}
\vecg{\lambda}} \cdot \op{{\color{black} \vecg{\tau}}} / 2$, by a
classical random process $\op{n} \rightarrow \xi (t) / 2$ {\footnote{We
introduce a factor 1/2 for convenience to adjust the amplitude of the
fluctuations to $\Delta \xi = 2$, as usually used for a random telegraph
process.}}. Applied to our case, the SQD introduces a fluctuating effective
magnetic field acting on the qubit [see \Eq{eq:field}]:
\begin{eqnarray}
  H_Q^{\tmop{eff}} & = & \frac{1}{2} \left( \vecg{\Omega} + \tfrac{1}{2} \xi
  (t)  {\color{black} \vecg{\lambda}} \right) \cdot \op{{\color{black}
  \vecg{\tau}}} .  \label{eq:hqstochastic}
\end{eqnarray}
Since our primary interest lies in the interplay of the coherent backaction
with the cotunneling-induced backaction, we can take $\Omega = 0$ and drop the
spin degree of freedom of the SQD electrons and therefore the effect of the
Coulomb interaction in the following. Neither of these assumptions are
critical for reproducing these effects (see {\color{black} Appendix
\ref{app:renormalization}}). The measurement then induces a dephasing of
superpositions of the qubit states ${\color{black} |} L {\color{black}
\rangle}$ and ${\color{black} |} R {\color{black} \rangle}$, physically
related to the double-QD electron residing either in the left or right QD.
This can be characterized by the decay of the off-diagonal elements of the
qubit density matrix, the {\tmem{visibility}}
{\cite{Bergli09,Grishin05,Abel08}}:
\begin{eqnarray}
  \left\langle \op{\tau}_+ (t) \right\rangle & = &  \left\langle
  {\color{black} \langle} L {\color{black} |} e^{- i \int_0^t d t_1
  H_Q^{\tmop{eff}} (t_1)} {\color{black} |} L {\color{black} \rangle} \right.
  \nonumber\\
  &  & \times \left. {\color{black} \langle} R {\color{black} |} e^{+ i
  \int_0^t d t_1 H_Q^{\tmop{eff}} (t_1)} {\color{black} |} R {\color{black}
  \rangle} \right\rangle .  \label{eq:tauplusav}
\end{eqnarray}
Here, the bracket $\langle \ldots \rangle$ denotes the average over many
realizations of the random process $\xi (t)$ with $\langle \xi (t)
\rangle = 2 p^1_{st}$. Splitting the qubit Hamiltonian
$H_Q = \langle H_Q \rangle + \delta H_Q$ according to \Eq{eq:field}
into a mean-field part, $\langle H_Q \rangle = \frac{1}{2} p^1_{\tmop{st}}
{\color{black} \vecg{\lambda}} \cdot {\color{black} \vecg{\tau}}$, and a
fluctuating part, $\delta H_Q = \frac{1}{4} \delta \xi (t) {\color{black}
\vecg{\lambda}} \cdot \op{{\color{black} \vecg{\tau}}}$, with $\delta \xi
(t) = \xi (t) - \langle \xi (t) \rangle$ and $\langle \delta \xi (t) \rangle =
0$, \Eq{eq:tauplusav} can be recast as
\begin{eqnarray}
  \left\langle \op{\tau}_+ (t) \right\rangle & = & \left\langle e^{- \frac{i
  \lambda}{2} \int_0^t d t_1 \delta \xi (t_1)} \right\rangle . 
  \label{eq:tplus}
\end{eqnarray}
Hence, the decay of the coherences depends only on the ``amplitude'' of the
fluctuations, while the average of the fluctuations, $\langle \xi (t)
\rangle$, is irrelevant. This mean-field part just induces a constant shift of
the qubit energy level, which has been absorbed into the average $\langle H_Q
\rangle$, see our discussion of the mean-field picture in {\color{black} Sec.
\ref{sec:meanfieldpicture}}.

For simplicity, let us first take $\delta \xi (t)$ to be a Gaussian random
process. In this case, \Eq{eq:tplus} can be rewritten as
{\cite{Bergli09}}
\begin{eqnarray}
  \left\langle e^{\frac{- i \lambda}{2} \int_0^t d t_1 \delta \xi (t_1)}
  \right\rangle  \ = \  e^{- \frac{\lambda^2}{8} \int_0^t d t_1 \int_0^t d t_2
  \langle \delta \xi (t_1) \delta \xi (t_2) \rangle}, & & \label{eq:reexp}
\end{eqnarray}
resulting in an exponential decay of $\left\langle \op{\tau}_+ (t)
\right\rangle = e^{- t / T_2}$. The dephasing time,
\begin{eqnarray}
  \frac{1}{T_2} \text{ \ } = \text{ \ } \frac{1}{T_{\phi}} & = & \left(
  \frac{\lambda}{2} \right)^2 \frac{S_{\delta \xi} (0)}{2}, 
  \label{eq:t2spectral}
\end{eqnarray}
can be related to the noise power spectrum $S_{\delta \xi} (\omega)$ of the
Gaussian random process $\delta \xi (t)$ {\cite{Bergli09}},
\begin{eqnarray}
  S_{\delta \xi} (\omega) & \assign & \int_{- \infty}^{+ \infty} d \tau e^{- i
  \omega \tau} \langle \delta \xi (0) \delta \xi (\tau) \rangle, 
  \label{eq:spectral}
\end{eqnarray}
where we used that the correlator $\langle \delta \xi (t_1) \delta \xi (t_2) \rangle $ only depends on the time difference $\tau = t_2 - t_1$ because a
Gaussian random process is time-translational invariant. We emphasize that the
relation (\ref{eq:reexp}) holds exactly for a Gaussian process, whose entire
statistics is fixed by two-point correlation functions.

For a general random process, however, higher-order time correlators may
contribute to \Eq{eq:reexp} {\cite{Galperin04,Bergli09,Makhlin04}} and
therefore the spectral function is not sufficient to characterize a random
process completely. A prominent example is a Poisson process inducing random
telegraph noise. Such a process would actually be a better model for a
capacitively coupled QD stochastically switching its occupation due to
tunneling processes. Random telegraph noise has also been extensively studied
to explain the origin of flicker ($1 / f$) noise in superconducting qubits
{\cite{Ithier05,Bergli09,Shnirman05,Schriefl06,Galperin04}} and Gaussian noise
in semiconductor QDs {\cite{Berthelot10,Kuhlmann13}}. This noise results from
an ensemble of fluctuating background charges, each of which may be compared
with our sensor QD, except that they do not carry a current \newer{on average}. The non-Gaussian
behavior of the Poisson process comes to light only if a few fluctuators
dominate the decoherence of the qubit state, which entails a nonexponential
decay of the coherences {\cite{Bergli09,Paladino02,Abel08,Neuenhahn09}} and
even coherence revivals have been predicted {\cite{Neuenhahn09,Abel08}}.

The impact of charge fluctuations of an individual QD have also been modeled
by classical random telegraph process {\cite{Bergli09,Itakura03}}. In general,
$\delta \xi$ switches between two values $\delta \xi_u$ (upper), and $\delta
\xi_l$ (lower) with different switching rates $1 / \tau_l$ for $\delta \xi_u
\rightarrow \delta \xi_l$, and $1 / \tau_u$ for $\delta \xi_l \rightarrow
\delta \xi_u$. This gives rise to different probabilities $p_u$ to find the
value $\delta \xi_u$, and $p_l$ to find the value $\delta \xi_l$,
respectively. In the weak-coupling limit, $\lambda \ll \gamma = 1 / \tau_u + 1
/ \tau_l$, it turns out that the dephasing is still exponential and the
dephasing rate reads as {\footnote{We adjusted the notation in {\color{black}
Ref. {\cite{Itakura03}}} to our notation by replacing $1 / \tau_0 = 1 / \tau_u
+ 1 / \tau_l = \gamma$ and $J_T \rightarrow \lambda / 2$.}}:
\begin{eqnarray}
  \frac{1}{T_2} & \approx & [1 - (p_u - p_l)^2]  \frac{(\lambda /
  2)^2}{\gamma} .  \label{eq:oneovert2telegraph}
\end{eqnarray}
Identifying $p_u \rightarrow p^1_{\tmop{st}}$ and $p_l \rightarrow
p^0_{\tmop{st}}$ with the stationary occupation probabilities of the SQD for
our case and noting $( \op{\vecg{\lambda}} \cdot \vec{e}_{| |} )^2 =
1$ for $\Omega = 0$, \Eq{eq:oneovert2telegraph} reproduces the the
dephasing rate (\ref{eq:tphi}), which we obtained in the high-temperature
limit.

Yet, in the classical fluctuator model, we have not explicitly assumed \new{
high temperatures}. Thus, if one naively extends the above result to
include cotunneling corrections, one gets a faulty result. The cotunneling
changes the occupation probabilities $p^0_{\tmop{st}}$ and $p^1_{\tmop{st}}$
and the switching rate $\gamma$, but there is no way in which additional
coherent backaction terms could appear [see the transition factor $r$, Eq.
(\ref{eq:r})]. As a consequence, one might overestimate the qubit dephasing
rate at the onset of Coulomb blockade. In addition, the classical approach
does also not account for the dissipative backaction, which is less important
for the dephasing times as compared to the coherent backaction {\footnote{The
reason is that the dissipative backaction does not affect the coupling between
the quasistationary and decay modes [see Eqs. (\ref{eq:transdq}) --
(\ref{eq:transqd})].}}.

The reason why the above classical approach is not able to reproduce the
coherent backaction is that the electron number $n$ is a classical variable
with a definite value at each instant of time, meaning the SQD is fluctuating
between the states $| 0 \rangle \langle 0 |$ and $| 1 \rangle \langle 1 |$,
where $| n \rangle$ denotes the SQD state with $n$ electrons on it. In this
way, it can only produce the stochastic backaction. Quantum coherences $| 0
\rangle \langle 1 |$ and $| 1 \rangle \langle 0 |$ involved in
\tmtextit{virtual} processes of the SQD (quantum fluctuations), which play a
role during the tunneling, are disregarded here. 
However, as we illustrate in
Appendix \ref{app:renormalization}, these coherences -- included
in our calculation -- are crucial for obtaining the coherent
backaction. (Note
that despite this, such coherences can not appear in the \tmtextit{real}
quantum state, the relevant density operator $\rho$ [see Eq.
(\ref{eq:reddens})] \new{due to} charge conservation.) \new{It turns out that
the related quantum fluctuations induce an additional
``phase kick'' while a charge transition in the SQD takes
place. This phase shift partially compensates the phase shift from the
stochastic backaction while the SQD is in a specific charge state. 
This makes
plausible why quite generally one may expect a coherent backaction that
mitigates measurement backaction when employing a \tmtextit{quantum} sensor
for the indirect detection of a qubit.}
Thus, a purely classical understanding of the indirect measurement backaction due to a sensor QD is
incomplete.

\subsection{Bloch-Redfield approach}\label{sec:br}

Another frequently employed method to study decoherence is the Bloch-Redfield
approach {\cite{Redfield57,Breuer}}, also in the context of qubits
{\cite{Chirolli08,Ithier05}}. In the original presentation
{\cite{Redfield57}}, this approach is first developed for a
semiclassical {\footnote{In Redfield's notation in {\color{black} Ref.
{\cite{Redfield57}}}, the random perturbation is for our model given by $G (t)
= \frac{1}{2} \xi (t)  {\color{black} \vecg{\lambda}} \cdot
\op{{\color{black} \vecg{\tau}}}$.}} and after this also for a quantum
perturbation. In both cases, one considers the limit of weak coupling between
a quantum system and its environment and an additional Born-Markov
approximation is made. Importantly, the approach generally predicts an
exponential decay into a stationary state irrespective of the statistics of
the environmental fluctuations. The corresponding relaxation matrix of the
reduced density matrix is furthermore related to the noise power spectrum of
the perturbation.
\newer{\subsubsection{Stochastic backaction in spectral function}}
For our case, the Bloch-Redfield approach confirms the dephasing rate
(\ref{eq:t2spectral}) also within a quantum treatment: one simply has to
replace in \Eq{eq:spectral} the classical average $\langle \delta \xi
(t) \delta \xi (t + \tau) \rangle$ by the quantum-ensemble average:
\begin{eqnarray}
  S_{\delta \op{n}} (\omega) & \assign & \int_{- \infty}^{+ \infty} d \tau
  e^{- i \omega \tau} 4 \left\langle \delta \op{n} (t) \delta \op{n} (t +
  \tau) \right\rangle .  \label{eq:sn}
\end{eqnarray}
For an indirect readout model of interest here, one treats the SQD as a
noninteracting two-level quantum system in contact with a thermal reservoir.
Applying the results of {\color{black} Ref. {\cite{Shnirman05}}} for a single
fluctuator, the spectral function reads {\footnote{To compare with Eq. (7) of
{\color{black} Ref. {\cite{Shnirman05}}}, we note that for our model there is
no direct hybridization $\Delta_j = 0$ between the orbital levels of the SQD,
i.e., $\cos (\theta_j) = 1$, and $\sin (\theta_j) = 0$. Furthermore, the
relaxation rate $\Gamma_{1, j}$ to the stationary state is identified with
$\gamma$ here.}}
\begin{eqnarray}
  S_{\delta \op{n}} (\omega) & = & [1 - (p^1_{\tmop{st}} - p^0_{\tmop{st}})^2]
  \frac{2 \gamma}{\gamma^2 + \omega^2}  \label{eq:sdn}
\end{eqnarray}
leading to \Eq{eq:oneovert2telegraph} when inserting $S_{\delta \op{n}}
(\omega = 0)$ into \Eq{eq:spectral}. Hence, also the Bloch-Redfield
approach -- even accounting for a quantum environment -- does not reproduce
the coherent backaction and may therefore be applied only in the
high-temperature limit to derive dephasing and relaxation times.

\subsubsection{Difficulty of capturing dissipative and coherent backaction}\label{sec:whynotbr}

The reason why the Bloch-Redfield approach fails to account for the
dissipative and coherent backaction is not specific to this approach. It is
rather a problem that appears in principle for any procedure that tries to
treat the sensor in an indirect measurement setup as a given environment
without a nonequilibrium dynamics of its own. Such an approach always needs to
make some assumptions about the sensor density operator, which, as we have
seen, does not seem to have a simple structure allowing for an educated guess
based on general physical principles. This means that one should simply
calculate the joint quantum state of sensor QD and qubit, which is what we have
done. It is, however, instructive to further understand the problems encountered
when one tries to avoid this \newer{by making (too) simple
approximations of this joint state.}\\
(i) The Bloch-Redfield approach (as well as many other density-operator
approaches) applied in the weak-coupling limit involves a factorization
assumption {\cite{Koller10}} for the state of the qubit and its environment
(here the sensor QD plus the reservoir): $\rho_{\tmop{tot}} (t) \approx \rho_Q
(t) \otimes \rho_{S R} (t)$. The factorization assumption ignores the
dissipative backaction valid only in the high-temperature limit.
Otherwise, the dissipative backaction leads to a nonzero stationary qubit Bloch
vector that reflects the nonfactorisability of the qubit-environment state
even in the stationary limit. By contrast, to find the decoherence rates in
the long-time limit $t - t_0 > 1 / \Gamma$, the factorization assumption may
still work.\\
(ii) The next critical point is then to find a proper description for the unknown
evolution $\rho_{S R} (t)$ of the environment. An assumption frequently made
is that the qubit environment is in a {\tmem{stationary}} state, $\rho_{S R}
(t) = \rho_{S R, \tmop{st}}$, for example, an equilibrium state. However,
stationarity \new{of the entire qubit environment (SQD plus electrodes)} is actually never reached whenever a measurement is performed,
for which the electrodes must be held at finite bias to produce a nonzero
measurement current. Here one should be careful to note that while the
{\tmem{reduced}} sensor QD density operator may become stationary after some
time, this is \tmtextit{not} true for the joint SQD-electrode state. In our
approach, the reduced sensor-qubit system can become stationary after
eliminating the electrodes, which are stationary.\\
(iii) Even if we further simplify the problem and assume zero bias voltage (e.g., to
compute the measurement backaction in linear response) and $\rho_{S R,
\tmop{eq}}$ is stationary, then it is still difficult to compute the
equilibrium state $\rho_{S R, \tmop{eq}}$ since the sensor QD is a strongly
interacting system. Naive assumptions made about $\rho_{S R, \tmop{eq}}$ are
prone to errors. Consider for instance the approximation $\rho_{S R,
\tmop{eq}} = \rho_{S, \tmop{st}} \otimes \rho_{R, 0}$, where $\rho_{S,
\tmop{st}}$ denotes the stationary SQD state and $\rho_{R, 0}$ is the
grand-canonical equilibrium state of the reservoirs. (One may be inclined to
make this approximation since weak coupling $\Gamma$ often implies such a
factorization.) If we use this state and average the two-point charge
correlator in \Eq{eq:sn}, this involves only charge-diagonal SQD states
$| 0 \rangle \langle 0 |$ and $| 1 \rangle \langle 1 |$ -- similar to the
semiclassical approach of {\Sec{sec:semiclassical}}. What
goes wrong here is that $\rho_{S, \tmop{st}} \otimes \rho_{R, 0}$ is not the
correct equilibrium state if we go beyond the lowest-order $\Gamma$
approximation in the tunneling. For larger $\Gamma$, the hybridization between
both systems cannot be neglected any more {\cite{Koller10}}.
During tunneling processes the total system can be in \tmtextit{virtual}
intermediate states involving sensor QD coherences $| 0 \rangle \langle 1 |$
and $| 1 \rangle \langle 0 |$ and corresponding charge coherences in the
electrode. These intermediate virtual states explicitly appear in the
calculation of the coherent backaction (see {\color{black} Appendix
\ref{app:renormalization}}). Thus, even the quantum-ensemble averaging
procedure that brings us from \Eq{eq:sn} to \Eq{eq:sdn} misses
the coherent backaction since it relies on a weak-coupling expansion between
SQD and electrodes.\\
This explains why for an {\tmem{indirect}} weak measurement setup, the
procedure employed in this paper \newer{and also in
Refs. \cite{Shnirman98,Makhlin01a}} seems unavoidable.
By integrating out the electrodes first one incorporates their effect on the
joint sensor-qubit system. 
As we have seen, this reveals that the qubit
experiences a stochastic, dissipative, as well as a coherent backaction
effect. This problem of describing the nonstationary environment is specific
to indirect measurement setups and not encountered when the qubit is directly
coupled to, e.g., a stationary environment. This is the case, for example, for
a bath of harmonic oscillators as in the spin-boson model
{\cite{Paladino08,Shnirman02,Chirolli08}}, where the environment may indeed be
described by a stationary equilibrium state. \new{Yet, previous studies for other types of
environments also show that the approximations made to integrate out the
entire qubit envronment can be too crude, including both a spin-boson model
\cite{Gutmann05} and a driven two-level fluctuator \cite{Gassmann02}. In
particular, Ref. \cite{Gassmann02} explicitly compares results for
different levels of approximations showing the breakdown of Markov and
secular approximation beyond the weak qubit-environment coupling.}

\subsection{Nonperturbative quantum solutions}\label{sec:compquantum}

Our study indicates that renormalization effects, based on quantum coherences
between the qubit and its detector, are vital for the description of the
measurement backaction even in the weak-coupling limit. We next compare our
results to prior studies that treat the qubit decoherence arising from single
fluctuators coupled to fermionic reservoirs fully quantum-mechanically.

Such studies employ various approaches, such as an exact numerical evaluation
of the visibility (\ref{eq:tauplusav}) using electrodes of finite size
{\cite{Abel08}}, a Heisenberg equation-of-motion technique
{\cite{Paladino02}}, or a Keldysh path-integral formalism
{\cite{Grishin05,Paladino02}}. All these approaches are nonperturbative both
in the measurement interaction $\lambda$ and the sensor tunneling $\Gamma$.
However, in contrast to our model these studies are limited to noninteracting
fluctuators (here: the SQD) in equilibrium with a single reservoir. Thus, they
cannot access the situation of a (nonequilibrium) signal current through the
fluctuator, an essential aspect of the indirect detection that we do consider.
Moreover, they only consider qubit energy splittings along the measurement
vector, in our notation $H_Q = \Omega \op{\tau}_z / 2$, that is, they only
study pure dephasing in which the qubit Bloch vector has no precessional motion.
This leads to drastic simplifications employed in the derivation of these
approaches which limit their applicability.

References {\cite{Abel08,Neuenhahn09}} highlight coherence revivals for the
short-time transients in the strong measurement regime $\lambda \gg \Gamma$.
This nonexponential decay reflects the non-Gaussian statistics of the quantum
telegraph process. Moreover, these studies also find oscillatory corrections
to the dephasing even in the weak-measurement regime $\lambda < \Gamma$ in
agreement with our work (see {\color{black} Fig. \ref{fig:elliptical}}). The
dependence of the dephasing rate on the level position is not reported and
thus cannot be compared.

In {\color{black} Ref. {\cite{Paladino02}}}, a path integral method is used to
study the qubit decoherence due to two-level fluctuators. Their dephasing rate
also includes terms containing the real part of the digamma function $\psi$,
which determines the renormalization function $\phi$ [\Eq{eq:phi}]. It
appears both in the cotunneling rates [see in \Eq{eq:dissrate}] as
well as in the coherent backaction [\Eq{eq:kappa}]. In contrast to
{\color{black} Ref. {\cite{Paladino02}}}, our expressions depend on the
derivative $\phi'$, which is arises from our expansions in $\lambda$ and
$\Gamma$. \ However, a direct comparison is not possible because tangible
results of {\color{black} Ref. {\cite{Paladino02}}} are actually given only
for the sequential tunneling regime where renormalization effects are
neglected {\footnote{This is a consequence of the approximation $K_{2, j}
(\lambda) \approx K_{2, j} (0)$ employed below Eq. (3) in {\color{black} Ref.
{\cite{Paladino02}}}.}}.

Reference {\cite{Grishin05}} also uses a path-integral approach and considers
also the case when the QD level $\varepsilon$ is tuned strongly away from
resonance $| \varepsilon - \mu | / T \gg 1$. Expanding their result (Eq. (15)
of {\color{black} Ref. {\cite{Grishin05}}}) for the long-time limit of the
dephasing rate in the weak-measurement limit, we obtain {\footnote{We express
the parameters $g$ and $\tilde{\varepsilon}$ in {\color{black} Ref.
{\cite{Grishin05}}} in terms of our notation as $g = \lambda / \Gamma$ and
$\tilde{\varepsilon} = 2 \varepsilon / \Gamma$.}}:
\begin{eqnarray}
  \frac{1}{T_2} & = & \frac{T}{4 \pi}  \frac{\lambda^2 \Gamma^2}{(\varepsilon
  - \mu)^4} + O \left( (\lambda/\Gamma)^4 \right) . 
  \label{eq:pathint}
\end{eqnarray}
Thus, the dephasing rate drops algebraically in the {\tmem{fourth}} power with
the level position. This is consistent with our results in the sense that also
here the decoherence does not scale as expected from the cotunneling noise. In
our case, we were only able to show that the algebraic scaling in $1 /
(\varepsilon - \mu)$ is suppressed at least to the {\tmem{second}} power (see
{\Sec{sec:gatevoltagedep}}). However, our analysis revealed
the physical origin of this behavior by \new{identifying what} cancels the
expected cotunneling noise contribution, namely the coherent backaction. Note
that the result (\ref{eq:pathint}) does not rely on the high-temperature
assumption $\Gamma / T \ll \lambda / \Gamma$ as in our case, which is why
their result can be expanded in orders $\lambda / \Gamma$ while accounting
also for higher-order tunneling corrections [e.g., $O (\Gamma^3 / T^2)$ and
higher, see {\App{app:validity}}]. This implies that to
compare concretely with their result (\ref{eq:pathint}), we would have to
include higher-order tunneling terms, which, however, we do not expect to
restore power laws with lower exponents. This is challenging for our model
since we account for Coulomb interactions \newer{and nonequilibrium conditions}. This is beyond the scope of the
present paper.

In summary, various aspects of our assumptions and findings seem to be in
accordance with previous approximate as well as exact quantum-mechanical
treatments and shed new light on them. Our study extends these approaches by
simultaneously dealing with a nonstationary, nonequilibrium, Coulomb-blockaded
sensor QD (fluctuator), which is fully quantum-correlated with the qubit
(non-factorizing density operator $\rho$, virtual off-diagonal charge
coherences during tunneling). The latter leads to the coherent backaction as
an integral part of the total backaction together with stochastic and
dissipative backaction, leading to the cancellation of cotunneling noise.



\section{Summary and Outlook}\label{sec:summary}
We studied an indirect detection setup, in which a charge qubit is
capacitively probed ($\lambda$) by a sensor quantum dot (SQD). The SQD is in
turn tunnel coupled ($\Gamma$) to electrodes in which the time-dependent
conductance is measured.
Electrons in the sensor QD occupy a single quantized orbital in which they strongly interact.

\emph{Kinetic equations.} We considered the weak-tunneling, weak-measurement limit $\Gamma / T \ll
\lambda / \Gamma \ll 1$, in which quantum fluctuation effects are important on
the time scale of the qubit-sensor interaction. We derived a kinetic equation
\ [\Eq{eq:kineq}] by integrating out the current-carrying electrodes to
obtain an effective theory for the composite qubit-SQD system. This revealed
three types of backaction on the qubit: (i) a stochastic backaction due
to random fluctuations of the qubit detuning, (ii) a dissipative backaction
 (coefficient $c$), the flip side of the modulation of the sensor
tunnel current by the qubit, and (iii) a coherent backaction due to the level
renormalization of the composite qubit-SQD system (coefficient $\kappa$).

We showed the importance of the effects of single-electron tunneling (SET), as
well as its cotunneling broadening and level-shift corrections. We also included
the leading non-Markovian correction from the electrodes induced by the
tunnel coupling ($\Gamma$) to the SQD (linear kernel frequency dependence).
Moreover, our approach captures {\tmem{all}} non-Markovian effects
introduced by the sensor QD on the qubit subsystem, which are mediated by the capacitive interaction.

\emph{Suppression of cotunneling-induced backaction.} By rewriting the kinetic equation in the basis of quasistationary and decay
modes (defined by the $\lambda = 0$ limit), we found that the interplay of
these types of backaction leads to a nontrivial cancellation; whereas the
dissipative backaction ($c$) independently couples these modes, the stochastic
backaction and coherent backaction ($\kappa$) partially cancel. In particular,
the \emph{change} in the stochastic backaction due to cotunneling
broadening is canceled by the coherent backaction [\Eq{eq:r}].
The expected algebraic decay $\propto 1 / (\varepsilon - \mu_r)$ of the backaction
(determining the decoherence rates) is thus suppressed, implying that the
actual power law must at least have a higher exponent. Experimentally, this is
important since it indicates that a SQD can be switched off by applying
a gate voltage
better than expected.
By identifying the underlying physics, we
suspect this to be a crucial difference with the backaction of sensors with
dense level spectra, such as single-electron transistors. \new{Thus, the
less-than-expected backaction due to the coherent backaction
is beneficial for switching the sensor on / off, provided one takes care to prepare
sensor state to avoid initial slip errors (see next paragraph).}

\emph{Initial slip.} We derived effective equations for the reduced qubit density operator [Eq.
(\ref{eq:xqt})], which are exact relative to the kinetic equations (\ref{eq:kineq}).
In particular, we keep all non-Markovian effects induced by the sensor
and account for a slip of the initial condition [\Eq{eq:xqefft}]. 
\newer{This slip depends on both the initial \emph{sensor QD}
state and the initial quantum \emph{correlations with the sensor QD}.}
It is important for the long-time qubit evolution and thus the
sensor QD needs also to be considered as part of the \emph{dynamical} quantum
circuit.
The dynamical state of the sensor QD is relevant for qubit error propagation,
e.g., the initial slip may introduce errors even for perfectly prepared
qubit initial states. This provides a concrete example for errors
usually phenomenologically introduced in quantum-error correction.
Such a sensor QD cannot be considered (without further evidence) as a
``black box'' in a quantum circuit which is merely characterized by static parameters (e.g., relaxation and dephasing times).
This is different from the treatment of the macroscopic electrodes
coupled to the sensor
QD. \newer{The electrodes, in which the current measurement is performed, can
instead be assumed to be stationary, which eliminates initial-slip effects on
qubit-SQD system (provided initial quantum correlations are neglected).}\\
\emph{High-temperature limit.} Specializing to the high-temperature limit $\Gamma / T \rightarrow 0$ and to
times larger than the SQD relaxation time $1 / \Gamma$, we obtained the
qubit evolution [\Eq{eq:leff}], which neglects non-Markovian effects
induced by the sensor QD. We connected dephasing and relaxation
times $\propto \lambda^2 / \Gamma$ with the component of the measurement
vector ${\vecg{\lambda}}$ along the mean-field qubit
axis $\tilde{\vecg{\Omega}}$  and perpendicular to it, respectively. We demonstrated the importance of the initial
slip $\propto \lambda / \Gamma$ [\Eq{eq:taueff0}] even in this simplest limit:
the set of initial states of the qubit-SQD system {\tmem{without}} slip is
only a subset of zero measure. Generally, the magnitude of the slip increases
the ``less factorizable'' the qubit-SQD state is and the ``more
nonstationary'' the sensor QD is \newer{before the detection is started.} Due to the latter, the initial slip should be reckoned
with in particular when the sensor QD is initialized in a fixed charge
state.
This happens, for example, when switching the sensor off by tuning the
gate voltage far away from resonance or
by reducing the tunnel coupling $\Gamma$ to the electrodes. By
contrast, switching off the capacitive interaction $\lambda$ leads to an
initially stationary and factorizable state, which is favorable for avoiding
an initial slip.
This difference in the backaction between various parameters for switching off a sensor QD is an important experimental implication of our study.
\newer{The need to control not only the qubit but also the readout
device carefully may not only be a nuisance
for engineering quantum circuits but could also provide additional means of
controlling qubits. For
example, one could consider to switch the SQD during the readout to
another readout point to compensate for manipulation errors detected in a
weak measurement process.}\\
\emph{Mode distortion.} The analysis of the isospin dynamics showed that additionally the qubit eigenmode
vectors are distorted due to the small but finite value of the decoherence
rates $\sim \lambda^2 / \Gamma$.  
\newer{This corrects the simple
mean-field picture, in which the qubit axis $\tilde{\vecg{\Omega}} =
\vecg{\Omega} + p^1_{\text{st}} \vecg{\lambda}$ is only influenced by
the average occupation of the SQD.}
\newer{The distortion reflects the breakdown of the secular
approximation -- often made in derivations -- because the
capacitive coupling $\lambda$ can be of the same order as the internal
qubit splitting $\Omega$. Importantly, the distortion must be
included to satisfy an isospin sum rule that follows from the
conservation of the isospin by the tunneling of sensor electrons
\cite{Hell14a,Salmilehto12}. The distortion is thus enforced by a general
principle. The experimentally relevant} consequence of this distortion for the qubit evolution is two-fold:
first, the usual circular Bloch-vector precession becomes slightly {\tmem{elliptical}}, with a shape that is not altered in time
as the size shrinks with dephasing rate $1 / T_2$.
Second, this ``precession plane'' is not orthogonal any more to the
``relaxation axis,'' along which the Bloch vector decays with the relaxation
rate $1 / T_1$. The relaxation axis is moreover slightly tilted relative to
the mean-field axis $\tilde{\vecg{\Omega}}$. Finally, the projections of the
isospin on the mean-field axis, the relaxation axis, and the precession plane
all show a superposition of relaxation and precessional dephasing motion. \newer{The
measurement backaction thus mixes the effects of relaxation and
dephasing, even in this simple Markovian limit. This effect may generally
appear in indirect-coupling setups, which are typical for detection
setups.\\}
The tilting of the qubit modes is related in a broader context to the concept of ``gauge qubits''
\cite{KribsBook}: two physically distinct qubits can be each subject
to strong decoherence but the joint Hilbert space formed by both qubits
may contain
a two-level subsystem with low decoherence. Locating such subsystems is interesting for developing strategies for quantum-error correction. In
our case, the qubit mode tilting reflects an ``admixing'' of the
sensor QD degrees of freedom to the low-decoherence subspace, which mostly
overlap with the qubit degrees of freedom.
This mixing 
can be strongly enhanced when
the coupling of the quasistationary and decay modes
becomes stronger, which effectively happens, e.g., due to enhanced
quantum-fluctuation effects at lower
temperatures. Investigating this mixing further would therefore be
interesting also in the context of quantum-error correction strategies.

\emph{Comparison with other approaches.} \newer{We compared the above results at various points with existing
approaches and explained why potential differences are expected within their
validity:}

(a) Semiclassical stochastic approaches cannot capture the
coherent backaction because the starting assumption of classical charge
fluctuations on the sensor QD already excludes relevant quantum coherences
of the qubit-sensor density operator. Including
these into the description leads to additional ``phase kicks'' that counteract the
stochastic ``phase kicks.'' Both together, when averaged, lead to a mitigated decoherence. It thus
seems that one should quite generally reckon with coherent backaction, which
can mitigate the measurement backaction when employing a \tmtextit{quantum}-dot
sensor for the indirect detection of a qubit.

(b) Density operator approaches that try to integrate out the sensor together with the attached electrodes
run into problems  as well because one needs to ``guess'' the
time-dependent, current-carrying sensor state as well as the quantum
correlations with the qubit. These can lead to non-Markovian behavior and
affect the initial slip of the qubit. All of this is systematically calculated
in our approach. The more advanced approach of {\color{black} Ref.
{\cite{Emary08}}}, which first calculates the nonequilibrium SQD state in the
absence of the qubit (an approximation carefully pointed out in Ref.
{\cite{Emary08}}), misses both the dissipative and coherent backaction.
Extending such an approach, e.g., to include cotunneling broadening would thus
lead to inconsistencies since only one part of two canceling effects is taken
into account.\\
(c) Nonperturbative quantum solutions of related models agree with the
cancellation in the backaction we found here as does a separate
calculation for a noninteracting limit of our model ($U=0$).\\
\braggio{
(d) It is also interesting to compare with prior studies not aiming at sensor backaction.
In particular, \Cite{Braggio06} highlighted the importance of a competition of next-to-leading-order effects and non-Markovian corrections,
which are closely related to the non-Markovian corrections of the first type affecting the dissipative switching rates \eq{eq:alpha} discussed in
\Sec{sec:nmdiscuss}.
However, what is imporant here are non-Markovian corrections of the second
type affecting the coherent backaction.  We stress that although these
non-Markovian corrections give an important part of the coherent
backaction, they are not identical to it: the coherent backaction already arises in the stationary limit that we studied in \Cite{Hell14a} where non-Markovian effects can be neglected.
In this way, we see that the general line of thought emphasized in
\Cite{Braggio06} extends to the coherent backaction, which is also a
second-order
effect, namely first order both in the tunneling ($\Gamma$) and the measurement
interaction ($\lambda$).
}\\
\emph{Outlook.} All this shows that the coherent backaction and other quantum
fluctuation effects (cotunneling, level renormalization) are intrinsic
effects of a \tmtextit{quantum} sensor: they can neither be ``added'' or
``controlled'' independently in an experiment, nor should they be neglected in a
calculation. In view of the above, further studies that can address the
experimentally relevant lower temperature dynamics of weak measurements using
quantum-dot sensors are necessary.
We expect the qualitative effects that we identified to be present and quantitatively stronger under
experimental conditions where, e.g., the capacitive interaction and tunnel
coupling may not be weak anymore (e.g., because of the trade-off between a
significant current signal and low backaction).
Besides the commonly discussed relaxation and dephasing backaction, also the
qubit initial slip, nonorthogonal mode distortion, and sensor-induced memory effects will be enhanced,
neither of which has received much attention so far.

\begin{acknowledgments}
  We acknowledge stimulating discussions with H. Bluhm, L. Schreiber, J.
  K{\"o}nig, V. Meden, C. Kl{\"o}ckner, D. Kennes, and
 P. Samuelsson. \braggio{We thank A. Braggio for helpful feedback on the
 preprint version.} We are
  grateful for support from the Alexander von Humboldt foundation.
\end{acknowledgments}


\appendix

\numberwithin{equation}{section}



\section{Importance of non-Markovian corrections induced by the electrodes}\label{sec:impnm}

In this Appendix, we explain how to account for non-Markovian corrections
arising from the memory induced by the electrodes on the \emph{qubit-sensor subsystem} and
illustrate their importance for a correct description of the detector
backaction. One should clearly distinguish from this further non-Markovian behavior
 induced by the SQD -- with the effect of the electrodes incorporated -- on
 the qubit
subsystem. This is discussed separately in \App{app:nm}.
In {\color{black} Appendix \ref{app:nonmarkov}}, we show how the
non-Markovian corrections due to the electrodes can be incorporated within the real-time diagrammatic formalism
based on a perturbative weak-tunneling ($\Gamma$) expansion. Within the
leading non-Markovian correction included in this paper, the kinetic equation remains
a time-local and first-order differential equation for the density operator.
Based on this, we then perform the weak-measurement expansion.
In {\color{black} Appendix \ref{app:positivity}}, we show that neglecting the
non-Markovian correction leads to a violation of the positivity of the SQD-qubit density
operator and to an overestimation of the measurement backaction. These
unphysical features are removed when the leading non-Markovian correction is
included.

\subsection{Incorporating non-Markovian effects in real-time
diagrammatics}\label{app:nonmarkov}

In {\color{black} Ref. {\cite{Hell14a}}}, we started from a kinetic equation
for the reduced density operator $\rho (t)$ of the joint system of SQD plus
qubit, obtained in the standard way by integrating out the electrodes' degrees of
freedom:
\begin{eqnarray}
  \dot{\rho} (t) & = & - i L_{Q S} \rho (t) + \int_0^t d t' W (t - t') \rho
  (t') .  \label{eq:gme2}
\end{eqnarray}
Here, $L_{Q S} \bullet = [H_Q + H_S + H_I, \bullet]$ is the internal
Liouvillian of the reduced system with ``$\bullet$'' denoting the operator the
Liouvillian acts on. The effects of the electrodes are incorporated in the kernel
$W$, which we evaluated using the real-time diagrammatic approach
{\cite{Schoeller09a,Leijnse08a}}. In {\color{black} Ref. {\cite{Hell14a}}}, we
were only interested in the stationary solution of \Eq{eq:gme2}, $\rho
(t) = \rho_{\tmop{st}}$, which obeys $\dot{\rho}_{\tmop{st}} = 0$.
This solution depends only on the time-integrated kernel
$\int_0^t d t' W (t - t') = \int_0^t d \tau W (\tau)$. In the long-time limit
$t \rightarrow \infty$, when the stationary state is approached, the
time-integrated kernel is given by the zero-frequency limit of the Laplace
transform
\begin{eqnarray}
  W (z) & = & \int_0^{\infty} d \tau e^{i z \tau} W (\tau) .  \label{eq:wz}
\end{eqnarray}
Hence, the exact $\rho_{\tmop{st}}$ is the same as the stationary solution of
the approximate, time-local Markovian kinetic equation,
\begin{eqnarray}
  \frac{d \rho}{d t} (t) & = & (- i L_{Q S} + W) \rho (t), 
  \label{eq:kineqmarkov}
\end{eqnarray}
with $W = W (z = i 0)$.

\emph{Weak-tunneling expansion.} Yet, in the present paper, we study the
nonstationary time evolution of the qubit-sensor density operator $\rho
(t)$ and we show now that this implies that
memory effects induced by the electrodes through the kernel $W$ have to be included. We will refer to all the
effects of the \tmtextit{frequency dependence} of $W (z)$, \Eq{eq:wz},
as \tmtextit{non-Markovian corrections induced by the electrodes} [even when effectively a time-local equation results, see \Eq{eq:timelocnm}]. An approach to include such
corrections within the real-time diagrammatic formalism has been given in
{\color{black} Ref. {\cite{Splettstoesser10}}}, and applied to study decay
rates in Refs. \cite{Contreras12,Schulenborg14a}. The basic idea is to perform a Taylor expansion of
$\rho (t')$ in the integral in \Eq{eq:gme2} around time $t$,
\begin{eqnarray}
  \rho (t') & = & \sum_{n = 0}^{\infty} \frac{1}{n!} \frac{d^n \rho (t)}{d
  t^n} (t' - t)^n,  \label{eq:rhoexpand}
\end{eqnarray}
and to subsequently perform the integration over $t'$. This results in a
well-defined expression if the kernel decays faster than any polynomial in $t
- t'$, which is usually fulfilled because the kernel decays exponentially $W
(t - t') \sim e^{- (t - t') / \tau_C}$ on the time scale of the inverse
temperature $\tau_C \sim 1 / T$ {\footnote{The correlation time is set by the
time dependence of the contraction functions in the diagrammatic expansion
(reservoir correlation functions), which drop on the time scale of the inverse
temperature $1 / T$, see Eqs. (93) and (100) in {\color{black} Ref.
{\cite{Saptsov14}}}. Note that the contribution from the stationary state
$\rho_{\tmop{st}}$ just vanishes, i.e., $W (t - t') \rho_{\tmop{st}} = 0$ for
any time $t'$.}}.

If we consider only times $t \gg \tau_c$, one
may compute $\int_0^t d t' W (t - t') (t' - t)^n = \int_0^t d \tau W (\tau) (-
\tau)^n$ by replacing $t \rightarrow \infty$ on the right-hand side since all
contributions from $\tau > t$ are negligibly small. Taking advantage of the
Laplace transform (\ref{eq:wz}), we obtain from \Eq{eq:gme2}
\begin{eqnarray}
  \frac{d \rho (t)}{d t} & = & - i L_{Q S} \rho (t) + \sum_n \frac{1}{n!}
  \partial^n W \frac{d^n \rho (t)}{d t^n},  \label{eq:kinexp}
\end{eqnarray}
with the $n$th derivative of the Laplace-transformed kernel (\ref{eq:wz}) at
zero frequency with respect to $- i z$:
\begin{eqnarray}
  \partial^n W & = & \left. \frac{\partial^n W (z)}{\partial ( -
  i z)^n} \right|_{z = i 0} .  \label{eq:dwz}
\end{eqnarray}
To include the leading-order non-Markovian corrections, we
truncate the sum on the right-hand side of \Eq{eq:kinexp} and keep only
the terms with $n \leqslant 1$. This yields the following time-local
differential equation of first order in time:
\begin{eqnarray}
  \frac{d \rho (t)}{d t} \ = \ \frac{1}{\mathbbm{1} - \partial W} (- i L_{Q S}
  + W) \rho (t) \ = \ - i L \rho (t),& &  \label{eq:timelocnm}
\end{eqnarray}
whose solution can be written as
\begin{eqnarray}
  \rho (t) & = & e^{- i L t} \rho (0)  \label{eq:nmsolve}
\end{eqnarray}
for initial state $\rho (0)$. When dropping the non-Markovian correction $1 /
(\mathbbm{1} - \partial W)$ in \Eq{eq:timelocnm}, one recovers the
Markovian generator of \Eq{eq:kineqmarkov}.

We next show that this solution with $L$ defined through Eq.
(\ref{eq:timelocnm}) accounts for all non-Markovian effects up to $O (\Gamma^2
/ T)$ provided one is seeking for exponential solutions of Eq.
(\ref{eq:kinexp}). We do not discuss algebraic or logarithmic time
dependencies here, which may also appear
{\cite{Pletyukhov10}}.
Thus, let us
substitute the exponential ansatz
\begin{eqnarray}
  \rho (t) & = & e^{A t} \rho (0),  \label{eq:expansatz}
\end{eqnarray}
into \Eq{eq:kinexp}. We obtain the following equation for $A$:
\begin{eqnarray}
  A & = & - i L_{Q S} + W + (\partial^1 W) A + \ldots \text{ } . 
  \label{eq:expanda}
\end{eqnarray}
When all derivatives $\partial^n W$ are dropped from the above equation, we
recover the Markovian generator $A = - i L_{Q S} + W$. Non-Markovian
corrections therefore enter through terms $\sim \partial^n W$ that capture the
frequency dependence of the kernel {\cite{Braggio06,Flindt08PRL,Flindt08PRB}}.

Equation (\ref{eq:expansatz}) shows that the derivatives of the density
operator scale as $d^n \rho / d t^n \sim A^n \sim \Gamma^n$ since $L_{Q S}
\sim \Delta \ll \Gamma$ and $W \sim \Gamma$ here. As a consequence, the
expansion (\ref{eq:rhoexpand}) of the density operator in
corrections from higher-order time derivatives is not independent of the perturbative expansion of the kernel $W$
in powers of $\Gamma$. When expanding $A$ in powers of $\Gamma$, we find that
the $n$th order derivative of the kernel scales as $\partial^n W A^n \sim
(\Gamma \text{/} T^n) \cdot \Gamma^n$ plus higher-order
corrections
{\footnote{This can be understood as follows: the finite frequency
kernel is given by Eq. (A6) in {\color{black} Ref. {\cite{Hell14a}}} by
replacing \ $i 0 \rightarrow z \text{/} T$. Each frequency derivative in
$\partial^n W$ acts on a propagator, i.e., it involves expressions such as
$\frac{\partial}{\partial z}
\frac{1}{z \text{/} T + X_i - (L_{Q S} - \mu_i)
\text{/} T}
\sim
\frac{1}{T}
\frac{\partial}{\partial (L_{Q S} - \mu_i)
\text{/} T}  \frac{1}{z / T + X_i - (L_{Q S} - \mu_i) \text{/} T}$, which can
be rewritten as a derivative of dimensionless energy ratios and a factor
$1 \text{/} T$ for each $z$-derivative. This yields schematically $\frac{1}{T}
\partial^n W \sim \frac{1}{T^n}  \left( \prod_i \frac{\Gamma_i}{T} \right)
\frac{\partial^n}{\partial [(L_{Q S} - \mu) \text{/} T]^n} I \left( \frac{L_{Q
S} - \mu}{T} \right)$, where $I$ is a dimensionless function that contains the
frequency integrals. Thus, $\partial^n W \sim \Gamma \text{/} T^n$ since $W
\sim \Gamma$ (plus higher-order corrections).}}.
Thus, when expanding the
kernel $W = W^{\Gamma} + \ldots$ only up to first order in $\Gamma$, Markovian
corrections must be ignored, i.e., $A^{\Gamma} = - i L + W^{\Gamma}$. Yet, when
expanding up to order $\Gamma^2 \text{/} T$ as considered here, one is allowed
to omit terms $(\partial^n W) A^n \sim (\Gamma \text{/} T^n) \cdot \Gamma^n$
only for $n \geqslant 2$ from \Eq{eq:expanda}, while one has to keep the
first derivative $\partial^1 W$. Solving for $A$, we obtain $A = - i L = (1 -
\partial^1 W)^{- 1} (- i L + W) + O (\Gamma^3 \text{/} T^2)$ in agreement with Eq.
(\ref{eq:timelocnm}). Further expanding the inverse and the kernel, we find
for the effective generator:
\begin{eqnarray}
  - i L & = & (1 + \partial^1 W^{\Gamma}) (- i L_{Q S} + W^{\Gamma}) +
  W^{\Gamma^2 / T} \nonumber\\
  &  & + (\Gamma^3 / T^2) .  \label{eq:aapprox}
\end{eqnarray}
\emph{Weak-measurement expansion.} The next step is the expansion of $- i
L$ in $\Delta / T$, where $\Delta \sim \lambda, \Omega$ denotes the small
energy scale of the detection and intrinsic qubit frequency. While we have to
keep the first-order terms in $\Delta / T$ for
\begin{eqnarray}
  W^{\Gamma} & = & W^{\Gamma, 0} + W^{\Gamma, \Delta} + O (\Gamma \Delta^2 /
  T^2),  \label{eq:w1}
\end{eqnarray}
we can neglect the $\Delta$ dependence of
\begin{eqnarray}
  \partial^1 W^{\Gamma} & = & \partial W^{\Gamma, 0} + O (\Gamma \Delta / T), 
  \label{eq:delw1}\\
  W^{\Gamma^2 / T} & = & W^{\Gamma^2 / T, 0} + O (\Gamma^2 \Delta / T^2), 
  \label{eq:w2}
\end{eqnarray}
and obtain for the effective, non-Markovian generator
\begin{eqnarray}
  - i L & = & (1 + \partial^1 W^{\Gamma, 0}) (- i L_{Q S} + W^{\Gamma,
   0}) \nonumber \\
  & & +
  W^{\Gamma, \Delta} + W^{\Gamma^2 / T, 0} \nonumber\\
  &  & + O (\Gamma^3 \text{/} T^2, \Gamma^2 \Delta \text{/} T^2, \Gamma
  \Delta^2 \text{/} T^2) .  \label{eq:aexp}
\end{eqnarray}
We now note that in {\color{black} Ref. {\cite{Hell13a}}}, we \ already
computed the kernels $W^{\Gamma, 0}$, $W^{\Gamma, \Delta}$, and $W^{\Gamma^2,
0}$ and therefore we have to merely evaluate $\partial W^{\Gamma, 0}$ to
obtain non-Markovian corrections up to the order considered here. The
frequency derivative of $\partial W^{\Gamma, 0}$ [see \Eq{eq:dwz}] can
be easily converted into energy derivatives ($\partial/\partial
\epsilon$)
{\footnote{The first-order kernel
reads schematically $W^{\Gamma} (z) \sim \int d x f^{\pm} (x) \text{/} [z + x
- L]$, i.e., $\partial^1 W^{\Gamma} (z) \sim \partial \text{/} \partial L \int
d x f^{\pm} (x) \text{/} [z + x - L]$.}},
which results in the kinetic equation
(\ref{eq:kineq}) in the main text. It should be noted
that $\partial W^{\Gamma, 0}$ is \emph{not} simply the derivative of
the SET
contribution: the imaginary factor in \Eq{eq:dwz} changes the role of
imaginary and real parts, which are related to $\delta$ functions and
principal-value parts in the frequency integrals, respectively. While in the SET
contributions only the $\delta$ functions remain and the principal-value parts
cancel out, this is opposite for the non-Markovian correction $\partial
W^{\Gamma, 0}$. These principal-value parts evaluate to the renormalization
function $\phi (x)$ [see \Eq{eq:phi}], which is of central importance
in our work and explains how the non-Markovian corrections can affect the
coherent backaction as noted in \Sec{sec:nmdiscuss}. To evaluate \ $\partial W^{\Gamma, 0}$, one thus must
first compute $W^{\Gamma, 0}$, then apply $\partial$ and only after that take
the relevant matrix elements (restricted by charge conservation). This
completes the derivation of our non-Markovian Liouvillian, accounting
consistently for all terms up to $O (\Delta, \Gamma, \Gamma^2 / T, \Gamma
\Delta / T)$.

To find the solution \eq{eq:expansatz} of the kinetic equation $\dot{\rho} (t) =
A \rho (t)$, we directly solve for $\rho (t)$ without expanding $\rho (t)$ in
$\Gamma / T$. Not solving the kinetic equation order-by-order
{\footnote{To
obtain a solution $\rho (t)$ that is also consistently expanded to all orders of
$\Delta$ and $\Gamma$, one should insert the expanded density operator $\rho
(t) = \rho^{0, 0} (t) + \rho^{\Gamma, 0} (t) + \rho^{0, \Delta} + \ldots$
together with the expanded kernels into the kinetic equation $\dot{\rho} (t) =
-i L \rho (t)$, and compare left- and right0hand side in all orders.}}
avoids well-known problems with ill-defined coherences {\cite{Weymann05}} and
nonequilibrium occupations close to the Coulomb blockade regime {\cite{Leijnse08a}}. As a consequence, our solution can
comprise of terms of higher order in $\Delta / T$ and $\Gamma / T$. In
particular, the stationary solution obtained from $A \rho_{\tmop{st}} = 0$
differs formally from that obtained by solving the corresponding Markovian equation
$A_{\partial W = 0} \rho_{\tmop{st}} = 0$. However, the deviations are of $O
(\Gamma^3 \text{/} T^2, \Gamma^2 \Delta \text{/} T^2, \Gamma \Delta^2 \text{/}
T^2)$ and therefore consistently negligible in the perturbative limit considered here.

\subsection{Retaining positivity and effect on coupling of
quasistationary and decay modes}\label{app:positivity}

\begin{center}
  \Figure{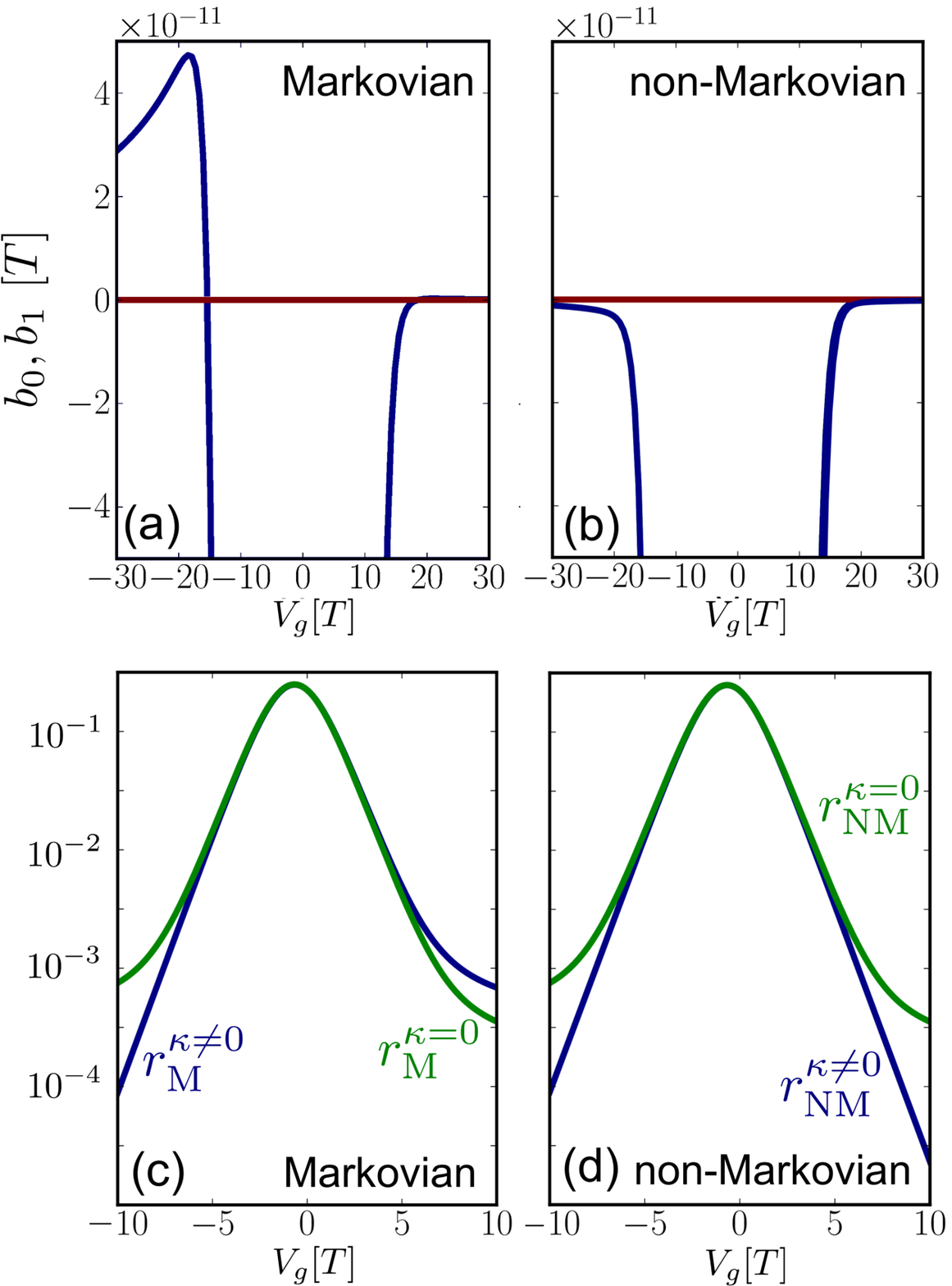}{(a)
  and (b): Two largest real parts $b_0, b_1$ of the eigenvalues of the
  SQD-qubit generator $- i L$, as a function of gate voltage $V_g$. We use the
  Markovian
  approximation (\ref{eq:markov}) of the generator $-i L_{\partial W =
 0}$ in (a) and the full non-Markovian generator (\ref{eq:aexp}) in (b).
  (c) and (d): Component of $\vec{r}$ [\Eq{eq:lqdm}]
  along $\vecg{\lambda}$. We show the projection 
  $r_{\text{M}} = \vec{r}_{\text{M}} \cdot \op{\vecg{\lambda}} =
  p^0_{\tmop{st}} p^1_{\tmop{st}} + \kappa \tfrac{1}{2} (1 + p^0_{\tmop{st}})$
  [see \Eq{eq:rm}] within the Markovian approximation in (c) and
$r_{\tmop{NM}} =
  \vec{r}_{\tmop{NM}} \cdot \op{\vecg{\lambda}} = p^0_{\tmop{st}}
  p^1_{\tmop{st}} - \kappa (p^1_{\tmop{st}} / 2 - p^0_{\tmop{st}})$ [see Eq.
  (\ref{eq:rnm})] including non-Markovian corrections in (d).
 In (c)  and (d), the blue curves include the coherent backaction ($\kappa \neq 0$),
  while the green curves exclude them by hand ($\kappa = 0$). The parameters
  in all plots are $V_b = 0$, $\Gamma_s = \Gamma_d = \bar{\Gamma} = 10^{- 2}
  T$, $\lambda = \Omega = 0.1 \bar{\Gamma} = 10^{- 3} T$, and $W = 1000
  T$.\label{fig:nm}}
\end{center}

In this appendix, we elucidate the importance of non-Markovian
corrections induced by the electrodes to describe the indirect detection setup. We explicitly
illustrate that without these corrections the kinetic equations
(\ref{eq:kineq}) for the SQD-qubit system possess exponentially increasing
solutions in time and the measurement backaction is qualitatively
different.

In {\color{black} Ref. {\cite{Hell14a}}}, we reported that the Markovian
kinetic equation when used to calculate a time-dependent solution for the
density operator $\rho (t)$ for the joint SQD-qubit system violate the
positivity condition, even though the stationary state showed no such
problems. (It should be noted that for the stationary state to be positive, in
addition to the dissipative and coherent backaction also $O (\Gamma^2 / T)$
effects were required, already hinting at the cancellation effect that we
discuss in the present paper.) The positivity problems for
time-dependent solutions arising in the Markovian approximation with
respect to the electrodes
 can be readily inferred  by diagonalizing the generator (\ref{eq:aexp}) in that \newer{approximation}:
\begin{eqnarray}
  - i L_{\partial W = 0} \bullet & = & \sum_i (b_i + i a_i) A_i \underset{S,
  Q}{\tmop{tr}} [\tilde{A}_i \bullet] .  \label{eq:markov}
\end{eqnarray}
Positivity violation occurs when at least one eigenvalue exists with a positive
real part, i.e., $b_i > 0$. This leads to exponentially growing contributions
$\rho (t) \sim \rho_{\tmop{st}} + e^{b_i t} \rho_i$ {\cite{Hell14a}}. Figure
{\color{black} \ref{fig:nm}}(a) shows the two largest real parts of the
eigenvalues of the Markovian generator (\ref{eq:markov}) for typical
parameters considered in this paper. It illustrates that a positivity
violation may appear when the SQD is tuned deep into the Coulomb blockade regime
for $V_g/T \gg 1$ [see Fig. 7(b) in {\color{black} Ref. {\cite{Hell14a}}}]. By
contrast, if we include the leading non-Markovian correction and use our full equation (\ref{eq:aexp})
instead, all eigenvalues have a nonpositive real part for all gate
voltages as shown in {\color{black} Fig. \ref{fig:nm}}(b). Thus, no positivity
violation can occur here. We investigated this thoroughly by numerically
exploring a large parameter regime. We note, as already pointed out in
{\color{black} Ref. {\cite{Hell14a}}}, that the inclusion of next-to-leading
order $\Gamma^2 / T$ corrections in the tunneling is also crucial to avoid
such positivity problems.

To assess the importance of non-Markovian corrections for the qubit
backaction, we next consider the effective Liouvillian (\ref{eq:leffzmain})
when starting from the Markovian approximation of the generator
(\ref{eq:markov}). As we discussed in {\color{black} Sec.
\ref{sec:mitigation1}}, the importance of the coherent backaction can be
assessed from the transition matrix from the quasistationary into the decay
modes. In both cases, the transition matrix takes the form
\begin{eqnarray}
  \Lambda^{d q} & = & \left(\begin{array}{cc}
    0 & (1 + p^0_{\tmop{st}}) c \vecg{\lambda} \cdot \vec{e}_{\alpha'}\\
    (1 + p^0_{\tmop{st}})  \vec{e}_{\alpha}^{\dag} \cdot \left( c
    \vecg{\lambda} \right) & - \vec{e}_{\alpha}^{\dag} \cdot \left( \vec{r}_{}
    \times \vec{e}_{\alpha'} \right)
  \end{array}\right) \nonumber\\
  &  &  \label{eq:lqdm}
\end{eqnarray}
where in the Markovian approximation the vector $\vec{r}$ reads
\begin{eqnarray}
  \vec{r}_{\text{M}} \ = \ \kappa (1 + p^0_{\tmop{st}}) \vecg{\Omega} + \left[
  p^0_{\tmop{st}} p^1_{\tmop{st}} + \kappa \tfrac{1}{2} (1 + p^0_{\tmop{st}})
  \right] \vecg{\lambda} & & \label{eq:rm}
\end{eqnarray}
and when including the leading non-Markovian correction
[see \Eq{eq:transdq}], the vector $\vec{r}$ reads
\begin{eqnarray}
  \vec{r}_{\text{NM}} & = & \left[ p^0_{\tmop{st}} p^1_{\tmop{st}} - \kappa
  \left( \tfrac{1}{2} p^1_{\tmop{st}} - p^0_{\tmop{st}} \right) \right]
  \vecg{\lambda},  \label{eq:rnm}
\end{eqnarray}
which is the central result \eq{eq:r} discussed in detail in the main text.
The non-Markovian corrections lead to two important differences between Eqs.
(\ref{eq:rm}) and (\ref{eq:rnm}).

First, the coherent backaction leads to a
different gate-voltage dependence of the transition matrix when non-Markovian
corrections are neglected. To illustrate this, we plot in {\color{black} Figs.
\ref{fig:nm}}(c) and \ref{fig:nm}(d) the component of $\vec{r}_{\text{M}}$ and
$\vec{r}_{\text{NM}}$ along the measurement vector $\vecg{\lambda}$. Due to the
cancellation effect explained in the main text, the coherent backaction
{\tmem{suppresses}} the transition factor in the non-Markovian case, leading
to an exponential gate-voltage dependence. This is different for the Markovian
case; here, the term arising from the coherent backaction, $\kappa (1 +
p^0_{\tmop{st}})$, changes sign at resonance and {\tmem{enhances}} the
transition rate into the decay modes for positive gate voltages [see
{\color{black} Fig. \ref{fig:nm}}(d)]. Moreover, the transition factor scales
there {\tmem{algebraically}} with $V_g$ in stark contrast to the exponential
dependence in the non-Markovian case.

The second difference between Eqs. (\ref{eq:rm}) and (\ref{eq:rnm}) is that the vector
$\vec{r}_{\text{M}}$ in the Markovian case also has a component along
$\vecg{\Omega}$. This component emerges because in the Markovian approximation the
coherent backaction appears as torque terms in the kinetic equations that are
proportional to $\kappa \left( \vecg{\Omega} + \vecg{\lambda} / 2 \right)$ [see Eqs. (25) and (26) in
{\color{black} Ref. {\cite{Hell14a}}}]. By contrast, in the non-Markovian case
the torque terms involve only the vector $\vecg{\lambda}$ \ [see Eq.
(\ref{eq:kineq})].

Both these differences clearly show that non-Markovian corrections
induced by the electrodes have
crucial consequences for the total backaction due to the capacitive
interaction $\lambda$ and must be accounted for consistently.




\section{Liouvillian perturbation theory }\label{app:liouville}

In this appendix, we derive the effective Liouvillian (\ref{eq:leffzmain})
that we investigate in the main text to study the measurement
backaction on the qubit evolution. We first keep our formulation as general as
possible to bring out the generic features and to indicate that this procedure
can be applied also to more complicated indirect detection models.
An important prerequisite is that one can identify a subspace of interest in
Liouville space whose dynamics takes place on a time scale that is well
separated from the dynamics in the complementary subspace. In our situation,
this is related to the slow dynamics $\sim 1 / \Delta$ of the subspace of the
quasistationary modes as compared to the fast dynamics $\sim 1 / \gamma$ in
the subspace of the decay modes. Using the well-established projection
technique {\cite{Nakajima58,Zwanzig60}} along the lines of Chapter 17 in
{\color{black} Ref. {\cite{Fickbook}}}, we outline how to obtain an effective
Liouvillian that mediates the dynamics in the subspace of interest only.
In contrast to most cases found in the literature, the unperturbed problem onto which we project already exhibits non-Hamiltonian dynamics, i.e., $L \bullet \neq [H,\bullet]$,
see discussion after \Eq{eq:some-r-equation}.
Therefore, we review these steps here to highlight that the initial slip is a general
phenomenon that appears when projecting out a complementary subspace. This in
fact prevents a full elimination of these degrees of freedom unless special
conditions apply. In the high-temperature limit, the projection can be
analytically performed and we can obtain the effective Liouvillian and initial
slip perturbatively in the coupling strength between the relevant subspace and
its complement.

We give here two complementary approaches. The first one is given in frequency
space in {\color{black} Appendix \ref{app:derivationfrequencyspace}}, which allows
for a very compact procedure. However, to \newer{understand} the physical meaning of the
involved approximations, we also show how to obtain our results in a
time-space formulation in {\color{black} Appendix \ref{app:derivationtimespace}}.
In {\color{black} Appendix \ref{app:leffcalc}}, we then apply our general
Liouville-space projection technique to the detection setup considered in the
main text. We finally derive the effective transient evolution of
the qubit, Eqs. (\ref{eq:taueq2})--(\ref{eq:taueff0}), by removing the SQD
degrees of freedom as far as possible.

\subsection{Derivation in frequency space}\label{app:derivationfrequencyspace}
We start with the derivation in frequency space by applying a general
projection technique. This results in the most general expression for
the time evolution, for which we give a perturbative expansion subsequently.

\subsubsection{Liouville-space projection technique}

Consider a system with density operator $\rho$ whose time evolution is
generated by a Liouvillian $L$, i.e.,
\begin{eqnarray}
  \dot{\rho} (t) & = & - i L \rho (t),  \label{eq:rhofull}
\end{eqnarray}
with initial condition $\rho (0)$ given at time $t = 0$. We refer to $L$ as
the Liouvillian although it might not have a simple commutator structure and
can have dissipative, non-Hermitian parts. To solve the above equation, one
can transform it to Laplace space. The Laplace transform of a function $f (t)$
is defined as $f (z) = \int_0^{\infty} d t e^{i z t} f (t)$. This yields using
\Eq{eq:rhofull}:
\begin{eqnarray}
  - i z \rho (z) - \rho (0) & = & - i L \rho (z) . 
  \label{eq:rhoz-eq}
\end{eqnarray}
Next, we are only interested in the evolution of $\rho$ in a subspace
$a$ of the entire Liouville space, defined by a projection superoperator
$P^a$, which satisfies $(P^a)^2 = P^a$. In contrast to {\color{black} Sec.
\ref{sec:projection}}, our formulation here is completely basis independent to
emphasize the generality.

Projecting the kinetic equation \eq{eq:rhoz-eq} onto the subspace $a$ and its
complement $b$ with projector $P^b = \mathbbm{1} - P^a$, we find
\begin{eqnarray}
  - i z \rho^a (z) - \rho^a (0) & = & - i L^{a a} \rho^a (z) - i L^{a b}
  \rho^b (z), \\
  - i z \rho^b (z) - \rho^b (0) & = & - i L^{b a} \rho^a (z) - i L^{b b}
  \rho^b (z) . 
\end{eqnarray}
Solving the second equation for $\rho^b (z)$,
\begin{eqnarray}
  \rho^b (z) & = & \frac{i}{z - L^{b b}} (\rho^b (0) - i L^{b a} \rho^a (z)), 
  \label{eq:rhob}
\end{eqnarray}
and inserting this into the first equation, we can formally write the
exact solution as
\begin{eqnarray}
  \rho^a (z) & = & \frac{i}{z - L^{a a}_{\tmop{eff}} (z)} \rho^a_{\tmop{eff}}
  (z) .  \label{eq:rhoaz}
\end{eqnarray}
This incorporates a frequency-dependent {\tmem{effective Liouvillian}},
\begin{eqnarray}
  L^{a a}_{\tmop{eff}} (z) & = & L^{a a} + L^{a b} \frac{1}{z - L^{b b}} L^{b
  a},  \label{eq:leffz}
\end{eqnarray}
and the frequency-dependent initial condition
\begin{eqnarray}
  \rho^a_{\tmop{eff}} (z) & = & \rho^a (0) + L^{a b} \frac{1}{z - L^{b b}}
  \rho^b (0) .  \label{eq:rhoeffz}
\end{eqnarray}
To transform \Eq{eq:rhoaz} back to time space we apply the inverse
Laplace transform:
\begin{eqnarray}
  \rho^a (t) & = & \int_{- \infty}^{+ \infty} \frac{d z}{2 \pi} e^{- i z t}
  \rho^a (z) .  \label{eq:backtransform}
\end{eqnarray}
Identifying the subspace $a$ ($b$) with the
subspace of the quasistationary (decay) modes labelled by $q$ ($d$), Eqs. (\ref{eq:rhoaz}) -- (\ref{eq:rhoeffz}) yield Eqs. (\ref{eq:xqz})
-- (\ref{eq:zinitial}) of the main part. To compute the solution in time
space, the integral (\ref{eq:backtransform}) has to be solved by applying the residue theorem. Since $L^{a
a}_{\tmop{eff}} (z)$ can be represented by a finite matrix in our case, we
only have isolated poles satisfying $z_p = L^{a a}_{\tmop{eff}}
(z_p)$. If the coupling between the subspaces
$a$ and $b$ is absent, these poles are given by the eigenvalues of $L^{a a}$.

Importantly, \Eq{eq:rhoeffz} shows that the component of the initial
state in the complementary $b$-subspace has been transformed into a
correction to the initial state in the targeted $a$ subspace. This is referred
to as a slippage of the initial condition {\cite{Gaspard99}}. In general, if
the coupling of the $b$-subspace to the $a$-subspace is nonzero ($L^{a b} \neq
0$), this initial slip is present unless the initial state has no projection
on the complementary space, $\rho^b (0) = P^b \rho (0) = 0$. In our problem,
$P^b$ projects onto density operators $\rho$, which are either not
factorizable or for which the reduced state of the SQD is not stationary (see
\App{app:slip}). This has to be compared with the
generalized master equation (\ref{eq:gme}) for the joint qubit-SQD state: As
noted there, when deriving the general kinetic equation one in fact projects
onto a stationary (equilibrium) state of the reservoirs. One then often
assumes (as we do) an initially stationary reservoir state \newer{that
factorizes with the SQD-qubit state. This} eliminates
the initial slip.

\subsubsection{Perturbative solution}

To next find a perturbative solution -- required to obtain the high-temperature equations \eq{eq:leff} and \eq{eq:taueff0} of the main text
-- we assume that the eigenvalues of $L^{a a}$
are well separated from those of $L^{b b}$ in the complex plane as compared to
the coupling mediated by $L^{a b}$ and $L^{b a}$. Thus, if
\begin{eqnarray}
  l \assign \max (| L^{a b} |, | L^{b a} |) \ \ll \ g \assign | | L^{b b} | -
  | L^{a a} | |,& &  \label{eq:condition}
\end{eqnarray}
we can neglect the coupling-induced shift of the poles $z_p$ in the
denominator of \Eq{eq:leffz} to lowest order in \ $l / g$. This is the
basis of our perturbative expansion of $L^{a a}_{\tmop{eff}}$ in orders of $l
/ g$.

For the measurement setup studied in this paper, the situation is even a bit
simpler because the eigenvalues in the quasistationary subspace satisfy
\begin{eqnarray}
  \omega \text{ \ } \assign \text{ \ } | L^{q q} | & \ll & g, 
  \label{eq:slowevolve}
\end{eqnarray}
which allows us to insert $z = 0$ into Eqs. (\ref{eq:leffz}) and
(\ref{eq:rhoeffz}). This means that we carry out a Markovian approximation (i)
for the effective Liouvillian
\begin{eqnarray}
  & \begin{array}{lllll}
    L^{a a}_{\tmop{eff}} & = & L^{a a}_{\tmop{eff}} (z = 0) & = & L^{a a} -
    L^{a b} \frac{1}{L^{b b}} L^{b a}
  \end{array} &  \label{eq:laaeff}
\end{eqnarray}
and (ii) for the effective initial condition:
\begin{eqnarray}
  \rho^a_{\tmop{eff}} (0) & = & \rho^a (0) - L^{a b}
  \frac{1}{L^{b b}} \rho^b (0) .  \label{eq:rhozmarkov}
\end{eqnarray}
One can next readily transform back to time space, which yields the effective
time-evolution equation,
\begin{eqnarray}
  \dot{\rho}^a (t) & = & - i L^{a a}_{\tmop{eff}} \rho^a (t), 
  \label{eq:efftime}
\end{eqnarray}
with the ``slipped initial condition'' (\ref{eq:rhozmarkov}). \newer{Equations
\eq{eq:laaeff} -- \eq{eq:efftime} reproduce Eqs. \eq{eq:taueq2} -- \eq{eq:taueff0} of the main text.}

We emphasize that this Markov approximation is profoundly different
from the Markov approximation discussed in {\color{black} Appendix
\ref{sec:impnm}}.
The latter accounts for memory effects of the
{\tmem{electrodes}} (kernel frequency dependence), which are already
contained in $L$ and therefore included into the
projection approach from the start.
In the present case we neglected the frequency dependence in Eqs. \eq{eq:rhoeffz} and \eq{eq:leffz}
 that is generated by considering a subsystem of the system described by $L$.
We address this point further in Appendix \ref{app:leffcalc} where we discuss this for our concrete
detection problem.

The Markovian approximation \eq{eq:efftime} neglects corrections to the effective Liouvillian of $O ([\max
(l, \omega)]^3 / \gamma^2)$ and therefore the solution of Eq.
(\ref{eq:efftime}), $\rho^a (t) = e^{- i L^{a a}_{\tmop{eff}} t}
\rho^a_{\tmop{eff}} (0)^{}$, is limited to times $t \ll \gamma^2 / [\max (l,
\omega)]^3$. Moreover, there is a restriction for \Eq{eq:efftime} to be
valid for small times: one has to require $t > 1 / \gamma$ basically to ensure
that the projection of the density operator on the decaying subspace, $\rho^b
(t)$, becomes negligibly small and effectively contributes only through the
slipped initial condition \Eq{eq:rhozmarkov}. This becomes clearer from
our complementary time-space derivation below.

\subsection{Derivation in time space}\label{app:derivationtimespace}

To gain further physical insight into the approximations made in the above
derivation, we re-derive the above results now in time space. We start here
from decomposing \Eq{eq:rhofull} into its components in the two
complementary subspaces $a$ and $b$:
\begin{eqnarray}
  \dot{\rho}^a (t) & = & - i L^{a a} \rho^a (t) - i L^{a b} \rho^b (t), \\
  \dot{\rho}^b (t) & = & - i L^{b a} \rho^a (t) - i L^{b b} \rho^b (t) . 
\end{eqnarray}
The second equation is formally solved by
\begin{eqnarray}
  \rho^b (t) & = & e^{- i L^{b b} t} \rho^b (0) \nonumber\\
  &  & - i \int_0^t d t' e^{- i L^{b b} (t - t')} L^{b a} \rho^a (t'), 
\end{eqnarray}
which inserted into the first equation yields the integro-differential
equation
\begin{eqnarray}
  \dot{\rho}^a (t) & = & - i L^{a a} \rho^a (t) \nonumber\\
  &  & - L^{a b}  \int_0^t d t' e^{- i L^{b b} (t - t')} L^{b a} \rho^a (t')
  \nonumber\\
  &  & - i L^{a b} e^{- i L^{b b} t} \rho^b (0) .  \label{eq:nakajimazwanzig}
\end{eqnarray}
This equation incorporates three terms: The first line represents the internal
evolution in subspace $a$, which reproduces the full time evolution of
$\rho^a$ if subspaces $a$ and $b$ are decoupled ($L^{a b} = L^{b a} = 0$).
However, if the coupling is nonzero, the second line accounts for virtual
transitions from the subspace $a$ into subspace $b$ at time $t' < t$, followed
by a period of internal evolution in the subspace $b$ (mediated by Liouvillian
$L^{b b}$) and a final transition back into subspace $a$ again at time
$t$. The third line is related to the initial slip: it accounts for the
contribution to $\rho^a (t)$ that stems from the initial projection of the
density operator $\rho^b (0)$ on subspace $b$, combined with a transition into
subspace $a$ at time $t$.

To next derive the effective Liouvillian \eq{eq:laaeff} and initial condition \eq{eq:rhozmarkov} from \Eq{eq:nakajimazwanzig}
for the measurement setup under study, we make use of the separation of time
scales governing the dynamics: This allows us to neglect the time
evolution in the quasistationary subspace in \Eq{eq:nakajimazwanzig},
i.e., we perform a Markov approximation with respect to the memory induced by
the decaying subspace (identifying now $a=q$ and $b=d$),
\begin{eqnarray}
  \rho^q (t') & \approx & \rho^q (t) .  \label{eq:mv}
\end{eqnarray}
Inserting \Eq{eq:mv} into Eq.
(\ref{eq:nakajimazwanzig}) yields
\begin{eqnarray}
  \dot{\rho}^q (t) & = & - i \left[ L_{\tmop{eff}}^{q q} + L^{q d}  \frac{e^{-
  i L^{d d} t}}{L^{d d}} L^{d q} \right] \rho^q (t) \nonumber\\
  &  & - i L^{q d} e^{- i L^{d d} t} \rho^d (0),  \label{eq:evolve}
\end{eqnarray}
with the effective Liouvillian
\begin{eqnarray}
  L^{q q}_{\tmop{eff}} & = & L^{q q} - L^{q d} \frac{1}{L^{d d}} L^{q d} + O
  \left( \tfrac{\omega l^2}{g^2}, \tfrac{l^3}{g^2} \right) . 
  \label{eq:leffgen}
\end{eqnarray}
The higher-order terms in \Eq{eq:leffgen} are due to non-Markovian
corrections $\Delta \rho^q (t') = \rho^q (t') - \rho^q (t)$ to \Eq{eq:mv}. They can be
estimated as follows: Since the exponential in the integral Eq.
(\ref{eq:nakajimazwanzig}) decays on a time scale $1 / g$, we only need to
account for corrections $\Delta \rho^q (t')$ for times $t'$ satisfying $t - t'
\lesssim 1 / g$. Integrating \Eq{eq:nakajimazwanzig}, this yields
corrections $\Delta \rho^q \sim \omega \text{/} g, l / g$. Multiplied with the
order $l^2 / g$ of the corrections from the decaying subspace, we obtain the
estimate of the higher-order terms in \Eq{eq:leffgen} in accordance
with the result obtained in frequency space.

The time-local, \Eq{eq:evolve} is not yet fully Markovian in the sense that it
still contains explicitly time-dependent terms in the time-evolution
generator, \newer{which therefore becomes frequency dependent in Laplace
space} {\footnote{See also the discussion of microscopic derivations of
master equations in {\color{black} Ref. {\cite{Breuer}}}.}}. In the long-time
limit $t \gg 1 / g$ this time dependence drops out: As we now argue, one can
omit the second term in the bracket in \Eq{eq:evolve} while the second
line of \Eq{eq:evolve} must not be neglected. One may drop the first
term since it gives a correction to the derivative $\sim l^2 \text{/} g$ on a
time scale $1 \text{/} g$, i.e., they result in an accumulated correction
$\Delta \rho^q (t)$ on the order of $\sim l^2 \text{/} g^2$, which can be
neglected. This is usually achieved in standard derivations of master
equations by setting $t \rightarrow \infty$ in the integral in Eq.
(\ref{eq:nakajimazwanzig}) {\cite{Breuer}}. By contrast, the corrections from
the second line of \Eq{eq:evolve} are of lower order $l / g$ as
integrating \Eq{eq:evolve} shows. In many cases, these terms do not
appear as one often assumes $\rho^d (0) = 0$ from the start. Here these terms must be kept and the solution of
\Eq{eq:evolve} can therefore be approximated as
\begin{eqnarray}
  \rho^q (t) & = & e^{- i L_{\tmop{eff}}^{q q} t}  \left[ {\color{white}
  \frac{}{}} \right. \rho^q (0) \nonumber\\
  &  & \left. - i \int_0^t d t' e^{+ i L_{\tmop{eff}}^{q q} t'} L^{q d} e^{-
  i L^{d d} t'} \rho^d (0) \right] \nonumber\\
  &  & + O \left( \tfrac{l^2}{g^2} \right) .  \label{eq:formalsolve}
\end{eqnarray}
Again, one can exploit that the exponentials in the second line of Eq.
(\ref{eq:formalsolve}) vary on a different time scale: While the factor $e^{-
i L^{d d} t'}$ is nonzero only on a short time scale $\sim 1 \text{/} g$, the
factor $e^{+ i L_{\tmop{eff}}^{q q} t'}$ changes on a much longer time scale
$\max (\omega, l^2 / g)$, and we may therefore expand $e^{+ i
L_{\tmop{eff}}^{q q} t'} \approx 1 + O (\omega / g, l^2 / g^2)$ in Eq.
(\ref{eq:formalsolve}). In the long-time limit $t \rightarrow \infty$, we can
then approximate the integral well by setting its upper bound $t \rightarrow
\infty$, resulting in
\begin{eqnarray}
  \rho^q (t) & = & e^{- i L^{q q}_{\tmop{eff}} t} \rho^q_{\tmop{eff}} (0) + O
  \left( \tfrac{\omega l}{g^2}, \tfrac{l^2}{g^2} \right), 
  \label{eq:leffready}
\end{eqnarray}
with the slipped initial state
\begin{eqnarray}
  \rho^q_{\tmop{eff}} (0) & = & \rho^q (0) - L^{q d} \frac{1}{L^{d d}} \rho^d
  (0) .  \label{eq:rhoqeff}
\end{eqnarray}
We have thus arrived again at \Eq{eq:laaeff} and (\ref{eq:rhozmarkov}),
respectively, particularly emphasizing that \Eq{eq:leffready} and
(\ref{eq:rhoqeff}) describe the time evolution correctly as long as $1 / g \ll
t \ll g^2 / [\max (l, \omega)]^3$; otherwise, the correction terms to the
effective Liouvillian (\ref{eq:leffgen}) can accumulate to a large error in
\Eq{eq:leffready}. This is also borne out by numerical checks that we performed.

\subsection{Effective Liouvillian for indirect detection}\label{app:leffcalc}

We next apply the above Liouville-space projection technique to the indirect
detection setup of the SQD-qubit system studied in the \Sec{sec:kineq} and \sec{sec:results} of the main text. In {\color{black} Appendix \ref{app:introproj}} we first make the
connection between the projections just discussed and the
dynamical variables considered in the main text. After this, we
provide in {\color{black} Appendix \ref{app:leffcompute}} some important intermediate
steps in the derivation of the effective Liouvillian (\ref{eq:leff}). We
comment \ in {\color{black} Appendix \ref{app:nm}} on the Markov approximation
with respect to the memory induced by the sensor QD on the qubit. Finally, we
discuss the validity of our effective Liouvillian in view of the perturbative
expansion of the kinetic equations in {\color{black} Appendix \ref{app:validity}}
and comment on the $U = 0$ limit of our problem.

\subsubsection{Definition of projections and representing matrices of the
Liouvillian}\label{app:introproj}

To make a connection between the projection $\rho^q = P^q \rho$ of the
SQD-qubit density operator on the quasistationary subspace and the dynamical
variables introduced in {\Sec{sec:couplingmodes}}, we
exploit the eigenbasis of the Liouvillian $L_0$ [\Eq{eq:L0}] with
left and right eigenvectors $\tilde{{V}}^k_p$ and ${V}^k_p$, respectively. In the following,
``$\dot{=}$'' denotes that we represent a basis-independent object on the
left hand side by its components on the right hand side with respect to the
eigenbasis of $L_0$. Representing, for example, the density operator $\rho$ in
this basis, we obtain
\begin{eqnarray}
  \rho & \dot{=} & \left(\begin{array}{c}
    \vec{X}^q\\
    \vec{X}^d
  \end{array}\right) \text{ \ = \ } \left(\begin{array}{c}
    1\\
    \tau^0_{\alpha} + \tau_{\alpha}^1\\
    p^1_{\tmop{st}} p^0 - p^0_{\tmop{st}} p^1\\
    p^1_{\tmop{st}} \tau_{\alpha}^0 - p^0_{\tmop{st}} \tau_{\alpha}^1
  \end{array}\right),  \label{eq:x}
\end{eqnarray}
with $\vec{X}^q$ and $\vec{X}^d$ given by \Eq{eq:rhoqrhod} in the main
text. In contrast to the representation (\ref{eq:kineq2}) in
{\Sec{sec:couplingmodes}}, working in the eigenbasis of
$L_0$ also fixes a particular basis for the isospin part, namely the
polarization basis which we order as \ $\vec{e}_- = \left( \vec{e}_1 + i
\vec{e}_2 \right) / \sqrt{2}$, $\vec{e}_0 = {\color{black} \vecg{\Omega}} /
\Omega$, and $\vec{e}_+ = \left( \vec{e}_1 - i \vec{e}_2 \right) / \sqrt{2}$.
The latter vectors are constructed from the right-handed orthonormal basis
$\vec{e}_0$, $\vec{e}_1 = {\color{black} \vecg{\lambda}} / \lambda$, and
$\vec{e}_2 = \vec{e}_0 \times \vec{e}_1 = {\color{black} \vecg{\Omega}}
\times {\color{black} \vecg{\lambda}} / \Omega \lambda$. This yields the
three components $\tau^n_{\alpha} = \vec{e}_{\alpha}^{\dag} \cdot
\vecg{\tau}^n$ $(\alpha = -, 0, +)$ in \Eq{eq:x} that make up the
isospins $\vecg{\tau}^n = \sum_{\alpha} \tau^n_{\alpha} \vec{e}_{\alpha}$. The
projection of $\rho$ on the quasistationary and the decaying subspace are next
represented as
\begin{eqnarray}
  \rho^q & = & P^q \rho \text{ \ } \dot{=} \text{ \ } \left(\begin{array}{c}
    \vec{X}^q\\
    \vec{0}
  \end{array}\right),  \label{eq:xq}\\
  \rho^d & = & P^d \rho \text{ \ } \dot{=} \text{ \ } \left(\begin{array}{c}
    \vec{0}\\
    \vec{X}^d
  \end{array}\right) .  \label{eq:xd}
\end{eqnarray}
The matrix $(L^{k k'}_0)_{p p'} = \tilde{\vec{V}}^{k \dag}_p \cdot L_0
\cdot \vec{V}^{k'}_{p'}$ of the unperturbed Liouvillian [see Eq.
(\ref{eq:l0})] \newer{expressed in terms of its
left and right eigenvectors $\tilde{{V}}^k_p$ and ${V}^k_p$,} respectively, is specified completely by the two diagonal blocks
\begin{eqnarray}
  - i L^{q q}_0 & = & \left(\begin{array}{cc}
    0 & 0\\
    0 & i \alpha \Omega \delta_{\alpha \alpha'}
  \end{array}\right),  \label{eq:l0qq}
        \\
  - i L^{d d}_0 & = & \left(\begin{array}{cc}
    - \gamma & 0\\
    0 & (- \gamma + i \alpha \Omega) \delta_{\alpha \alpha'}
  \end{array}\right),  \label{eq:l0dd}
\end{eqnarray}
whence the off-diagonal blocks vanish: $L_0^{q d} = L_0^{d q} = 0$. Here and
below we use the shorthand notation $M_{\alpha \alpha'}$ for a $3 \times 3$
matrix $M$ in the above-mentioned polarization basis. In this polarization basis the cross
product $\vecg{\Omega} \times$ is diagonal:
\begin{eqnarray}
  & \begin{array}{lll}
    \vecg{\Omega} \times & = & \sum_{\alpha \alpha'} \left[
    \vec{e}^{\dag}_{\alpha} \cdot (\vecg{\Omega} \times \vec{e}_{\alpha'})
    \right]  \vec{e}_{\alpha} \vec{e}_{\alpha'}^{\dag}
      \\
      \\
    & \dot{=} & \left(\begin{array}{ccc}
      - i \Omega & 0 & 0\\
      0 & 0 & 0\\
      0 & 0 & + i \Omega
    \end{array}\right),
  \end{array} &  \label{eq:omegacross}
\end{eqnarray}
Next expressing the four blocks of the perturbation $\Lambda = L - L_0$ in
this basis, one finds
\begin{eqnarray}
  \Lambda^{q q} & = & \left(\begin{array}{cc}
    0 & 0\\
    0 & \lambda p^1_{\tmop{st}} s_{\alpha \alpha'}
  \end{array}\right),  \label{eq:lqq}\\
  \Lambda^{q d} & = & \left(\begin{array}{cc}
    0 & 0\\
    0 & - \lambda s_{\alpha \alpha'}
  \end{array}\right),  \label{eq:lqd}\\
  \Lambda^{d q} & = & \left(\begin{array}{cc}
    0 & (1 + p^0_{\tmop{st}}) \lambda c_{\alpha'}^{\ast}\\
    (1 + p^0_{\tmop{st}}) \lambda c_{\alpha} & - r \lambda s_{\alpha \alpha'}
  \end{array}\right),  \label{eq:ldq}\\
  \Lambda^{d d} & = & \left(\begin{array}{cc}
    0 & \lambda c_{\alpha'}^{\ast}\\
    \lambda c_{\alpha} & (p^0_{\tmop{st}} - 3 \kappa / 2) \lambda s_{\alpha
    \alpha'}
  \end{array}\right) .  \label{eq:ldd}
\end{eqnarray}
The components of the matrix $s_{\alpha \alpha'}$ are proportional to the
(signed) volume of the parallel epiped spanned by the vectors
$\vec{e}_{\alpha}$, $\vec{e}_{\alpha'}$, and $\op{{\color{black}
\vecg{\lambda}}} = \vecg{\lambda} / \lambda$,
\begin{eqnarray}
  s_{\alpha \alpha'} & = & \vec{e}_{\alpha}^{\dag} \cdot \left(
  \op{{\color{black} \vecg{\lambda}}} \times \vec{e}_{\alpha'} \right)
  \nonumber \\
  & = & \frac{i \lambda}{\sqrt{2}} \left(\begin{array}{ccc}
    0 & + 1 & 0\\
    + 1 & 0 & - 1\\
    0 & - 1 & 0
  \end{array}\right),  \label{eq:saa}
\end{eqnarray}
and the remaining factor $c_{\alpha}$ is given by the projected
isospin-to-charge conversion vector
\begin{equation}
  \begin{array}{lllll}
    c_{\alpha} & = & c \left( \vec{e}_{\alpha}^{\dag} \cdot \op{{\color{black}
    \vecg{\lambda}}} \right) & = & \frac{c}{\sqrt{2}} \left(\begin{array}{c}
      1\\
      0\\
      1
    \end{array}\right),
  \end{array}
\end{equation}
with $c$ given by \Eq{eq:conversion}. Finally, \Eq{eq:ldq}
incorporates the transition factor
\begin{eqnarray}
  r & = & p^1_{\tmop{st}} p^0_{\tmop{st}} - \kappa \left(
  \tfrac{1}{2} p^1_{\tmop{st}}-  p^0_{\tmop{st}}\right),
          \label{eq:some-r-equation}
\end{eqnarray}
whose dependence on the SQD parameters we thoroughly discussed in
{\Sec{sec:mitigation1}} in the main text.

Equation (\ref{eq:lqq}) together with \Eq{eq:saa} reveals that the
``direct'' perturbation of the quasistationary modes for $\lambda \neq 0$ is
not diagonal in the unperturbed eigenbasis. This expresses the fact that the
mean-field $\tilde{\vecg{\Omega}} = \vecg{\Omega} + p^1_{\tmop{st}}
\vecg{\lambda}$ is tilted with respect to $\vecg{\Omega}$. Moreover, since the
Liouvillian is not Hermitian, $L \neq L^{\dag}$, the transition matrices are
also not simply related by Hermitian conjugation, i.e., 
$\Lambda^{d q} \neq (\Lambda^{q d})^{\dag}$. This means that transitions between
quasistationary and decay modes are not always possible in both directions
(see {\color{black} Fig. \ref{fig:modes}}), in contrast to Hamiltonian dynamics. In particular, all transitions
{\tmem{into}} the stationary charge modes $(q, c)$ are forbidden (the first
rows of $\Lambda^{q q}$ and $\Lambda^{q d}$ are zero). This is a consequence
of the sum rules guaranteeing probability and isospin conservation by
tunneling (see {\color{black} Ref. {\cite{Hell14a}}}). Physically,
the eigenvalue of the stationary charge mode must stay pinned to zero.

We further find from the above equations [see Eqs.
(\ref{eq:condition})-(\ref{eq:slowevolve})]:
\begin{eqnarray}
  g & = & | | L^{b b} | - | L^{a a} | | \text{ \ \ \ \ } \text{ } \sim \text{
  \ } \gamma, \\
  l & = & \max (| L^{a b} |, | L^{b a} |) \text{ \ } \sim \text{ \ } \lambda,
  \\
  \omega & = & | L^{q q} | \text{ \ \ \ \ \ \ \ \ \ \ \ \ \ \ \ \ \ } \sim
  \text{ \ } \Delta \text{ \ } \sim \text{ \ } \lambda, \Omega . 
\end{eqnarray}
Thus, the time scales on the quasistationary subspace and the decaying
subspace separate as required by \Eq{eq:slowevolve} and satisfied by
our assumption $\Delta \ll \gamma$. Moreover, the weak-coupling assumption
(\ref{eq:condition}) is fulfilled in the weak-measurement limit $\lambda
\ll \gamma$. For further discussion of the consistency see also {\color{black}
Appendix \ref{app:validity}}.

\subsubsection{Computation of effective Liouvillian}\label{app:leffcompute}

Exploiting Eqs. (\ref{eq:xq}) and (\ref{eq:efftime}), the effective evolution
of the quasistationary modes can be expressed as
\begin{eqnarray}
  \dot{\vec{X}}^q (t) \text{ = } \left(\begin{array}{c}
    0\\
    \dot{\tau}_{\alpha} (t)
  \end{array}\right) & = & - i L^{q q}_{\tmop{eff}}  \vec{X}^q (t) \text{ \ =
  \ } L^{q q}_{\tmop{eff}} \left(\begin{array}{c}
    1\\
    \tau_{\alpha} (t)
  \end{array}\right) . \nonumber\\
  &  &  \label{eq:xqdot}
\end{eqnarray}
Consistently accounting for terms up to second order in $\Delta$ in the
high-temperature limit, the effective Liouvillian reads
\begin{eqnarray}
  L^{q q}_{\tmop{eff}} & = & L_0^{q q} + \Lambda^{q q} - i \Lambda^{q d}
  \frac{1}{\gamma} \Lambda^{q d} + O \left( \tfrac{\Delta^3}{\gamma^2} \right)
  .  \label{eq:lqqeff}
\end{eqnarray}
Inserting Eqs. (\ref{eq:l0qq}) -- (\ref{eq:ldd}), we obtain
\begin{eqnarray}
  - i L_{\tmop{eff}}^{q q} & \dot{=} & \left(\begin{array}{cc}
    0 & 0\\
    I_{\alpha} & - i (L_{\tmop{eff}})_{\alpha \alpha'}
  \end{array}\right),  \label{eq:leffmatrix}
\end{eqnarray}
with
\begin{eqnarray}
  I_{\alpha} & = & O \left( \tfrac{\Delta^3}{\gamma^2} \times \tfrac{\Gamma}{T}
  \right)  \label{eq:ialpha}\\
  (L_{\tmop{eff}})_{\alpha \alpha'} & = & i \alpha \Omega \delta_{\alpha
  \alpha'} + \lambda p^1_{\tmop{st}} s_{\alpha \alpha'} - \frac{\lambda^2
  r}{\gamma} (s \cdot s)_{\alpha \alpha'} \nonumber\\
  &  & + O \left( \tfrac{\Delta^3}{\gamma^2} \right) .  \label{eq:laapr}
\end{eqnarray}
The injection term $I_{\alpha}$ of \Eq{eq:ialpha} is thus negligible in the
high-temperature limit considered here {\footnote{We checked that when
accounting for the leading-order expression in \Eq{eq:ialpha} for lower
temperatures, one reproduces the stationary state of {\color{black} Ref.
{\cite{Hell14a}}} with nonzero stationary isospin $\vecg{\tau}_{\tmop{st}}$.}. In Eq.
(\ref{eq:laapr}) the first two terms are responsible for the qubit precession
with frequency $\sim \Delta$, while the third term induces the isospin decay
with the a rate $\sim \lambda^2 \text{/} \gamma$, which coincides with Eq.
(\ref{eq:r}) in the main text.

Inserting \Eq{eq:leffmatrix} back into \Eq{eq:xqdot}, we obtain
for the total isospin evolution:
\begin{eqnarray}
  \dot{\vecg{\tau}} (t) & = & \sum_{\alpha} \dot{\tau}_{\alpha}
  \vec{e}_{\alpha} \text{ \ = \ } \sum_{\alpha} \vec{e}_{\alpha} (- i
  L_{\tmop{eff}})_{\alpha \alpha'} \tau_{\alpha'} (t) \nonumber\\
  & = & - i L_{\tmop{eff}} \vecg{\tau} (t)  \label{eq:taudot}
\end{eqnarray}
with
\begin{eqnarray}
  L_{\tmop{eff}} & = & (L_{\tmop{eff}})_{\alpha \alpha'} \vec{e}_{\alpha}
  \vec{e}^{\dag}_{\alpha'} \nonumber\\
  & = & \sum_{\alpha = 0, \pm} (i \tilde{\Omega}_{\alpha} - \gamma_{\alpha})
  \vec{e}_{\tmop{eff}, \alpha}  \tilde{\vec{e}}^{\dag}_{\tmop{eff},
  \alpha}, 
\end{eqnarray}
with $\tilde{\Omega}_{\alpha}$, $\gamma_{\alpha}$, and $\vec{e}_{\tmop{eff},
\alpha}$, thus establishing Eqs. (\ref{eq:leff}) -- (\ref{eq:tphi}) in the main part of
the paper.

Finally, the slipped initial state following from \Eq{eq:rhoqeff}
reads
\begin{eqnarray}
  &  & \vec{X}^q_{\tmop{eff}} (0) \text{ \ = \ } \left(\begin{array}{c}
    1\\
    \tau_{\tmop{eff}, \alpha} (0)
  \end{array}\right) \\
  &  & \text{ \ = \ } \vec{X}^q (0) + \Lambda^{q d} \frac{1}{\gamma} 
  \vec{X}^d_{\tmop{eff}} (0) + O \left( \tfrac{\Delta^2}{\gamma^2} \right) \\
  &  & \text{ \ = \ } \left(\begin{array}{c}
    1\\
    (\tau^0 + \tau^1)_{\alpha} (0) - \frac{\lambda}{\gamma} s_{\alpha \alpha'}
    (p^0_{\tmop{st}} \tau^1 - p^1_{\tmop{st}} \tau^0)_{\alpha} (0)
  \end{array}\right) \nonumber\\
  &  & \text{ \ \ \ \ \ } + O \left( \tfrac{\Delta^2}{\gamma^2} \right) 
  \label{eq:xqeffapp}
\end{eqnarray}
with $\tau^n_{\alpha} = \vec{e}^{\dag}_{\alpha} \cdot \vecg{\tau}^n$. Computing
$\vecg{\tau}_{\tmop{eff}} (0) = \sum_{\alpha} \tau_{\tmop{eff}, \alpha} (0)
\vec{e}_{\alpha}$ from \Eq{eq:xqeffapp}, we arrive at Eq.
(\ref{eq:taueff0}) from the main text. The set of states with zero slip is
discussed in detail in {\color{black} Appendix \ref{app:slip}} and it is shown that the
``size'' of this set grows with the ``distance'' of the sensor from the
stationary state.

This completes the derivation of the effective isospin evolution
discussed in {\Sec{sec:leffexpand}}.

\subsubsection{Markov approximation relative to the sensor QD}\label{app:nm}

Regarding the discussion of non-Markovian effects, we have to
distinguish between two different memory effects, namely those arising from
the {\tmem{electrodes}} (imposed on the qubit-SQD system) and those arising
from the {\tmem{sensor quantum dot}} (imposed on the qubit only). As
explained in {\color{black} Appendix \ref{app:positivity}}, we account for the
leading non-Markovian effect on the composite qubit-sensor QD system from the
{\tmem{electrodes}}, which is induced by the tunnel coupling; these effects
are contained in the effective Liouvillian $L$ in the kinetic equations
(\ref{eq:kineq}), which forms the starting point of the above analysis. Their
effect is to modify the coupling of the quasistationary and decay modes
through the coherent backaction terms in the kinetic equation (\ref{eq:kineq})
as explored in {\color{black} Appendix \ref{app:positivity}} and \ {\color{black}
Sec. \ref{sec:nmdiscuss}}.

By contrast, the frequency dependence of the effective
Liouvillian $L_{\tmop{eff}}$ for the reduced qubit system, Eq.
(\ref{eq:leffz}), incorporates non-Markovian effects on the qubit due to
memory of the {\tmem{sensor quantum dot}} after the electrodes were
integrated out (thus effectively of SQD plus electrodes). These are accounted for by
our general Laplace-space approach given in {\color{black} Appendix
\ref{app:derivationfrequencyspace}} in an exact way relative to \Eq{eq:kineq}. In principle, these non-Markovian effects
can be already studied based on the solutions of our full kinetic equations
(\ref{eq:kineq}). However, the expressions \eq{eq:xqt} and \eq{eq:xqefft} provide
 a more convenient starting point to gain further insight on an analytical level.

In the main text, we focused for further illustration on the
high-temperature limit that allows us to make the Markov approximation
(\ref{eq:mv}) when expanding the effective Liouvillian (\ref{eq:leff}) to
lowest order in $\lambda / \gamma$. In this limit, SQD-induced memory effects on the
qubit are thus neglected. Moreover, we neglect $O (\Gamma^2 / T)$ corrections to
the SQD tunneling rates and, thus, non-Markovian corrections from the
electrodes are also consistently neglected.
\cut{and consistent with this we drop the
corresponding non-Markovian corrections from the sensor QD.}
This is reasonable because if memory effects of
the SQD are not accounted for, then memory effects from the electrodes should
have no effect \emph{a fortiori}.

\subsection{Validity of perturbation expansion}\label{app:validity}

In this part of the Appendix, we collect various remarks on
the validity and the limitations of our perturbation theory.

\subsubsection{Effective Liouvillian (\ref{eq:leffzmain})}As stated
in {\Sec{sec:weakcoupling}}, our kinetic equations are
applicable as long as $\Gamma / T \ll \Delta / \Gamma \ll 1$. One may now
wonder whether it is permissible to expand the denominator in Eq.
(\ref{eq:leffzmain}) in orders of $\Delta / \Gamma$ and to truncate this
expansion. In general, the answer is no, for the following reason: To lowest
order in $\Delta / \Gamma$, the effective Liouvillian scales as $\lambda^2 /
\Gamma$. The coherent backaction and and the cotunneling terms then yield
corrections of order $\lambda^2 / \Gamma \times \Gamma / T$; however,
higher-order corrections in $\Delta / \Gamma$ are at least of order $\lambda^2
/ \Gamma \times \Delta / \Gamma$ and are therefore not negligible once we
include terms of order $\lambda^2 / \Gamma \times \Gamma / T$ {\footnote{We note that we still consider the weak-measurement regime
($\lambda / \Gamma \ll 1$). The validity of our kinetic equations is
just limited to high temperatures implying $\Gamma / T \ll \lambda / \Gamma$.
}}. Therefore, even
though we start from the weak-measurement limit, one must not expand the
denominator in \Eq{eq:leffzmain} from the start.

However, if we first neglect all $\Gamma / T$ corrections
(dissipative and coherent backaction and $\Gamma^2 / T$ corrections to the SQD
rates), then one can consistently expand the effective Liouvillian in $\Delta
/ T$. The lowest-order approximation to this is investigated in {\color{black}
Sec. \ref{sec:results}} and requires high temperatures so that the $\Gamma / T$
corrections are sufficiently small.

\subsubsection{Kinetic equation (\ref{eq:kineq})}
The reader may also wonder
whether one should not include terms of order $\Delta^2$ into the kinetic
equation (\ref{eq:kineq}) since the lowest order to the effective Liouvillian
scales at least quadratically in $\Delta$. However, such terms appear only in
combination with additional tunneling processes as the internal interaction $L
\sim \Delta$ is treated without approximation in the kinetic equations. Thus,
terms including $\Delta^2$ must be at least of order $\Gamma \Delta^2 / T^2$
and are therefore strongly suppressed in the high-temperature limit that we
consider in this paper. For example, if these terms appear in $\Lambda^{q q}$,
they lead to corrections of order $\Gamma \Delta^2 / T^2 \ll \Delta^2 / \Gamma
\times \Gamma / T$ since $\Gamma / T \ll 1$. They are even less important if
they appear in $\Lambda^{q d}, \Lambda^{d q}, \Lambda^{q q}$, where they would
lead to terms of order $\Delta^2 / \Gamma (\Delta \Gamma / T^2)$ when inserted
into \Eq{eq:leffzmain}.

Moreover, one may also wonder whether terms of order $\Gamma^3 / T^2$ should
be included into the kinetic equation (\ref{eq:kineq}) since they can also
modify the stochastic backaction. However, in the weak-coupling,
weak-measurement limit considered in this paper, those result in corrections
of even higher-order $\Delta^2 / \Gamma (\Gamma^2 / T^2) \ll \Delta^2 / T$.
Yet, deep in the Coulomb blockade regime (which is beyond the present
scope), these terms should play an important role since the transition factor
$r$ \ [\Eq{eq:r}] --- which controls the strength of the total
backaction together with $c$ --- is exponentially suppressed by virtue of the
effective cancellation of $\Gamma^2 / T$ and $\Gamma \lambda / T$ terms
explained in {\Sec{sec:mitigation1}}. We expect
higher-order corrections to cause deviations from this exponential suppression
of the transition factor $r$, resulting in an algebraic scaling $\sim 1 /
(\varepsilon - \mu_r)^n$ but with an exponent of higher
than that of cotunneling broadening, $n > 1$, in agreement with other
theoretical work (see {\Sec{sec:compquantum}}). We
emphasize that the main point of our work is just to explain physically the
cancellation of the expected leading power law $n = 1$. A much more elaborate
analysis is needed to find the actual power law (including the complications
of a nonstationary detector with strong local Coulomb interaction, etc.).

\subsubsection{Exact result for $U = 0$}Finally, to further support the
cancellation in \Eq{eq:r} that we found here for large Coulomb
interaction $U$ and perturbatively in $\Gamma$, we computed in a separate
calculation the effective Liouvillian for a {\tmem{non}}interacting sensor QD
$(U = 0)$ nonperturbatively in $\Gamma$ but only to leading order in
$\lambda$. This releases the condition $\Gamma / T \ll \lambda / \Gamma$ and
can be used to investigate also the opposite regime $\lambda / \Gamma \ll
\Gamma / T (\ll 1)$. These results actually confirm the mitigation of the
cotunneling-induced stochastic backaction by the coherent backaction also in
this case. Yet, it remains an interesting future question to understand the
role of the coherent backaction in the low-temperature, \new{strong
qubit-sensor} coupling regime
when also the strong interaction $U$ is accounted for.




\section{Initial states without slip}\label{app:slip}

In this Appendix, we characterize the set of initial qubit-sensor QD density
operators with zero initial slip [see Eqs. \eq{eq:slipqd} and \eq{eq:taueff0}] in the reduced dynamics of the qubit in the
high-temperature limit. We show that this imposes a strong condition: Initial states with
zero slip form a set of measure zero, thus making the initial slip a relevant
source of errors in indirect detection unless the qubit-SQD quantum state (\emph{not} just the qubit state) is
under accurate control. In addition, increasing the nonstationarity of the
sensor reduces these sets of zero slip. We also formulate the constraint in
terms of relations between the reduced qubit state and the composite qubit-SQD
state and discuss how it relates to the factorizability of the qubit-SQD
state.

\begin{center}
  \Figure{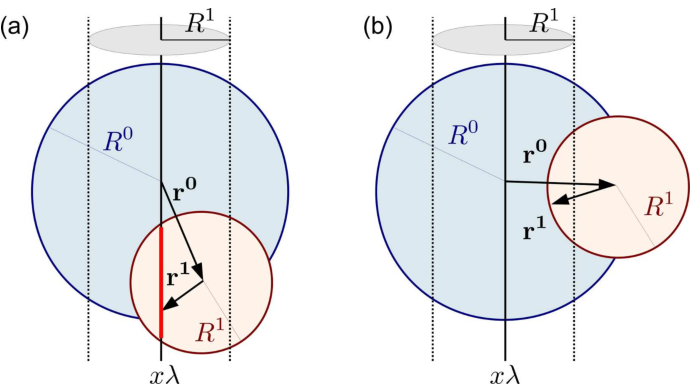}{Geometric
  restrictions on the qubit-sensor density operator imposed by requiring zero
  initial slip. We show a 2D cross section of the 3D construction described in
  the text, i.e., the circles represent spheres and the dotted lines indicate
  the boundaries of a cylinder around the line $x
  \vecg{\lambda}$.
  \label{fig:slipgeom} \ }
\end{center}

\emph{Kinematic restrictions.} It is most convenient for the following
considerations to work with the representation of the initial qubit-SQD state,
\begin{eqnarray}
  \rho (0) & = & \sum_n \hat{P}^n \otimes \tfrac{1}{2} \left( p^n (0)
  \op{\mathbbm{1}_{}}_Q + \vecg{\tau}^n (0) \cdot \hat{\vecg{\tau}} \right),
\end{eqnarray}
in terms of the occupation probabilities $p^n (0)$ and the charge-specific
isospins ${\color{black} \vecg{\tau}}^n (0)$. We recall that for this state
to be a valid quantum state, $p^n (0)$ and ${\color{black} \vecg{\tau}}^n
(0)$ have be real (for $\rho^{\dag} = \rho$), $\sum_n p^n (0) = 1$ (for
$\tmop{tr} \rho = 1$), and the magnitudes
\begin{equation}
  \begin{array}{lll}
    \left| \vecg{\tau}^n (0) \right| & \leqslant & p^n (0)
  \end{array} \label{eq:poscons}
\end{equation}
have to be restricted (for positivity, $\rho \geqslant 0$). We note that the
reduced state of the SQD and the qubit,
\begin{eqnarray}
  \rho_S \text{ \ } \assign \text{ \ } \underset{Q}{\tmop{tr}} \rho (0) & = &
  \sum_n p^n (0) \hat{P}^n, \\
  \rho_Q \text{ \ } \assign \text{ \ }  \underset{S}{\tmop{tr}} \rho (0) & = &
  \tfrac{1}{2} \left( \op{\mathbbm{1}_{}}_Q + \vecg{\tau} (0) \cdot
  \hat{\vecg{\tau}} \right), 
\end{eqnarray}
are completely specified by the \newer{occupation probabilities} $p^n (0)$ (in fact by only one) and the Bloch
vector $\vecg{\tau} (0) = {\color{black} \vecg{\tau}}^0 (0) + {\color{black}
\vecg{\tau}}^1 (0) \nocomma$, respectively. When $p^n(0) = p_{\tmop{st}}^n$, the
SQD is stationary, but this does not imply that \new{the} qubit-SQD state state $\rho
(0)$ is factorisable (see below).

\emph{Charge-specific isospins.} We first investigate the slip based on the
charge-specific isospins $\vecg{\tau}^n(0)$ since these have simple kinematic constraints,
allowing for an easy, complete characterization. The zero-slip condition
obtained from \Eq{eq:taueff0},
\begin{eqnarray}
  \vec{0} & = &  {\color{black} \vecg{\lambda}} \times \left( \vec{r}^0 +
  \vec{r}^1 \right),  \label{eq:slipr}
\end{eqnarray}
is expressed using rescaled isospin vectors $\vec{r}^0 = p^1_{\tmop{st}}
{\color{black} \vecg{\tau}}^0 (0)$ and $\vec{r}^1 = - p^0_{\tmop{st}}
{\color{black} \vecg{\tau}}^1 (0)$. For an arbitrary initial state, the two
three-dimensional vectors $\vec{r}^0$ and $\vec{r}^1$ are taken from a
six-dimensional set that is constrained only by the positivity of $\rho (0)$:
the vectors $\vec{r}^0$ and $\vec{r}^1$ have to lie within spheres of
different radii given by $\left| \vec{r}^0 \right| \leqslant R^0 \assign
p^1_{\tmop{st}} p^0 (0)$ and $\left| \vec{r}^1 \right| \leqslant R^1 \assign
p^0_{\tmop{st}} p^1 (0)$, respectively. The radii are equal if and only if the
reduced sensor state $\rho_S$ is stationary: the condition $p^0_{\tmop{st}}
p^1 (0) = p^1_{\tmop{st}} p^0 (0)$ is equivalent to $p^n (0) =
p^n_{\tmop{st}}$ ($n = 0, 1$) due to the normalization condition $\sum_n p^n
(0) = \sum_n p^n_{\tmop{st}} = 1$.

From this set of valid initial states let us now construct those which have
zero slip. According to \Eq{eq:slipr}, this requires the sum of the
vectors $\vec{r}^0$ and $\vec{r}^1$ to lie on the line defined by the
measurement vector: $\vec{r}^0 + \vec{r}^1 = x \vecg{\lambda}$ with any $x \in
\mathbbm{R}$. For the construction, \new{sketched in
Fig. \ref{fig:slipgeom}}, first draw a sphere with radius $R^0$
\new{(blue)} and
draw the line $x \vecg{\lambda}$ through its origin \new{(black)}. For each vector
$\vec{r}^0$ in this sphere draw a \new{second} sphere of radius $R^1$
centered at its tip \new{(red)}.
The set of vectors $\vec{r}^1$ that give zero-slip state are just given by the
intersection of this \new{second} sphere with the line $x \vecg{\lambda}$. This is a set of measure zero.
Moreover, from the figure it is clear that the construction is possible only
if $\vec{r}^0$ is inside a cylinder of radius $R^1$ with the line $x
\vecg{\lambda}$ as its axis. Whenever $\vec{r}^0$ lies outside this cylinder,
it is not possible at all to construct a zero-slip initial state.

The radii $R^n$ are controlled by the initial reduced sensor state through
$p^n (0)$ and there are two extreme limits:

(i) Stationary state $p^n (0) = p^n_{\tmop{st}}$: $R^0 = R^1$. In this case
for every $\vec{r}^0$ one can find an $\vec{r}^1$ giving a zero-slip state.
Still, the subset has measure zero in the total set of possible states.

(ii) Integral charge state $p^{0, 1} (0) = 0, p^{1, 0} (0) = 1$:
$R^{0, 1} = 0$, $R^{1, 0} = p^{1, 0}_{\tmop{st}}$. In these two cases, the
charge-specific isospin $\vecg{\tau}^{0, 1} (0) = \vecg{\tau} (0)$ coincides
with the total isospin, while $\vecg{\tau}^{1, 0}(0) = \vec{0}$. The slip is
${\color{black} \vecg{\tau}}_{\tmop{eff}} (0) - \vecg{\tau} (0) = \mp
{\color{black} \vecg{\lambda}} \times p^{1, 0}_{\tmop{st}} {\color{black}
\vecg{\tau}} (0) / \gamma$ and can be avoided only if $\vecg{\tau}^{0, 1}
(0)$ is collinear to $\vecg{\lambda}$ so that the tip of $\vec{r}^{0,1}$
lies on the line $x \vecg{\lambda}$ \newer{to allow for $\vec{r}^{1,0}=\vec{0}$} as sketched in \Fig{fig:slipgeom}(a). 

As the sensor deviates from the stationary state, the radii $R^1$ and $R^0$
differ and the above possibilities for constructing initial zero-slip states are reduced.
 Since in any case the zero-slip states are sets of measure zero, it is clear that most
preparation errors of the \emph{sensor QD} will lead to an initial slip on the \emph{qubit}, i.e., the backaction-induced initial
slip generates \emph{additional} errors beyond the control over the qubit.

\emph{Restriction on the qubit isospin.}
It is instructive to further clarify the restrictions that the
above imposes on the \emph{reduced qubit state}, a density operator completely characterized by ${\color{black} \vecg{\tau}} (0)$, relative to the composite qubit-sensor state $\rho (0)$. To this end we change the variables to
\begin{eqnarray}
  {\color{black} \vecg{\tau}} (0) & = & {\color{black} \vecg{\tau}}^0 (0)
  + {\color{black} \vecg{\tau}}^1 (0) \\
  \delta_{\tmop{st}} (0) & = & p^1_{\tmop{st}} p^0 (0) - p^0_{\tmop{st}} p^1
  (0) \\
  \vecg{\delta} (0) & = & p^1 (0) \vecg{\tau}^0 (0) - p^0 (0) \vecg{\tau}^1 (0) 
\end{eqnarray}
The conditions imposed by the positivity $\rho (0)$ on ${\color{black}
\vecg{\tau}} (0)$ and $\vecg{\delta} (0)$ are not easy to formulate and will
be ignored in the following (note that $| {\color{black} \vecg{\tau}} (0) |
\leqslant 1$ is only necessary, not sufficient). Rewriting the zero-slip
condition \eq{eq:taueff0} by inserting $\vecg{\tau}^{0, 1} (0) =
p^{0, 1} (0) \vecg{\tau} (0) \pm \vecg{\delta} (0)$ yields
\begin{eqnarray}
  \vec{0} & = &  {\color{black} \vecg{\lambda}} \times \left[ \vecg{\delta}
  (0) + \delta_{\tmop{st}} (0) \vecg{\tau} (0) \right]  \label{eq:slipdelta}
\end{eqnarray}
Noting that $\vecg{\delta} (0) = \vec{0}$ corresponds to a factorizable state
[see \Eq{eq:faccond}] and $\delta_{\tmop{st}} (0) = 0$ to a stationary
reduced sensor state (see above), we have four cases in which there is zero initial slip:

(i) nonfactorizable, nonstationary initial state $\vecg{\delta} (0) \neq
\vec{0}$, $\delta_{\tmop{st}} (0) \neq 0$: The qubit state and the qubit-SQD
correlations must be fine tuned such that $\vecg{\delta} (0) +
\delta_{\tmop{st}} (0) \vecg{\tau} (0) = x \vecg{\lambda}$ or some $x \in
\mathbbm{R}$.

(ii) nonfactorizable, stationary initial state $\vecg{\delta} (0) \neq
\vec{0}$, $\delta_{\tmop{st}} (0) = 0$: The qubit state can be arbitrary, but the qubit-SQD state must have have very special correlations such that $\vecg{\delta} (0) \propto \vecg{\lambda}$.

(iii) factorizable, nonstationary initial state $\vecg{\delta} (0) = \vec{0}$,
$\delta_{\tmop{st}} (0) \neq 0$: The qubit state must be prepared in \new{a}
measurement-basis state $\vecg{\tau} (0) \propto \vecg{\lambda}$.

(iv) factorizable, stationary initial state $\vecg{\delta} (0) = \vec{0}$,
$\delta_{\tmop{st}} (0) = 0$: no conditions, there is no slip.

Equation (\ref{eq:slipdelta}) emphasizes that the set of states without slip
does not simply coincide with factorizable initial states (``uncorrelated
states''). It further shows that the more the sensor deviates from the
stationary state, now quantified by $\delta_{\tmop{st}} (0)$, the larger the
initial slip becomes. This concludes our discussion of the states with zero
slip and, as the previous analysis showed, the zero-slip states are subsets of measure zero of the set of possible initial states.
The experimental possibilities to avoid the slip [case (iv)] are discussed in \Sec{sec:effliouville} and \Sec{sec:summary}.




\section{Renormalization-induced qubit phase
kicks}\label{app:renormalization}

In this final Appendix, we explain how the coherent backaction appears in the
visibility (\ref{eq:tauplusav}) as an additional contribution that cannot be
understood in a semi-classical stochastic picture discussed in {\color{black}
Sec. \ref{sec:semiclassical}}. We show that, loosely speaking, the coherent backaction
induces additional ``phase kicks'' on the qubit that partially ``undo'' the phase
kicks induced by the stochastic backaction that result in decoherence. As a
result, the decoherence is mitigated. We emphasize from the start that these
phase kicks are completely unrelated to the initial slip: They do not lead to
a phase shift of coherent isospin precession but instead only affect the net
qubit decoherence.

Our objective in the following is to merely further illustrate the physical
origin of the coherent backaction by calculations that are certainly not
rigorous. The proper treatment is achieved by our kinetic equation
(\ref{eq:kineq}) which is based on the systematic real-time diagrammatic
approach. Still, we believe the following may be instructive.

We start from the expression (\ref{eq:tauplusav}) for the visibility,
\begin{eqnarray}
  D (t) &= & {\tmop{tr}} \left[ (|+\rangle \langle - | \otimes
			  \mathbbm{1}_{SR}) \right. \nonumber\\
 & & \left. \times e^{- i H t} (|+\rangle\langle-| \otimes \rho_{S R}) e^{i H
  t} \right], \label{eq:d}
\end{eqnarray}
assuming the initial qubit-environment state factorizes and $H$ is given by
\Eq{eq:H}. We furthermore neglect the spin degree of freedom for the
SQD and thereby also the Coulomb interaction effect, i.e., the SQD takes two
charge states ${\color{black} |} 0 {\color{black} \rangle}$ and
${\color{black} |} 1 {\color{black} \rangle}$. As noted in {\color{black}
Sec. \ref{sec:semiclassical}} and {\color{black} Appendix \ref{app:validity}},
neither is essential for the coherent backaction. We next split up the
Hamiltonian $H = H_0 + H_T$ and expand \Eq{eq:d} in the tunneling $H_T$
of electrons between the SQD and the reservoir and vice versa. In $O (\Gamma)
\sim O (H_T^2)$, the following term contributes (there are more terms; we just
consider a relevant one):
\begin{eqnarray}
  D (t) & \sim & \underset{0 \leqslant t_1, t_2 \leqslant t}{\int d t_1 d t_2}
  \underset{S, R}{\tmop{tr}} [\langle +| e^{-
  i H_0 (t - t_1)} H_T e^{- i H_0 t_1}| +
  \rangle \nobracket \nonumber\\
  &  & \text{ \ \ \ \ \ } \otimes \rho_{S R} \langle -
  | e^{i H_0 (t - t_2)} H_T e^{i H_0 t_2}| -
  \rangle \nobracket]  \label{eq:drealtime}
\end{eqnarray}
Importantly, the tunneling process for the ket-evolution $
\langle +  | \leftarrow \langle +|$ may happen at a time $t_1$ different from the time $t_2$
for the tunneling process of the bra evolution $\langle -| \rightarrow | - \rangle$.
The coherent evolution in between is responsible for an additional phase
shift. To see this, we assume the SQD to be initially unoccupied, i.e.,
$\rho_{S R} = {\color{black} |} 0 {\color{black} \rangle} {\color{black}
\langle} 0 {\color{black} |} \otimes \rho_{R, 0}$, and to end up in a singly
occupied state ${\color{black} |} 1 {\color{black} \rangle} {\color{black}
\langle} 1 {\color{black} |}$. We assume again for simplicity ${\color{black}
\vecg{\Omega}} = \vec{0}$. Evaluating the trace over the electrodes, we get
\begin{eqnarray}
  D (t) & \sim & \underset{0 \leqslant t_2 \leqslant t_1 \leqslant t}{\int d
  t_1 d t_2} \sum_r \int d \omega f^+_r (\omega) \Gamma_r (\omega) e^{- i
  \frac{\lambda}{2} (t - t_1)} \nonumber\\
  &  & \times \left[ e^{- i \left( \varepsilon + \frac{\lambda}{2} - \omega
  \right) (t_1 - t_2)} + e^{- i \left( - \varepsilon + \frac{\lambda}{2} +
  \omega \right) (t_1 - t_2)} \right], \nonumber\\
  &  &  \label{eq:drec}
\end{eqnarray}
where the two terms in the bracket relate to the two cases, $t_1 \geqslant
t_2$, and $t_1 \leqslant t_2$ in \Eq{eq:drealtime}, respectively. It is
easy to show that the terms in the second line arise from SQD-qubit
coherences of the type ${\color{black} |} 1, + {\color{black} \rangle}
{\color{black} \langle} 0, - {\color{black} |}$ and ${\color{black} |} 0, +
{\color{black} \rangle} {\color{black} \langle} 1, - {\color{black} |}$,
respectively. Finally, the SQD reaches the singly occupied real state
${\color{black} |} 1 {\color{black} \rangle} {\color{black} \langle} 1
{\color{black} |}$, giving rise to the phase shift $e^{- i \lambda (t -
t_1)}$. The latter is related to the stochastic backaction \newer{(since the
time $t_1$ is random)}, while the former
phase shift contains the coherent backaction as we explain next. For this
purpose, we redefine $\tau \assign t_1 - t_2$, assume large $t$, and introduce
a small imaginary part into the exponentials to ensure convergence. Since
$\tau$ can take nearly all positive values for large $t$, the coherent phase
shift becomes approximately
\begin{eqnarray}
  & \sim & \int_0^{\infty} d \tau \left[ e^{- i \left( \varepsilon +
  \frac{\lambda}{2} - \omega - i 0 \right) \tau} + e^{- i \left( - \varepsilon
  + \frac{\lambda}{2} + \omega - i 0 \right) \tau} \right] \nonumber\\
  & = & (- i) \left[ \frac{1}{\varepsilon + \lambda / 2 - \omega - i 0} -
  \frac{1}{\varepsilon - \lambda / 2 - \omega + i 0} \right] \nonumber\\
  & \approx & - 2 \tmop{Im} \frac{1}{\varepsilon - \omega + i 0} - i
  \lambda \frac{\partial}{\partial \varepsilon} \tmop{Re} \frac{1}{\varepsilon
  - \omega + i 0}, 
\end{eqnarray}
where the last step holds in first order in $\lambda$. We ignore the first
part, which is independent of $\lambda$, and insert the second part into Eq.
(\ref{eq:drec}), which yields a term proportional to the renormalization
function. When taking the time derivative of \Eq{eq:drec}, it
reproduces the coherent backaction:
\begin{eqnarray}
  \dot{D} & \sim & (- i) \sum_r \frac{\Gamma_r \lambda}{T} \phi'_r \left(
  \frac{\varepsilon - \mu_r}{T} \right)  \text{ \ } = \text{ \ } - i \kappa 
\end{eqnarray}
which coincides with the effect of the term \ $\dot{{\color{black}
\vecg{\tau}}}^1 = \kappa \vecg{\lambda} \times {\color{black}
\vecg{\tau}}^0$ in the kinetic equation (\ref{eq:kineq}).\\
What our heuristic argument clarifies is why a semiclassical approach as
discussed in {\Sec{sec:semiclassical}} is not capable of
reproducing the coherent backaction: this approach crucially relies on the
assumption that the \tmtextit{charge} state of the SQD is a
\tmtextit{classical variable}, i.e. $n (t)$ is just a fixed (but random)
function of $t$. Thus, the transition $0 \rightarrow 1$ happens on both the
bra- and the ket-part of the evolution at the same time $t_1 = t_2$ (while the
time $t_1$ is random). In this case, the coherent phase shift just vanishes.
This amounts formally to replacing $1 / x \rightarrow - i \pi \delta (x)$ and
``by hand'' dropping the principal value integral term $P (1 / x)$ that
contains renormalization effects. In the single-step Born-Markov approach
discussed in {\Sec{sec:br}}, the problem is to find a
proper description of the environmental state that contains the intermediate
coherences ${\color{black} |} 1 {\color{black} \rangle} {\color{black}
\langle} 0 {\color{black} |}$ and ${\color{black} |} 0 {\color{black} \rangle}
{\color{black} \langle} 1 {\color{black} |}$.
\new{Our above nonrigorous derivation of the coherent backaction terms thus illustrates
that virtual fluctuations in nanoscale sensors can have a real impact on
the measurement backaction.}


\bibliographystyle{apsrev}
\end{document}